\newcommand{\CLASSINPUTinnersidemargin}{20mm}
\documentclass[journal,draftclsnofoot,onecolumn,12pt,twoside]{IEEEtran}
\usepackage{xcolor}
\definecolor{Mycolor}{RGB}{0,0,0}
\definecolor{Mycolor2}{RGB}{0,0,0}
\definecolor{mycolor3}{RGB}{0,0,0}
\definecolor{mycolor4}{RGB}{0,0,0}
\definecolor{mycolor5}{RGB}{0,0,0}

 \usepackage{cite}
  \usepackage[center]{caption}
\usepackage{algorithm} %ctan.org\pkg\algorithms
\usepackage{algpseudocode}
\usepackage[neverdecrease]{paralist}
\usepackage{graphicx}
\usepackage{amssymb}
\usepackage{amsfonts}
\usepackage[cmex10]{amsmath}
\usepackage{dsfont}
\usepackage{float}
\usepackage{url}
\usepackage{bm}
\usepackage{hhline}
\usepackage{enumerate}
\usepackage{pgf,tikz}
\usepackage{mathrsfs}
\usepackage{mathtools}
\usetikzlibrary{arrows}
\usetikzlibrary[patterns]

 \usepackage[colorlinks=true]{hyperref}
 \hypersetup{
    bookmarks=true,         % show bookmarks bar?
    unicode=false,          % non-Latin characters in Acrobat's bookmarks
    pdftoolbar=true,        % show Acrobat's toolbar?
    pdfmenubar=true,        % show Acrobat's menu?
    pdffitwindow=false,     % window fit to page when opened
    pdfstartview={FitH},    % fits the width of the page to the window
    pdftitle={My title},    % title
    pdfauthor={Author},     % author
    pdfsubject={Subject},   % subject of the document
    pdfcreator={Creator},   % creator of the document
    pdfproducer={Producer}, % producer of the document
    pdfkeywords={keyword1, key2, key3}, % list of keywords
    pdfnewwindow=true,      % links in new PDF window
    linkbordercolor=green,
    anchorcolor=black,filecolor=black,menucolor=black,runcolor=black,
    colorlinks=true,       % false: boxed links; true: colored links
    linkcolor=red,          % color of internal links (change box color with linkbordercolor)
    citecolor=blue,        % color of links to bibliography
    filecolor=magenta,      % color of file links
    urlcolor=cyan           % color of external links
}

\IEEEoverridecommandlockouts
\newtheorem{Theorem}{Theorem}
\newtheorem{Proposition}{Proposition}
\newtheorem{Corollary}{Corollary
}
\newtheorem{Definition}{Definition}
\newtheorem{Example}{Example}

\newtheorem{lemma}[Theorem]{Lemma}

\newtheorem{remark}{Remark}

\ifCLASSINFOpdf
\else
\fi
\begin{document}
\title{Design and Practical Decoding of Full-Diversity Construction A Lattices for Block-Fading Channels}
%-----------------------------------------------------------main affiliation
%\author{
%\IEEEauthorblockN{Hassan~Khodaiemehr}
%\IEEEauthorblockA{Department of Mathematics and Computer \\ Science,
%Amirkabir University of Technology\\
% E-mail: h.khodaiemehr@aut.ac.ir}
% \and
%\IEEEauthorblockN{Mohammad-Reza~Sadeghi}
%\IEEEauthorblockA{Department of Mathematics and Computer \\ Science,
%Amirkabir University of Technology\\
%Email: msadeghi@aut.ac.ir}
%\and
%\IEEEauthorblockN{Amin~Sakzad}
%\IEEEauthorblockA{Department of ECSE, Monash\\
%University, Victoria, Australia\\
%E-mail: amin.sakzad@monash.edu}
%}
%--------------------------------------------------------------
\author{Hassan Khodaiemehr,  Daniel Panario,~\IEEEmembership{Senior Member,~IEEE} and Mohammad-Reza~Sadeghi
%\author{
%,~\IEEEmembership{Member,~IEEE,}
\thanks{Hassan Khodaiemehr is with the Department of Computer Science and Statistics, Faculty of Mathematics, K. N. Toosi University of Technology, P. O. Box: 16765-3381,  Tehran, Iran, and also with the School of Mathematics, Institute for Research in Fundamental Sciences,
P. O. Box: 19395-5746, Tehran, Iran. Email: ha.khodaiemehr@kntu.ac.ir.

Daniel Panario is with the School of Mathematics and Statistics, Carleton University, Ottawa, Canada. Email: daniel@math.carleton.ca.

Mohammad-Reza~Sadeghi is with the Department
of Mathematics and Computer Science, Amirkabir University of Technology (Tehran Polytechnic), Tehran, Iran. Email: msadeghi@aut.ac.ir.

Part of this work has been presented in \cite{myISIT} at ISIT 2016, Spain. The research of the first author was supported by a grant from IPM
(No. 99050115). The second author is partially funded by NSERC of Canada.
}}

%\and
%\IEEEauthorblockN{Emanuele~Viterbo}
%\IEEEauthorblockA{Dept of ECSE, Monash University, Australia\\
%E-mail: emanuele.viterbo@monash.edu}}

%\author{Amin~Sakzad, and Mohammad-Reza~Sadeghi\thanks{A. Sakzad is with the ECSE Department at Monash University, Melbourne, Australia. M. R. Sadeghi is with the Faculty of Mathematics and Computer Science, Amirkabir University of Technology, Tehran, Iran.
%E-mails: amin.sakzad@monash.edu and msadeghi@aut.ac.ir.}}
\maketitle
\begin{abstract}
%LDPC lattices were the first family of lattices which have an efficient decoding algorithm in high dimensions over an AWGN channel.  Considering Construction D' of lattices with one binary LDPC code as underlying code gives the well known Construction A LDPC lattices or   LDPC lattices.
Block-fading channel (BF) is a useful model for various
wireless communication channels in both indoor and outdoor
environments. Frequency-hopping schemes and orthogonal
frequency division multiplexing (OFDM) can conveniently
be modelled as BF channels. Applying lattices in this type of channel entails dividing a lattice point into
multiple blocks such that fading is constant within a block
but changes, independently,  across blocks. The design of lattices for BF channels offers a
challenging problem, which differs greatly from its counterparts like
AWGN channels. Recently, the original binary Construction A for lattices, due to Forney, has been generalized to a lattice construction from totally real and complex multiplication (CM) fields. This generalized algebraic Construction A of lattices provides signal space diversity, intrinsically, which is the main requirement for the signal sets designed for fading channels.
In this paper, we construct full-diversity algebraic lattices for BF channels using Construction A over totally real number fields. We propose two new decoding methods for these family of lattices which have complexity that grows linearly in the dimension of the lattice.
The first decoder is proposed for full-diversity algebraic LDPC lattices which are generalized Construction A lattices with a binary LDPC code as underlying code. This decoding method contains iterative and non-iterative phases.
In order to implement the iterative phase of our
decoding algorithm, we propose the definition of a parity-check matrix and Tanner graph for full-diversity algebraic Construction A lattices. We also prove that \textcolor{mycolor5}{using an underlying LDPC code that achieves the  outage probability limit over one-block-fading channel,} the constructed algebraic LDPC lattices together with the proposed decoding method admit diversity order $n$ over an $n$-block-fading channel.  Then, we modify the proposed algorithm  by removing its iterative phase
which enables full-diversity practical decoding of all generalized Construction
A lattices without any assumption about their underlying code.
In contrast with the known results on AWGN channels in which non-binary Construction A lattices \textcolor{mycolor3}{always outperform the binary ones, we  provide some instances showing that algebraic Construction A lattices obtained from binary codes outperform the ones based on non-binary codes in block fading channels}.
Since available lattice construction methods from totally real and complex multiplication (CM) fields do not provide diversity in the binary case,  we generalize algebraic Construction A lattices over a wider family of
number fields namely monogenic number fields.
\end{abstract}
%\IEEEpeerreviewmaketitle
\section{Introduction}
\IEEEPARstart{A}{} lattice in $\mathbb {R}^{N}$ is an additive subgroup of $\mathbb {R}^{N}$ which is isomorphic to $\mathbb {Z}^{N}$ and spans the real vector space $\mathbb {R}^{N}$ \cite{2}. Lattices have been extensively addressed for the problem of coding in  additive white Gaussian noise (AWGN) channels.
\textcolor{mycolor4}{
In these cases, we regard an infinite lattice  as a code without restrictions employed for the AWGN channel~\cite{polytrev}.
%In communication over AWGN channels using lattices, there is no power constraint.
%Such unconstrained communication has been investigated by Poltyrev~\cite{polytrev}
}
%In such a communication system, instead of coding rate and capacity, normalized logarithmic density (NLD) and generalized capacity $C_{\infty}$ are used, respectively.

There exist different methods to construct lattices. One of the most distinguished ones is constructing lattices based on codes, where Construction A, D and D' have been proposed (for details see, for example, \cite{2}). In \cite{forneyspherebound}, it is shown that the sphere bound can be approached by a large class of coset codes or multilevel coset codes with multistage decoding,
including Construction D lattices and other certain binary lattices. Their results are based on  channel coding theorems of information theory.
%In one side, Forney {\em et al.} proved  the existence of sphere-bound-achieving and capacity-achieving lattices via Construction D,
As a result of their study, the concept of volume-to-noise (VNR) ratio was introduced as a parameter for measuring the efficiency of lattices  \cite{forneyspherebound}.
The subsequent challenge in lattice theory has been to find structured classes of lattices that can be encoded
and decoded with reasonable complexity in practice, and with performance that can approach the sphere-bound. This results in the transmission with arbitrary small error probability whenever VNR approaches to $1$. A capacity-achieving lattice can raise to a capacity-achieving lattice code by selecting a proper shaping region~\cite{erez,urbanke}.

Applying maximum-likelihood (ML) decoding for lattices in high dimensions  is infeasible and forced researchers to apply other low complexity decoding methods for lattices to obtain practical capacity-achieving lattices.
Integer lattices built by Construction A, D and D' can be decoded with linear complexity based on soft-decision decoding of their underlying linear binary and non-binary codes \cite{sloane,sadeghi,19,20,21,IWCIT2015,QC_LDPC_lattice,Hassan2,HassanISC1}. %Therefore, finding the lattices of high enough dimension $N$ which have a decoder with reasonable complexity, enabling transmission with arbitrary small error probability whenever VNR approaches $1$, is a promising research area.
The search for sphere-bound-achieving and capacity-achieving lattices and lattice codes followed by proposing low density parity-check (LDPC) lattices \cite{sadeghi},  low density lattice codes (LDLC)~\cite{LDLC}, integer low-density lattices based on Construction A (LDA)~\cite{19} and polar lattices \cite{polar}.
\textcolor{mycolor3}{In \cite{Leech}, the authors have introduced Leech-shaped LDA
constellations by employing the direct sum of
a low-dimensional sublattice as a shaping region for LDA lattices to get significant shaping gain and reaching a gap to capacity of $0.8$ dB with $2.7$
bits/dim.}
%Turbo lattices, based on Construction D \cite{sakzad10}, and
 %are other families of lattices with practical decoding methods.

%Among the above family of lattices, LDPC lattices are those that have sparse parity-check matrices, obtained by using a set of nested binary LDPC codes as underlying codes, together with Construction D'. If the number of underlying LDPC codes (or the level of construction) is one, Construction D'  coincides with Construction A  and   LDPC lattices are obtained \cite{IWCIT}.
The theory behind Construction A is well understood.
There is a series of dualities between theoretical properties of the underlying codes and their resulting lattices. For
example there are connections between the dual of the code and the dual of the lattice, or between the weight enumerator of the code and the theta series of the lattice \cite{2,ebeling}. Construction A has been generalized
in different directions; for example a generalized construction
from the cyclotomic field $\mathbb{Q}(\xi_p)$, $\xi_p=e^{2\pi i/p}$ and $p$ a prime, is presented
in \cite{ebeling}. Then, in \cite{ConstA}, a generalized construction of
lattices over a number field from linear codes is proposed.  There is consequently a rich
literature studying Construction A over different alphabets and for different tasks.

%
%Forney {\em et al.} ~\cite{forneyspherebound} proved the existence of sphere-bound-achieving and capacity-achieving lattices via Construction D lattices theoretically. They also established the concept of volume-to-noise (VNR) ratio as a parameter for measuring the efficiency of lattices. Therefore, finding the lattices of high enough dimension $n$ which have a decoder with reasonable complexity, enabling transmission with arbitrary small error probability whenever VNR approaches $1$ is a promising research area. A capacity-achieving lattice can raise to a capacity-achieving lattice code by selecting a proper shaping region~\cite{erez,urbanke}.

%The search for sphere-bound-achieving and capacity-achieving lattices and lattice codes has begun with~\cite{sadeghi}. Low density parity-check (LDPC) lattices are those that have sparse parity-check matrices. In this class of lattices, a set of nested LDPC codes are used to generate lattices with sparse parity-check matrices. Another class of lattices, so-called low density lattice codes (LDLC) introduced and investigated in~\cite{loeligerLDLC} and \cite{LDLC}. Turbo lattices employed Construction D along with turbo codes to achieve capacity gains~\cite{sakzad10} and \cite{sakzadmanuscript}. Integer low-density lattices based on construction A which are known as LDA lattices \cite{19} and polar lattices~\cite{polar}, are another families of lattices with practical decoding method.

Lattices have  been also considered for transmission over fading channels. Specifically, algebraic lattices, defined as
lattices obtained via the ring of integers of a number field, provide efficient modulation schemes~\cite{alglattice1} for fast Rayleigh fading channels. Families of algebraic lattices are known
to reach full-diversity, the first design criterion for fading
channels; see the definition of full-diversity in Section \ref{system_model_subsec_2}. Algebraic lattice codes are then natural candidates for the design of codes for block-fading (BF) channels.

The block-fading channel \cite{blockfading} is a useful channel model for a class of slowly-varying wireless communication channels. Frequency-hopping schemes and orthogonal frequency division multiplexing
(OFDM), applied in many wireless communication systems standards, can conveniently
be modelled as BF channels.  In a BF channel a codeword spans a finite number
$n$ of independent fading blocks. As the channel realizations are constant within blocks, no codeword
is able to experience all the states of the channel; this implies that the channel is non-ergodic and therefore it is
not information stable. It follows that the Shannon capacity of this channel is zero \cite{rootLDPC}.

\textcolor{Mycolor}{Based on Poltyrev's work on infinite lattices for AWGN  channels, a Poltyrev outage limit (POL) in presence of block fading has been presented in
\cite{outage,punekar} for lattices. The diversity order of this POL is the same as the number of fading blocks in the
channel. In addition, a family of full-diversity low-density lattices (LDLC)  suited under maximum-likelihood
decoding has been presented in \cite{outage}. Next, the authors proposed a full-diversity lattice construction
for sparse integer parity-check matrices capable to use iterative probabilistic decoding \cite{punekar}. In both cases, the full-diversity property has been proven theoretically. Construction methods in  \cite{punekar} are provided for diversity order at most $4$.
Using optimal decoders for decoding lattices on BF channels implies exponential complexity in the worst-case.
%As far as we are aware, all available lattice based schemes on BF channels were proposed by using optimal decoders \cite{outage} which have exponential complexity in the worst-case.
}

\textcolor{Mycolor}{In this paper we propose a general framework to design full-diversity (binary and non-binary) Construction A lattices and their practical decoding methods. In the binary case, in which the underlying code is a binary LDPC code, our proposed decoding is a combination of optimal decoding in small dimensions and iterative decoding \cite{myISIT}. Next we generalize this decoding algorithm to the non-binary case in which we also remove any assumption about the underlying code. Indeed, by using the proposed framework in this paper, not only the LDPC codes but any linear code can be employed to construct full-diversity Construction A lattices for which decoding is provided with linear complexity in the dimension of lattice.
The proposed decoding algorithms preserve the diversity order of the lattice and make it tractable to decode high-dimension full-diversity lattices on the BF channel.}

The rest of this paper is organized as follows. In Section~\ref{Preliminaries}, we provide preliminaries about lattices and algebraic number theory.
In Section~\ref{lattices_and_codes}, we present the available methods for constructing full-diversity lattices from totally
real number fields.
The introduction of the full-diversity   algebraic LDPC lattices is also given.
In Section~\ref{system_model}, the system model is described for the Rayleigh BF channel. The available methods for evaluating the performance of finite and infinite lattice constellations over fading and block-fading channels are also discussed in this section. The design criteria of Construction A lattices with good error performance over BF channels is also given in this section.
In Section~\ref{monogenic_sec}, the introduction of  monogenic number fields, as the tools for constructing  full-diversity   Construction A lattices with binary underlying code, is provided.
In Section~\ref{new_construction}, our construction of full-diversity lattices is given.
In Section~\ref{decod2}, a new iterative decoding method is proposed for full-diversity   algebraic LDPC lattices in high dimensions. The analysis of the proposed decoding method is also given in this section.
In Section~\ref{non_binary_decoding_sec}, a non-iterative decoder is proposed which enables
full-diversity practical decoding of all generalized Construction
A lattices without any assumption about their underlying code.
In Section~\ref{Numerical_Results}, we give computer simulations, providing decoding performance of both algorithms and a comparison against available bounds and other counterparts like LDLCs.
%We conclude with Section~\ref{conclusion}.
Section~\ref{conclusion} contains  concluding remarks.

\textbf{Notation}: Matrices and vectors are denoted by bold upper
and lower case letters, respectively. The $i$th element of vector $\mathbf{a}$ is denoted
by $a_i$ or $\mathbf{a}(i)$ and the  entry $(i,j)$ of a matrix $\mathbf{A}$ is denoted by
$A_{i,j}$; $[\,\,]^t$ denotes the transposition for vectors and matrices. \textcolor{mycolor3}{For a vector $\mathbf{x}$ of length $n$ and $1\leq i<j\leq n$, the notation $\mathbf{x}(i:j)$ is used throughout the paper to indicate the subvector of $\mathbf{x}$ made of its coordinates from $i$~to~$j$.}
\section{Preliminaries on Lattices and Algebraic Number Theory}~\label{Preliminaries}
In order to make this work self-contained, general notations and basic definitions of algebraic number theory and lattices are given next. We reveal the connection between lattices and algebraic number theory at the end of this section.
\subsection{Algebraic number theory}
Let $K$ and $L$ be two fields. If $K\subset L$, then $L$ is a field extension of $K$ denoted by $L/K$. The dimension of $L$ as vector space over $K$ is  the degree of $L$ over $K$, denoted by
$[L : K]$. Any finite extension of $\mathbb{Q}$ is  a number field.

Let $L/K$ be a field extension, and let $\alpha\in L$. If there
exists a non-zero irreducible monic  polynomial $p_{\alpha} \in K[x]$ such that $p_{\alpha}(\alpha) = 0$,  $\alpha$ is algebraic over $K$. Such a polynomial is  the minimal polynomial of $\alpha$ over $K$. If all the elements of $L$ are algebraic over $K$,  $L$ is an algebraic extension of $K$.
\begin{Definition}
Let $K$ be an algebraic number field of degree $n$; $\alpha\in K$ is an algebraic integer if it is a root of a monic polynomial with coefficients in $\mathbb{Z}$. The set of algebraic integers of $K$ is  the ring of integers of $K$, denoted by $\mathcal{O}_K$. The ring $\mathcal{O}_K$ is also called the maximal order of $K$.
\end{Definition}

If $K$ is a number field, then $K = \mathbb{Q}(\theta)$ for an algebraic integer $\theta\in \mathcal{O}_K$ \cite{Stewart}. For a number field $K$ of degree $n$, the ring of integers $\mathcal{O}_K$ forms a free $\mathbb{Z}$-module of rank $n$.
\begin{Definition}
Let $\left\{\omega_1,\ldots,\omega_n\right\}$ be a basis of the $\mathbb{Z}$-module $\mathcal{O}_K$, so that
we can uniquely write any element of $\mathcal{O}_K$ as $\sum_{i=1}^n a_i\omega_i$ with $a_i \in \mathbb{Z}$ for all $i$. Then, $\left\{\omega_1,\ldots,\omega_n\right\}$ is an integral basis of $K$.
\end{Definition}
\begin{Theorem}{\cite[ p. 41]{Stewart}}
Let $K = \mathbb{Q}(\theta)$ be a number field of degree $n$ over $\mathbb{Q}$. There are exactly $n$ embeddings $\sigma_1,\ldots, \sigma_n$ of $K$ into $\mathbb{C}$ defined by $\sigma_i(\theta) = \theta_i$, for $i = 1,\ldots, n$, where the $\theta_i$'s are the distinct zeros in $\mathbb{C}$ of the minimal polynomial of $\theta$ over $\mathbb{Q}$.
\end{Theorem}
\begin{Definition}
Let $K$ be a number field of degree $n$ and $x\in K$. The elements $\sigma_1(x),\ldots, \sigma_n(x)$ are
the conjugates of $x$ and
\begin{equation}\label{norm}
  N_{K/\mathbb{Q}}(x)=\prod_{i=1}^n \sigma_i(x),\quad \mathrm{Tr}_{K/\mathbb{Q}}(x)=\sum_{i=1}^n \sigma_i(x),
\end{equation}
are the norm and the trace of $x$, respectively.
\end{Definition}

For any $x \in K$, we have $N_{K/\mathbb{Q}}(x),\mathrm{Tr}_{K/\mathbb{Q}}(x)\in\mathbb{Q}$. If $x\in \mathcal{O}_K$, we have $N_{K/\mathbb{Q}}(x),\mathrm{Tr}_{K/\mathbb{Q}}(x)\in\mathbb{Z}$.
\begin{Definition}
%Let $\left\{\omega_1,\ldots,\omega_n\right\}$ be an integral basis of $K$. The discriminant of $K$ is defined as
%\begin{equation}\label{disc}
%d_K = (\det[(\sigma_j(\omega_i))_{i,j=1}^n])^2.
%\end{equation}
Let $\left\{\omega_1,\ldots,\omega_n\right\}$ be an integral basis of $K$. The discriminant of $K$ is defined as
\begin{equation}\label{disc}
d_K =\det(\mathbf{A})^2,
\end{equation}
where $\mathbf{A}$ is the matrix $A_{i,j}=\sigma_j(\omega_i)$, for $i,j=1,\ldots,n$.
\end{Definition}

The discriminant of a number field belongs to $\mathbb{Z}$ and it is independent of the choice of
 basis.
\begin{Definition}
Let $\left\{\sigma_1,\ldots, \sigma_n\right\}$ be the $n$ embeddings of $K$ into $\mathbb{C}$. Let $r_1$ be the number of embeddings with image in $\mathbb{R}$, the field of
real numbers, and $2r_2$ the number of embeddings with image in $\mathbb{C}$ so
that $r_1 + 2r_2 = n$. The pair $(r_1, r_2)$ is  the signature of $K$. If $r_2 = 0$ we have a totally
real algebraic number field. If $r_1 = 0$ we have a totally complex algebraic number field.
\end{Definition}
\begin{Definition}
Let us order the $\sigma_i$'s so that, for all $x\in K$, $\sigma_i(x)\in\mathbb{R}$, $1 \leq i \leq r_1$, and $\sigma_{j+r_2}(x)$ is the complex conjugate of $\sigma_j(x)$ for
$r_1 + 1 \leq j \leq r_1 + r_2$. The canonical embedding $\sigma : K \rightarrow \mathbb{R}^{r_1} \times \mathbb{C}^{r_2}$ is the homomorphism defined by
\begin{equation}\label{embeding}
  \sigma(x)=(\sigma_1(x),\ldots,\sigma_{r_1}(x),\sigma_{r_1+1}(x),\ldots, \sigma_{r_1+r_2}(x)).
  %\in \mathbb{R}^{r_1}\times \mathbb{C}^{r_2}.
\end{equation}
\end{Definition}

If we identify $\mathbb{R}^{r_1}\times \mathbb{C}^{r_2}$ with $\mathbb{R}^n$, the canonical embedding can be
rewritten as $\sigma :K \rightarrow\mathbb{R}^n$
\begin{eqnarray}\label{embeding2}
  \sigma(x)&=&(\sigma_1(x),\ldots,\sigma_{r_1}(x),\Re\sigma_{r_1+1}(x),\Im\sigma_{r_1+1}(x),\ldots,\Re\sigma_{r_1+r_2}(x),\Im\sigma_{r_1+r_2}(x)),
  %\in \mathbb{R}^{n}.
\end{eqnarray}
where $\Re$ denotes the real part and $\Im$ the imaginary part.
\begin{Definition}
A ring $A$ is  integrally closed in a field $L$ if every element
of $L$ which is integral over $A$ in fact lies in $A$. A ring is
integrally closed if it is integrally closed in its quotient field.
\end{Definition}
\begin{Theorem}{\cite[p. 18]{serglang}}\label{dedekind}
\textcolor{mycolor3}{Let $D$ be a Noetherian ring, that is, there is no infinite strictly ascending sequence of ideals in $D$. In addition, let $D$ be integrally closed and such that every non-zero prime ideal of $D$ is maximal. Then every ideal of $D$ can be uniquely factored into prime ideals.}
\end{Theorem}

A ring satisfying the properties of Theorem~\ref{dedekind} is \textcolor{mycolor3}{called} a Dedekind ring.
The ring of algebraic integers in a number field is a Dedekind ring.
\begin{Definition}
Let $A$ be a ring and $x$ an element of some field $L$ containing $A$. Then, $x$ is integral over $A$ if either one of the following two conditions is satisfied:
\begin{enumerate}
  \item there exists a finitely generated non-zero $A$-module $M\subset L$ such
that $xM \subset M$;
  \item the element $x$ satisfies an equation
  \begin{equation*}
    x^n+a_{n-1}x^{n-1}+\cdots +a_0=0,
  \end{equation*}
with coefficients $a_i\in A$, and $n\geq 1$. Such an equation is an integral equation.
\end{enumerate}
\end{Definition}

Let $A$ be a Dedekind ring, $K$ its quotient field, $L$ a finite separable
extension of $K$ (that is, for every $\alpha \in L$, the minimal polynomial of  $\alpha$
 over $K$ has non-zero formal derivative), and $B$ the integral closure of $A$ in $L$. If $\mathfrak{p}$ is a prime ideal
of $A$, then $\mathfrak{p}B$ is an ideal of $B$ and has a factorization
\begin{equation}\label{factorization}
\mathfrak{p}B=\mathfrak{P}_1^{e_1}\cdots \mathfrak{P}_r^{e_r},
\end{equation}
into primes of $B$, where $e_i\geq 1$. It is clear that a prime $\mathfrak{P}$ of $B$ occurs in this factorization if and only if $\mathfrak{P}$ lies above $\mathfrak{p}$. Each $e_i$ is  the ramification index of $\mathfrak{P}_i$ over $\mathfrak{p}$, and is also written $e(\mathfrak{P}_i/\mathfrak{p})$.
If $\mathfrak{P}$ lies above $\mathfrak{p}$ in $B$, we denote by $f(\mathfrak{P}/\mathfrak{p})$ the degree of the residue class field extension $B/\mathfrak{P}$ over $A/\mathfrak{p}$, and call it the residue class degree or inertia degree.
\begin{Theorem}{\cite[p. 24]{serglang}}
Let $A$ be a Dedekind ring, $K$ its quotient field, $L$ a
finite separable extension of $K$, and $B$ the integral closure of $A$ in $L$. Let
$\mathfrak{p}$ be a prime ideal  of $A$. Then
\begin{equation}\label{ramification}
[L:K]=\sum_{\mathfrak{P}|\mathfrak{p}}e(\mathfrak{P}/\mathfrak{p}) f(\mathfrak{P}/\mathfrak{p}).
\end{equation}
\end{Theorem}

When $L/K$ is a Galois extension of degree $n$, \eqref{ramification} simplifies to $n =efg$, where $g$ is the number of primes $\mathfrak{P}$ of $B$ above $\mathfrak{p}$. In other words, $e(\mathfrak{P}/\mathfrak{p}) = e$ and $f(\mathfrak{P}/\mathfrak{p}) = f$ for all $\mathfrak{P}|\mathfrak{p}$. If $e_{\mathfrak{P}} =f_{\mathfrak{P}}=1$  for all $\mathfrak{P}|\mathfrak{p}$, then  $\mathfrak{p}$ splits completely in $L$. In that case, there are exactly $[L: K]$ primes of $B$ lying above $\mathfrak{p}$.
A prime $\mathfrak{p}$ in $K$ is \emph{ramified} in a number field $L$  if the prime ideal factorization~(\ref{factorization}) has some $e_i$ greater than $1$. If every $e_i$ equals $1$, $\mathfrak{p}$ is \emph{unramified} in $L$.
If $[L: K] = e(\mathfrak{P}/\mathfrak{p})$, $\mathfrak{P}$ is \emph{totally ramified} above $\mathfrak{p}$. In this case, the residue class degree is equal to $1$. Since $\mathfrak{P}$ is the only prime of $B$ lying above $\mathfrak{p}$,  $L$ is totally ramified over $K$.  If the characteristic $p$ of the residue
class field $A/\mathfrak{p}$ does not divide $e(\mathfrak{P}/\mathfrak{p})$, then $\mathfrak{P}$ is \emph{tamely ramified} over $\mathfrak{p}$ (or $L$ is tamely ramified over $K$). If it does, then $\mathfrak{P}$ is \emph{strongly ramified}.
\subsection{Lattices}~\label{A.Lattices}
%============================================================new
Any discrete additive subgroup $\Lambda$ of the $m$-dimensional real space $\mathbb{R}^m$ is a lattice.  Every lattice $\Lambda$ has a basis $\mathcal{B}=\{{\mathbf b}_1,\ldots,{\mathbf b}_n\}\subseteq\mathbb{R}^m$,  $n \leq m$, where
\textcolor{mycolor3}{the vectors of the basis (the $\mathbf{b}_i$'s) are linearly independent and}
 every ${\mathbf x}\in\Lambda$ can be represented as an integer linear combination of vectors in $\mathcal{B}$. The $n\times m$ matrix $\mathbf{M}$ with ${\mathbf b}_1,\ldots,{\mathbf b}_n$ as rows, is a generator matrix for the lattice.
The rank of the lattice is $n$ and its dimension is $m$.
If $n = m$, the lattice is  a full-rank lattice. In this paper, we consider only full-rank lattices.
A lattice $\Lambda$  can be described in terms of a generator matrix $\mathbf{M}$ by
\begin{equation}\label{lattice_def}
  \Lambda=\left\{\mathbf{x}=\mathbf{uM}\,|\,\mathbf{u}\in\mathbb{Z}^{n}\right\}.
\end{equation}
When using lattices for coding, their Voronoi cells and volume always play an important role. For any lattice
point $\mathbf{p}$ of a lattice $\Lambda\subset \mathbb{R}^m$, its Voronoi cell is defined by
\begin{equation}
\mathcal{V}_{\Lambda}(\mathbf{p})=\left\{\mathbf{x}\in \mathbb{R}^{m},\,d(\mathbf{x},\mathbf{p})\leq d(\mathbf{x},\mathbf{q})\,\, \textrm{for}\,\, \textrm{all} \,\, \mathbf{q}\in\Lambda\right\},
\end{equation}
where $d(\mathbf{x},\mathbf{y})$, for $\mathbf{x},\mathbf{y}\in\mathbb{R}^{m}$ denotes the Euclidean distance between $\mathbf{x}$ and $\mathbf{y}$.  All Voronoi cells are translates of the Voronoi cell  around the origin which is denoted by $\mathcal{V}_{\Lambda}(\mathbf{0})\vcentcolon = \mathcal{V}(\Lambda)$.
The matrix $\mathbf{G} = \mathbf{M}\mathbf{M}^t$ is  a Gram matrix for the lattice.
\begin{Definition}
An integral lattice $\Gamma$ is a free $\mathbb{Z}$-module of finite rank together with a positive definite symmetric bilinear form $\left\langle ,\right\rangle:\Gamma\times \Gamma \rightarrow \mathbb{Z}$.
\end{Definition}
\begin{Definition}
The discriminant of a lattice $\Gamma$, denoted
$\mathrm{disc}(\Gamma)$, is the determinant of $\mathbf{G}=\mathbf{M}\mathbf{M}^t$ where $\mathbf{M}$ is a generator matrix for $\Gamma$. The volume $\textrm{vol}(\Gamma)$ of a lattice $\Gamma$ is defined as $|\det(\mathbf{M})|=\sqrt{\det(\mathbf{G})}$.
\end{Definition}

The discriminant is related to the volume of a lattice by
\begin{equation}\label{disc}
  \textrm{vol}(\Gamma)=\sqrt{\mathrm{disc}(\Gamma)}.
\end{equation}
Moreover, when $\Gamma$ is integral, we have $\mathrm{disc}(\Gamma) = |\Gamma^{*}/\Gamma|$, where $\Gamma^{*}$ is the dual of the lattice $\Gamma$ defined by
\begin{equation}\label{dual}
  \Gamma^{*}=\left\{y\in \mathbb{R}^m\,\,|\,\,y\cdot x\in \mathbb{Z} \,\,\,\mbox{for}\,\,\mbox{all}\,\,\, x\in \Gamma\right\}.
\end{equation}
When $\Gamma = \Gamma^{*}$, the lattice $\Gamma$ is unimodular.

The canonical embedding~(\ref{embeding2})  gives a geometrical representation of a number
field and makes the connection between algebraic number fields and lattices.
\begin{Theorem}{\cite[p. 155]{Stewart}}
Let $\left\{\omega_1, \ldots,\omega_n\right\}$ be an integral basis of a number field $K$. The $n$ vectors $\mathbf{v}_i = \sigma(\omega_i) \in \mathbb{R}^n$, $i = 1,\ldots,n$ are linearly independent,
so they define a full rank \textcolor{mycolor3}{algebraic} lattice $\Lambda = \Lambda(\mathcal{O}_K) = \sigma(\mathcal{O}_K)$.
\end{Theorem}
\begin{Theorem}{\cite{samuel}}\label{vol_theorem}
Let $d_K$ be the discriminant of a number field $K$. The volume of the fundamental parallelotope of $\Lambda(\mathcal{O}_K)$ is given by
\begin{equation}\label{vol_alg}
  \textrm{vol}(\Lambda(\mathcal{O}_K))=2^{-r_2}\sqrt{|d_K|}.
\end{equation}
\end{Theorem}
\section{Lattice Constructions using Codes} \label{lattices_and_codes}
There exist many ways to construct lattices based on codes~\cite{2}. Here we mention  a lattice construction from totally real and complex multiplication fields \cite{ConstA}, which naturally generalizes
Construction A of lattices from $p$-ary codes obtained from the
cyclotomic field $\mathbb{Q}(\xi_p)$, with $\xi_p=e^{2\pi i/p}$ and $p$ a prime number \cite{ebeling}. This contains the
so-called Construction A of lattices from binary codes as a particular case.
\subsection{Algebraic Construction A lattices}~\label{algebraic_const_A}
Given a number field $K$ and a prime $\mathfrak{p}$  of $\mathcal{O}_K$ above $p$ where $\mathcal{O}_K/\mathfrak{p}\cong \mathbb{F}_{p^f}$, let $\mathcal{C}$ be an $(N, k)$ linear code over $\mathbb{F}_{p^f}$. The \textcolor{mycolor3}{algebraic} Construction A of lattices for block fading coding using the underlying code $\mathcal{C}$ and a number field $K$  is given in \cite{ConstA}.
\begin{Definition}
Let $\rho : \mathcal{O}_K^N\rightarrow \mathbb{F}_{p^f}^{N}$ be the mapping defined by the reduction modulo the ideal $\mathfrak{p}$ in each of the $N$ coordinates. Define  \textcolor{mycolor3}{algebraic Construction A lattice}  $\Gamma_{\mathcal{C}}$ to be the preimage of $\mathcal{C}$ in $\mathcal{O}_K^N$, that is,
\begin{equation}\label{alg_costA}
\Gamma_{\mathcal{C}}=\left\{\mathbf{x}\in \mathcal{O}_K^N\,\,|\,\, \rho(\mathbf{x})=\mathbf{c},\,\,  \mathbf{c}\in \mathcal{C}\right\}.
\end{equation}
\end{Definition}

We conclude that $\Gamma_{\mathcal{C}}$ is a $\mathbb{Z}$-module of rank $nN$.
%There are variations of the bilinear form $\left\langle ,\right\rangle :\Gamma_{\mathcal{C}}\times \Gamma_{\mathcal{C}}\rightarrow \mathbb{Z}$ depending on the field $K$ and the code $\mathcal{C}$.
When $K$ is totally real, $\rho^{-1}(\mathcal{C})$ forms a lattice with the following symmetric bilinear form \cite{ConstA}
\vspace{-0.2cm}
\begin{equation}\label{bilinear}
  \left\langle x,y\right\rangle=\sum_{i=1}^N \mbox{Tr}_{K/\mathbb{Q}}(\alpha x_i y_i),
\end{equation}
where $\mathbf{x}=(x_1,\ldots,x_N)$ and $\mathbf{y}=(y_1,\ldots, y_N)$ are vectors in $\mathcal{O}_K^N$, $\alpha\in \mathcal{O}_K$ is a totally positive element, meaning that $\sigma_i(\alpha)> 0$ for all $i$, and $\mbox{Tr}_{K/\mathbb{Q}}$ is defined in (\ref{norm}). Thus, $\Gamma_{\mathcal{C}}$ together with the bilinear
form~(\ref{bilinear}) is an integral lattice. A similar construction is obtained from a CM-field \cite{ConstA}. A
CM-field is a totally imaginary quadratic extension of a totally
real number field. If $K$ is a CM-field and $\alpha \in  \mathcal{O}_K \cap \mathbb{R}$ is totally positive, then $\rho^{-1}(\mathcal{C})$ forms a lattice with the following symmetric bilinear form
\begin{equation}\label{bilinear2}
  \left\langle x,y\right\rangle=\sum_{i=1}^N \mathrm{Tr}_{K/\mathbb{Q}}(\alpha x_i \bar{y_i}),
\end{equation}
where $\bar{y_i}$ denotes the complex conjugate of $y_i$. If $K$ is totally real, then $\bar{y_i}=y_i$, and this
notation treats both cases of totally real and CM-fields at the same time. It has been shown \cite{ConstA} that if $\mathcal{C}\subset \mathcal{C}^{\bot}$, then $\sum_{i=1}^N \mathrm{Tr}_{K/\mathbb{Q}}(x_i \bar{y_i})\in p\mathbb{Z}$, and thus  the symmetric bilinear form can be normalized by a factor $1/p$, or equivalently, by choosing $\alpha = 1/p$.

Other variations of the above construction have been considered
in the literature. The case $N = 1$ is considered in \cite{ref18} where  the problem reduces to understanding which lattices can be obtained on the ring of integers of a number field. The case  that $K$ is the cyclotomic field $\mathbb{Q}(\xi_p)$ has been considered in \cite{ebeling}. In \cite{ref12}, the prime ideal $\mathfrak{p}$ is considered to be $(2m)$, yielding codes over a ring of polynomials
with coefficients modulo $2m$. In \cite{ref19}, $\mathfrak{p}$ is considered to be $(2-\xi_p+\xi_p^{-1})$ and the resulting codes are over $\mathbb{F}_p$. Quadratic extensions $K = \mathbb{Q}(\sqrt{-l})$ are
considered in \cite{ref13} and \cite{ref20} where the reduction is done by the ideal $(p^e)$ and the resulting codes are over the ring $\mathcal{O}_K/p^eO_K$.

A generator matrix for the lattice $\Gamma_{\mathcal{C}}$ is computed in \cite{ConstA}. Let $K$ be a Galois extension and
the prime $\mathfrak{p}$ be chosen so that $\mathfrak{p}$ is totally ramified. Therefore, we have $p\mathcal{O}_K = \mathfrak{p}^n$. Now, let $\mathcal{C} \subset \mathbb{F}_p^N$ be a linear code over $\mathbb{F}_p$ of length $N$.
Since $\Gamma_{\mathcal{C}}$ has rank $nN$ as a free $\mathbb{Z}$-module, we  obtain the $\mathbb{Z}$-basis of $\Gamma_{\mathcal{C}}$. Let $\left\{\omega_1,\ldots,\omega_n\right\}$ be a $\mathbb{Z}$-basis of $\mathcal{O}_K$. Then,
a generator matrix for the lattice formed by $\mathcal{O}_K$ together with
the standard trace form $\left\langle w, z\right\rangle = \mathrm{Tr}_{K/\mathbb{Q}}(wz)$, $w, z\in \mathcal{O}_K$, is given by
\begin{equation}\label{gen_OK}
  \mathbf{M}=[\sigma_j(\omega_i)]_{i,j=1}^{n}.
\end{equation}
The prime ideal $\mathfrak{p}$ is a $\mathbb{Z}$-module of rank $n$. It then has a $\mathbb{Z}$-basis $\left\{\mu_1,\ldots, \mu_n\right\}$ where $\mu_i=\sum_{j=1}^n \mu_{i,j}\omega_j$. Thus
\begin{equation}\label{gen_P}
  [\sigma_j(\mu_i)]_{i,j=1}^{n}=\mathbf{D}\mathbf{M},
\end{equation}
where $\mathbf{D}=[\mu_{i,j}]_{i,j=1}^{n}$.
\begin{Theorem}{\cite[Proposition 1]{ConstA}}\label{theorem6}
The \textcolor{mycolor3}{algebraic lattice} $\Gamma_{\mathcal{C}}$ is a sublattice of $\mathcal{O}_K^N$
with discriminant
\begin{equation}\label{disc}
  \mathrm{disc}(\Gamma_{\mathcal{C}})=d_K^N(p^f)^{2(N-k)},
  \end{equation}
where $d_K = (\det([\sigma_i(\omega_j)]_{i,j=1}^n))^2$ is the discriminant of $K$. The
lattice $\Gamma_{\mathcal{C}}$ is given by the generator matrix
\begin{equation}\label{Gamma_C_gen}
  \mathbf{M}_{\Lambda}=\left[
                 \begin{array}{cc}
                   \mathbf{I}_k\otimes \mathbf{M} & \mathbf{A}\otimes \mathbf{M} \\
                   \mathbf{0}_{n(N-k)\times nk} & \mathbf{I}_{N-k}\otimes \mathbf{DM} \\
                 \end{array}
               \right],
\end{equation}
where $\otimes$ is the tensor product of matrices, $\left[
                                                      \begin{array}{cc}
                                                        \mathbf{I}_k & \mathbf{A} \\
                                                      \end{array}
                                                    \right]
$ is a generator matrix of ${\mathcal{C}}$, $\mathbf{M}$ is the matrix of embeddings of a $\mathbb{Z}$-basis of $\mathcal{O}_K$ given in~(\ref{gen_OK}), and $\mathbf{DM}$ is the matrix of embeddings of a $\mathbb{Z}$-basis of $\mathfrak{p}$ in~(\ref{gen_P}).
\end{Theorem}

\subsection{Algebraic LDPC lattices}~\label{ConstructionAversusConstructionD'}
Assume that $\mathcal{C}$ is a linear code over $\mathbb{F}_p$ where $p$ is a prime number, so $\mathcal{C}\subseteq\mathbb{F}_{p}^N$.
%Let $d_{\min}$ denote the minimum distance of $\mathcal{C}$.
A lattice $\Lambda$ constructed based on Construction A \cite{2} can be derived from $\mathcal{C}$ by:
\begin{equation}\label{constA}
\Lambda=p\mathbb{Z}^N+\epsilon\left(\mathcal{C}\right),
\end{equation}
where $\epsilon\colon\mathbb{F}_p^N\rightarrow\mathbb{R}^N$ is an embedding function which sends a vector in $\mathbb{F}_p^N$ to its real version. %In this work, we are particularly interested in binary codes and lattices with $p=2$.

\begin{Definition}
An LDPC lattice $\Lambda$ is a lattice based on a binary LDPC code $\mathcal{C}$ as its underlying code. Equivalently, ${\mathbf x}\in\mathbb{Z}^N$ is in $\Lambda$ if $\mathbf H_{\mathcal{C}}{\mathbf x}^t=\mathbf{0} \pmod{2}$, where ${\mathbf H}_{\mathcal{C}}$ is the parity-check matrix of $\mathcal{C}$ \cite{IWCIT2015,QC_LDPC_lattice}.
\end{Definition}

This LDPC lattice can also be constructed via Construction A using the same underlying code $\mathcal{C}$.

\begin{Example}{\cite{ConstA}}\label{example1}
%This example is discussed in \cite{ConstA}.
\textcolor{mycolor4}{Let $p$ be a prime number and $\xi_p$ be a primitive $p$th root of unity.} Consider the cyclotomic field $K = \mathbb{Q}(\xi_p)$ with the ring of integers $\mathcal{O}_K = \mathbb{Z}[\xi_p]$. The
degree of $K$ over $\mathbb{Q}$ is $p - 1$, and $p$ is totally ramified, with $p\mathcal{O}_K = (1-\xi_p)^{p-1}$. Thus, taking the prime ideal $\mathfrak{p} = (1-\xi_p)$ with the residue field $\mathcal{O}_K /\mathfrak{p}\cong \mathbb{F}_p$,  the bilinear form $\left\langle x,y\right\rangle=\sum_{i=1}^N \mathrm{Tr}_{K/Q}(x_iy_i)$ and a linear code $\mathcal{C}$ over $\mathbb{F}_p$, then $\Gamma_{\mathcal{C}}$ yields the so-called Construction A as described above.
Since $\mathbb{Q}(\xi_p)$ is a CM-field, we can  use the  bilinear form corresponding to~(\ref{bilinear2}) with $\alpha = 1/p$. By using this bilinear form, the generator matrix is as follows
\begin{eqnarray}\label{constA_p}
  \mathbf{M}_{\Lambda} &=& \frac{1}{\sqrt{p}}\left[
                    \begin{array}{cc}
                      \mathbf{I}_{k}& \mathbf{P}_{k\times (N-k)} \\
                      \mathbf{0}_{(N-k)\times k} & p\mathbf{I}_{N-k} \\
                    \end{array}
                  \right].
\end{eqnarray}
It has been proved in \cite{ConstA} that if $\mathcal{C} \subset \mathcal{C}^{\perp}$, then $\Gamma_{\mathcal{C}}$ is an integral lattice of rank $N(p - 1)$.
Our particular case is based on Construction A of lattices from codes  when $p = 2$.
In such case, $\xi_p = -1$, $\mathcal{O}_K = \mathbb{Z}$, and $\mathfrak{p} = 2\mathbb{Z}$.
$\hfill\square$\end{Example}

Next, we present the definition of full-diversity  \textcolor{mycolor3}{algebraic} LDPC lattices using algebraic number fields.
\begin{Definition}\label{parity_def}
Let $\mathcal{C}$ be a binary LDPC code of length $N$ and dimension $k$.  Consider the number field $K$ with the ring of integers $\mathcal{O}_K$. Let $n$ be the degree of $K$ over $\mathbb{Q}$ and $\mathfrak{p}$ be a prime in $\mathcal{O}_K$ with residue field $\mathcal{O}_K /\mathfrak{p}\cong \mathbb{F}_2$. Define $\rho:\mathcal{O}_K^N\rightarrow \mathbb{F}_2^N $ as the componentwise reduction modulo $\mathfrak{p}$ and $\sigma^i:\mathcal{O}_K^i\rightarrow \mathbb{R}^{in}$, for positive integer $i$, as
$$\sigma^i(x_1,\ldots,x_i)=(\sigma(x_1),\ldots,\sigma(x_i)),$$
where $\sigma$ is the canonical embedding in~(\ref{embeding2}). Let $\left\{\omega_1,\ldots,\omega_n\right\}$ be the integral basis for $\mathcal{O}_K$. Define $\sigma^{-1}:\sigma(\mathcal{O}_K)\rightarrow \mathcal{O}_K$ such that for $x=\sum_{l=1}^n u_{l}\omega_l$ in $\mathcal{O}_K$
$$\sigma^{-1}(\sigma_1(x),\ldots,\sigma_n(x))=x.$$
Define $(\sigma^i)^{-1}$ similarly to $\sigma^i$ but replacing $\sigma$ with $\sigma^{-1}$.
Then, $\Lambda=\sigma^N(\Gamma_{\mathcal{C}})=\sigma^N(\rho^{-1}(\mathcal{C}))$ is  the \textcolor{mycolor3}{algebraic}  LDPC lattice based on the number field $K$. The parity-check matrix $\mathbf{H}_{\Lambda}$ for $\Lambda$ is an $n(N-k)\times nN$ matrix over $\mathbb{F}_2$ of rank $n(N-k)$ such that
%$\mathbf{x}\in \Lambda$ if and only if
%$$\mathbf{H} (\sigma^{-1}(\mathbf{x}))^t\equiv \mathbf{0}_{(N-k)\times 1},\,\, \bmod \,\, \mathfrak{p}.$$
\begin{equation}\label{parity_check}
  \Lambda=\left\{\mathbf{x}\in\sigma^{N}(\mathcal{O}_K^N)\,\,|\,\,\rho((\sigma^{N-k})^{-1}(\mathbf{x}\mathbf{H}^t))=\mathbf{0}_{1\times (N-k)}\right\}.
\end{equation}
\end{Definition}
\begin{Theorem}\label{theorem_parity}
Let $\mathcal{C}$ be a binary LDPC code of length $N$ and dimension $k$. Let $\mathbf{H}$ and $\mathbf{G}=\left[
                           \begin{array}{cc}
                              \mathbf{I}_k & \mathbf{A}\\
                           \end{array}
                         \right]$
be the parity-check and generator matrices of $\mathcal{C}$, respectively. Consider the Galois extension $K/\mathbb{Q}$ with the ring of integers $\mathcal{O}_K$. Let $n$ be the degree of $K$ over $\mathbb{Q}$ and let $2$ be totally ramified in $\mathcal{O}_K$. The prime $\mathfrak{p}$  is chosen above $2$ so that $2\mathcal{O}_K=\mathfrak{p}^n$ with  residue field $\mathcal{O}_K /\mathfrak{p}\cong \mathbb{F}_2$. Then, $\mathbf{H}_{\Lambda}=\mathbf{H}\otimes\mathbf{I}_n$ is a parity-check matrix for  \textcolor{mycolor3}{algebraic} LDPC lattice $\Lambda=\sigma^N(\Gamma_{\mathcal{C}})=\sigma^N(\rho^{-1}(\mathcal{C}))$.
\end{Theorem}
\begin{IEEEproof}
Based on the assumed conditions and Theorem~\ref{theorem6}, the generator matrix of $\Lambda$ has the following form
 \begin{equation*}
  \mathbf{M}_{\Lambda}=\left[
                 \begin{array}{cc}
                   \mathbf{I}_k\otimes \mathbf{M} & \mathbf{A}\otimes \mathbf{M} \\
                   \mathbf{0}_{n(N-k)\times nk} & \mathbf{I}_{N-k}\otimes \mathbf{DM} \\
                 \end{array}
               \right].
\end{equation*}
Let $\mathbf{u}=(u_1,\ldots,u_{nN})$ be an integer vector. First we show that $\rho((\sigma^{N-k})^{-1}(\mathbf{u}\mathbf{M}_{\Lambda}\mathbf{H}_{\Lambda}^t))=\mathbf{0}$. To this end,
\begin{eqnarray*}
   \mathbf{M}_{\Lambda}\mathbf{H}_{\Lambda}^t &=&
   \left[
                 \begin{array}{c}
                   \left[\mathbf{I}_k \,\,\,\,\, \mathbf{A}\right]\otimes \mathbf{M} \\
                   \left[\mathbf{0}_{(N-k)\times k} \,\,\, \mathbf{I}_{N-k}\right]\otimes \mathbf{DM}\\
                 \end{array}
   \right](\mathbf{H}\otimes\mathbf{I}_n)^t \\
   &=& \left[
                 \begin{array}{c}
                   \left[\mathbf{I}_k \,\,\,\,\, \mathbf{A}\right]\mathbf{H}^t\otimes \mathbf{M} \\
                   \left[\mathbf{0}_{(N-k)\times k} \,\,\, \mathbf{I}_{N-k}\right]\mathbf{H}^t\otimes \mathbf{DM}\\
                 \end{array}
   \right].
\end{eqnarray*}
The $\mathbb{Z}$-linearity of $(\sigma^{N-k})^{-1}$ implies the sufficiency of proving  $\rho((\sigma^{N-k})^{-1}(\mathbf{b}_i))=\mathbf{0}$, where $\mathbf{b}_i$ is the $i$th row of $\mathbf{M}_{\Lambda}\mathbf{H}_{\Lambda}^t$, for $i=1,\ldots,nN$. Since $\mathbf{H}$ and $\left[\mathbf{I}_k \,\, \mathbf{A}\right]$ are the parity-check matrix and the generator matrix of the binary code $\mathcal{C}$, respectively, $\left[\mathbf{I}_k \,\, \mathbf{A}\right]\mathbf{H}^t=2\mathbf{Z}$ for a $k\times(N-k)$ integer matrix $\mathbf{Z}$. On the other hand, $\left[\mathbf{0}_{(N-k)\times k} \,\, \mathbf{I}_{N-k}\right]\mathbf{H}^t= \mathbf{H}_{N-k}$, where $\mathbf{H}_{N-k}$ is the last $N-k$ rows of $\mathbf{H}^t$. For $1\leq i\leq kn$, let $r_i=\left\lfloor\frac{i}{n}\right\rfloor+1$, where $\left\lfloor c\right\rfloor$ is the floor of a real number $c$, and $s_i=i-(r_i-1)n$. Then
$$\mathbf{b}_i=\left( 2z_{r_i,1}\mathbf{M}_{s_i} , 2z_{r_i,2}\mathbf{M}_{s_i},  \ldots , 2z_{r_i,N-k}\mathbf{M}_{s_i}
%                 \begin{array}{cccc}
%                   2z_{r_i,1}\mathbf{M}_{s_i} & 2z_{r_i,2}\mathbf{M}_{s_i} & \ldots & 2z_{r_i,N-k}\mathbf{M}_{s_i} \\
%                 \end{array}
               \right),$$
in which $\mathbf{Z}_{r_i}=\left(z_{r_i,1},\ldots,z_{r_i,N-k}\right)$ and $\mathbf{M}_{s_i}=\left(\sigma_1(\omega_{s_i}),\ldots,\sigma_n(\omega_{s_i})\right)$ are $r_i$th and $s_i$th rows of $\mathbf{Z}$ and $\mathbf{M}$, respectively. Finally,
%\small
\begin{IEEEeqnarray*}{rCl}
  && \rho((\sigma^{N-k})^{-1}(\mathbf{b}_i)) \\
  &=& \rho((\sigma^{N-k})^{-1}\left(
                   2z_{r_i,1}\mathbf{M}_{s_i} ,  \ldots , 2z_{r_i,N-k}\mathbf{M}_{s_i}
               \right)) \\
   &=& \rho\left(
                   2z_{r_i,1}\sigma^{-1}(\mathbf{M}_{s_i}),  \ldots , 2z_{r_i,N-k}\sigma^{-1}(\mathbf{M}_{s_i})
               \right)  \\
   &=& \rho\left(
                   2z_{r_i,1}\omega_{s_i} ,  \ldots , 2z_{r_i,N-k}\omega_{s_i}
               \right) \\
   &=&\mathbf{0},
\end{IEEEeqnarray*}
%\normalsize
where the last equation follows from the fact that
$$\left(2z_{r_i,1}\omega_{s_i},  \ldots,  2z_{r_i,N-k}\omega_{s_i}
               \right) \in (2\mathcal{O}_K)^{N-k}\subset \mathfrak{p}^{N-k}.$$
For $kn+1\leq i\leq nN$,  let $r_i=\left\lfloor\frac{i}{n}\right\rfloor-k+1$,  and $s_i=i-(r_i+k-1)n$. Consider $\left\{\mu_1,\ldots,\mu_n\right\}$ as the $\mathbb{Z}$-basis of $\mathfrak{p}$.  Then
$$\mathbf{b}_i=\left( h_{r_i,1}\mathbf{P}_{s_i} , h_{r_i,2}\mathbf{P}_{s_i} , \ldots , h_{r_i,N-k}\mathbf{P}_{s_i}
%                 \begin{array}{cccc}
%                   h_{r_i,1}\mathbf{P}_{s_i} & h_{r_i,2}\mathbf{P}_{s_i} & \ldots & h_{r_i,N-k}\mathbf{P}_{s_i} \\
%                 \end{array}
               \right),$$
where $(h_{r_i,1},\ldots,h_{r_i,N-k})$ and $\mathbf{P}_{s_i}=(\sigma_1(\mu_{s_i}),\ldots,\sigma_n(\mu_{s_i}))$ are the $r_i$th and $s_i$th rows of $\mathbf{H}_{N-k}$ and $\mathbf{DM}$, respectively. In this case
%\small
\begin{IEEEeqnarray*}{rCl}
  &&\rho((\sigma^{N-k})^{-1}(\mathbf{b}_i))\\&=& \rho((\sigma^{N-k})^{-1}\left(
                   h_{r_i,1}\mathbf{P}_{s_i} ,  \ldots , h_{r_i,N-k}\mathbf{P}_{s_i}
               \right)) \\
   &=& \rho\left(
                   h_{r_i,1}\sigma^{-1}(\mathbf{P}_{s_i}),  \ldots , h_{r_i,N-k}\sigma^{-1}(\mathbf{P}_{s_i})
               \right)  \\
   &=& \rho\left(
                   h_{r_i,1}\mu_{s_i} ,  \ldots , h_{r_i,N-k}\mu_{s_i}
               \right) \\
   &=&\mathbf{0}.
\end{IEEEeqnarray*}
%\normalsize
Now, let $\mathbf{x}\in\sigma^{N}(\mathcal{O}_K^N)$ such that $\rho((\sigma^{N-k})^{-1}(\mathbf{x}\mathbf{H}_{\Lambda}^t))=\mathbf{0}$. We  show that $\mathbf{x}\in\Lambda$. For the sake of this, we have
$$\mathbf{x}=\left(\sigma_1(x_{1}),\ldots,\sigma_n(x_{1}),\ldots,\sigma_1(x_{N}),\ldots,\sigma_n(x_{N})\right),$$
where $\tilde{\mathbf{x}}=(x_1,\ldots,x_N)\in \mathcal{O}_K^N$. Then
\begin{IEEEeqnarray*}{rCl}
  \mathbf{x}\mathbf{H}_{\Lambda}^t &=& \mathbf{x}\left[\mathbf{h}_1 , \mathbf{h}_2 , \ldots , \mathbf{h}_{n(N-k)}
%                                                                         \begin{array}{cccc}
%                                                                           \mathbf{h}_1 & \mathbf{h}_2 & \cdots & \mathbf{h}_{n(N-k)} \\
%                                                                         \end{array}
                                                                       \right]^t
   \\
   &=& \left(\mathbf{x}\cdot \mathbf{h}_1^t,\mathbf{x}\cdot \mathbf{h}_2^t,\ldots,\mathbf{x}\cdot \mathbf{h}_{n(N-k)}^t\right),
\end{IEEEeqnarray*}
where $\mathbf{x}\cdot \mathbf{h}_i^t$ is the inner product of $\mathbf{x}$ and the $i$th column of $\mathbf{H}_{\Lambda}^t$, $\mathbf{h}_i$, for $i=1,\ldots,n(N-k)$. The computation of the $i$th component is as follows
\begin{IEEEeqnarray*}{rCl}
  \mathbf{x}\cdot \mathbf{h}_i^t &=& \sum_{k=1}^n\sum_{j=0}^{N-1}h_{jn+k,i}\sigma_{k}(x_{j+1}) \\
   &=& \sum_{k=1}^n\sigma_{k}\left(\sum_{j=0}^{N-1}h_{jn+k,i}x_{j+1}\right) \\
   &=& \sigma_s\left(\sum_{j=1}^N h_{j,r}^c x_j\right) \\
   &=& \sigma_s\left(\tilde{\mathbf{x}}\cdot \mathbf{h}_r^c \right),
\end{IEEEeqnarray*}
where $r=\left\lfloor\frac{i}{n}\right\rfloor+1$, $s=i-(r-1)n$ and $ \mathbf{h}_r^c=(h_{1,r}^c,\ldots,h_{N,r}^c)^t$ is the $r$th column of $\mathbf{H}^t$.
It should be noted that the two last equations in the above follow from the fact that $\mathbf{h}_i$ is of the form
$\mathbf{h}_i=\left(\mathbf{h}_i^1,\mathbf{h}_i^2,\ldots,\mathbf{h}_i^{N}\right)^t$, where
\begin{equation*}
 \mathbf{h}_i^j= \left(\overbrace{0,\cdots,0}^{(s-1)-\mathrm{times}},h_{j,r}^c,\overbrace{0,\cdots,0}^{(n-s)-\mathrm{times}}\right),\quad j=1,\ldots,N.
\end{equation*}
%in which $\mathbf{h}_r^c=\left(h_{1,r},h_{2,r},\ldots,h_{N-k,r}\right)^t$ is the $r$th column of $\mathbf{H}$.
Thus
\begin{IEEEeqnarray*}{rCl}
  \mathbf{x}\mathbf{H}_{\Lambda}^t &=& \left( \sigma_1(\tilde{\mathbf{x}}\cdot \mathbf{h}_1^c),
  \ldots,\sigma_n(\tilde{\mathbf{x}}\cdot \mathbf{h}_1^c),\ldots, \right. \\
 &&  \> \left. \sigma_1(\tilde{\mathbf{x}}\cdot\mathbf{h}_{N-k}^c), \ldots,\sigma_n(\tilde{\mathbf{x}}\cdot \mathbf{h}_{N-k}^c) \right) \\
   &=& (\sigma(\tilde{\mathbf{x}}\cdot \mathbf{h}_1^c),\ldots,\sigma(\tilde{\mathbf{x}}\cdot \mathbf{h}_{N-k}^c)) \\
   &=&  \sigma^{N-k}\left(\tilde{\mathbf{x}}\cdot \mathbf{h}_1^c,\ldots,\tilde{\mathbf{x}}\cdot \mathbf{h}_{N-k}^c\right)\\
   &=& \sigma^{N-k}\left(\tilde{\mathbf{x}}\mathbf{H}^t\right).
\end{IEEEeqnarray*}
Thus, $\rho((\sigma^{N-k})^{-1}(\mathbf{x}\mathbf{H}_{\Lambda}^t))=\mathbf{0}$ implies $\rho\left(\tilde{\mathbf{x}}\mathbf{H}^t\right)=\mathbf{0}$ which indicates $\rho(\tilde{\mathbf{x}})\in C$, and so $\mathbf{x}\in \Lambda$.
\end{IEEEproof}

Theorem~\ref{theorem_parity} is also valid in the non-binary case,  where the conditions of Theorem~\ref{theorem6} are fulfilled. The authors of \cite{ConstA} proposed Construction~A based on number fields for non-binary linear codes. They have used cyclotomic number fields $\mathbb{Q}(\xi_{p^r})$ and their maximal totally real subfields $\mathbb{Q}(\xi_{p^r}+\xi_{p^r}^{-1})$, $r\geq 1$, as examples for their construction method. Using their method for the binary case $p=2$ does not provide diversity and gives us the well known Construction A \cite{2} that we describe in this section. In Section~\ref{new_construction}, we
 propose a new method for using Construction A over number fields in the binary case.
 \section{System Model and Performance Evaluation on Block-Fading Channels}\label{system_model}
\textcolor{Mycolor}{In this section, we describe our system model for communication over  BF channels using algebraic lattices.
%We consider the communication channel as a BF (BF) channel. We also assume the perfect channel state information (CSI) at the receiver end.
%===============================================================================
%First, we describe communication over fading channels using algebraic lattices of the form $\sigma(\mathcal{O}_K)$ where $K$ is a number field of degree $n$, $\mathcal{O}_K$ the ring of integers of $K$ and $\sigma$ is the canonical embedding.  We also present the available design  criteria and performance measurements in fading channels.  Then, by using  Construction A lattices with an underlying $[N,k]$-linear code $\mathcal{C}$, this model is converted to a model that describes communication over a BF channel with fading block length $N$.
In communication over a flat fading channel, the received discrete-time  signal vector is given by
\begin{equation}\label{channel}
  \mathbf{y}_i^t=\mathbf{H_F}\mathbf{x}_i^t+\mathbf{n}_i,\quad i=1,\ldots, N,
\end{equation}
where $\mathbf{y}_i\in \mathbb{R}^n$ is the received $n$-dimensional real  signal vector, $\mathbf{x}_i\in\mathbb{R}^n$ is the transmitted $n$-dimensional real signal vector, $\mathbf{H_F} = \textrm{diag}(\mathbf{h})$ with $\mathbf{h} = (h_1,\ldots, h_n)\in \mathbb{R}^n$ is the $n\times n$  flat fading diagonal matrix, and $\mathbf{n}_i \in \mathbb{R}^n$ is the noise vector whose samples are i.i.d. with Gaussian distribution $\sim \mathcal{N}(0, \sigma_{\mathcal{N}}^2)$.}

\textcolor{Mycolor}{Let $\mathbf{x}\in\mathbb{R}^{nN}$ be a frame composed of $N$ modulation symbols $\mathbf{x}_i$, each one  with dimension $n$,  or composed of $nN$ channel uses.
In this paper, $\mathbf{x}$ is chosen from a  Construction A lattice $\Lambda=\sigma^N\left(\rho^{-1}(\mathcal{C})\right)$ based on a number field $K$  of degree $n$,  with an underlying $[N,k]$-linear code $\mathcal{C}$. This setting  describes communication over a BF channel with fading block length $N$.
We define $\gamma$ the signal-to-noise ratio (SNR) for an infinite lattice constellation $\Lambda$ as follows:
\begin{IEEEeqnarray}{rCl}\label{SNR_def}
\gamma=\frac{\mathrm{vol}(\Lambda)^{\frac{2}{nN}}}{\sigma_{\mathcal{N}}^2}.
\end{IEEEeqnarray}
The case of complex signals obtained from $2$ orthogonal real signals can be similarly modeled by (\ref{channel}) by replacing $N$
with $N' = 2N$. In communication over a BF channel, we assume that the fading matrix $\mathbf{H_F}$ is constant during one frame and it changes independently from frame to frame. This corresponds to a BF channel with $n$ blocks \cite{blockfading}. We further assume perfect channel state information (CSI) at the receiver, that is, the receiver perfectly knows the fading coefficients.}
%Therefore, for a given fading realization, the channel transition probabilities are given by
%\begin{equation}\label{channel_prob}
%  p(\mathbf{y}|\mathbf{x},\mathbf{H_F})=(2\pi\sigma_{\mathcal{N}}^2)^{-\frac{n}{2}}\exp\left(-\frac{1}{2\sigma_{\mathcal{N}}^2}\|\mathbf{y}-\mathbf{H_F}\mathbf{x}\|^2\right).
%\end{equation}

\textcolor{Mycolor}{In this paper, we consider Rayleigh fading channels as our communication model. Rayleigh fading is a reasonable model when there are many objects in the environment that scatter the radio signal before it arrives at the receiver. Due to the central limit theorem,
%if there is sufficiently much scatter,
\textcolor{mycolor3}{if there are many scatterers in the environment,
the channel impulse response can be} modelled as a Gaussian process. If the scatters have no dominant components, then such a process has zero mean and phase evenly distributed between $0$ and $2\pi$ radians. Thus, the envelope of the channel response is Rayleigh distributed. Often, the gain and phase elements of such channel's distortion are  represented as complex numbers. In this case, Rayleigh fading is exhibited by a complex random variable with real and imaginary parts  modelled by independent and identically distributed zero-mean Gaussian processes.
With the aid of an in-phase/quadrature component interleaver \cite{ConstA,alglattice1}, it is possible to
remove the phase of the complex fading coefficients to obtain a real fading which is Rayleigh distributed and guarantee that the fading coefficients are independent from one real symbol to the next.}

\textcolor{Mycolor}{Thus, the received  vector $\mathbf{y}$ from Rayleigh BF channel with $n$ fading blocks and coherence time $N$ can be written as follows:
\begin{IEEEeqnarray}{rCl}\label{AWGN_output1}
\mathbf{y}^{t}=(\mathbf{I}_N\otimes\mathbf{H_F})\mathbf{x}^t+\mathbf{n}^t,
\end{IEEEeqnarray}
where $\mathbf{H_F}=\textrm{diag}(|h_1|,\ldots,|h_n|)$ and the fading coefficients $h_i$'s are complex Gaussian random
variables with variance $\sigma_b^2$, so that $|h_i|$ is Rayleigh distributed with parameter $\sigma_b^2$, for all $i = 1,\ldots , n$, and $\mathbf{n}=(\mathbf{n}_1,\ldots,\mathbf{n}_N)=(\nu_1,\ldots,\nu_{nN})$ in which $\nu_i\sim \mathcal{N}(0,\sigma_{\mathcal{N}}^2)$ for $i=1,\ldots,nN$, is the Gaussian noise.}
%Moreover, we assume that the real fading coefficients follow a Rayleigh distribution
%\begin{equation}\label{nakagami}
%  p_h(x)=\frac{2m^mx^{2m-1}}{\Gamma(m)}e^{-mx^2},
%\end{equation}
%\begin{equation}\label{nakagami}
%  p_h(x)=2xe^{-x^2}.
%\end{equation}
%where $m > 0$ and $\Gamma(x)\triangleq \int_{0}^{+\infty}t^{x-1}e^{-t}dt $ is the Gamma function. In the literature $m \geq 0.5$ is usually considered  \cite{porakis}; however, the fading
%distribution is well defined and reliable communication is possible for any
%$0 < m < 0.5$.
%Define the coefficients $\gamma_i = h_i^2$  for $i =1,\ldots , n$, which correspond to the fading power gains with
%probability density function  (PDF) $p_{\gamma}(x) = e^{-x}$.
%and cumulative distribution function (CDF) $P_{\gamma}(x) = 1 - \overline{\Gamma}(x,1)$,
%respectively, where $\overline{\Gamma}(a,x)\triangleq\frac{1}{\Gamma(a)}\int_{x}^{+\infty}t^{a-1}e^{-t}dt$
%is the normalized incomplete Gamma function~\cite{statistic} and $\Gamma(x)\triangleq \int_{0}^{+\infty}t^{x-1}e^{-t}dt $ is the Gamma function.

%Analyzing the Nakagami$-m$ fading channels, in which the fading coefficients have Nakagami$-m$ distribution,  recovers the analysis for other fading channels, including Rayleigh fading by setting $m = 1$ and Rician fading with parameter $\kappa$ by setting
%$m = (\kappa + 1)^2/(2\kappa + 1)$~\cite{statistic2}.
%========================================================================
%\subsection{Multidimensional lattice constellations}
\subsection{Error performance of lattices over block-fading channels}\label{system_model_subsec_2}
\textcolor{mycolor3}{In communication using lattices, the transmitted signal vector $\mathbf{x}$ belongs to an $nN$-dimensional infinite lattice $\Lambda\subset \mathbb{R}^{nN}$.
We consider
%signal constellations $\mathcal{S}$ that are generated as a finite subset of points carved from
the  lattice $\Lambda=\left\{\mathbf{u}\mathbf{M}_{\Lambda}|\mathbf{u}\in\mathbb{Z}^{nN}\right\}$
with full rank generator matrix $\mathbf{M}_{\Lambda}\in \mathbb{R}^{nN\times nN}$. For a given channel
realization, we define the faded lattice seen by the receiver as the lattice $\Lambda'$ whose generator matrix
is given by $ \mathbf{M}_{\Lambda}'=(\mathbf{I}_N \otimes \mathbf{H_F})\mathbf{M}_{\Lambda}$.
%In order to simplify the labeling
%operation, constellations are of the type $\mathcal{S} = \left\{\mathbf{u}\mathbf{M}_{\Lambda} + \mathbf{x}_0 |\mathbf{u}\in\mathbb{Z}_M^{nN}
%\right\}$, where $\mathbb{Z}_M = \left\{0, 1,\ldots,M - 1\right\}$ represents an
%integer pulse-amplitude modulation (PAM) constellation, $\log_2(M)$ is the number of bits per dimension and $\mathbf{x}_0$ is an offset vector which minimizes the average transmitted energy. The rate of such constellations
%is $R = \log_2(M)$ $\textrm{bit/s/Hz}$. This is usually referred to as full-rate uncoded transmission \cite{SLB}.
}

\textcolor{mycolor3}{Lattices can be considered as infinite cases of multidimensional signal sets. The performance evaluation of multidimensional signal sets has attracted significant attention due to the special type of diversity that these constellations present \cite{SSD} and
the fact that they can be efficiently used to combat the signal
degradation caused by fading. The diversity order of a multidimensional signal set is the minimum number of distinct components
between any two constellation points. In a similar fashion, the diversity order of an infinite lattice is the minimum Hamming distance between
any two coordinate vectors of the lattice points.} To distinguish from other well-known types of diversity
(time, frequency, space, code) this type of diversity is called \emph{modulation diversity} or \emph{signal space diversity} (SSD) \cite{SSD}.
The design of constellations with signal space diversity has been extensively studied in \cite{alglattice1,SLB,viterbo,Pappi}.
\textcolor{mycolor3}{In this paper, we consider the error performance of maximum likelihood (ML) decoder of infinite lattices  as the benchmark of our performance analysis. Moreover, we only consider Construction A lattices.
Let $\mathcal{C}\subset \mathbb{F}_p^N$ be an $[N,k]$ linear code, where $p$ is a prime number, and $\mathcal{O}_K$ be the integers ring of a totally real number field $K$ of degree $n$. Let $\mathfrak{p}$ be a prime ideal of $\mathcal{O}_K$ such that $\mathcal{O}_K/\mathfrak{p}\cong \mathbb{F}_p$. Also, consider $\sigma_1,\ldots,\sigma_n$ to be $n$ real embeddings of $K$. Every lattice vector $\mathbf{x}$ in $\Lambda=\sigma^N(\Gamma_{\mathcal{C}})=\sigma^N(\rho^{-1}(\mathcal{C}))\subset\mathbb{R}^{nN}$ has the following  form
%\begin{eqnarray*}
%% \nonumber to remove numbering (before each equation)
%  \mathbf{x} &=& \sigma(\mathbf{c}+\mathbf{p}) \\
%   &=& \left(\sigma(c_1+p_1),\ldots,\sigma(c_N+p_N)\right) \\
%   &=& \left(\sigma_1(c_1+p_1),\ldots,\sigma_n(c_1+p_1)\ldots,\sigma_1(c_N+p_N),\ldots,\sigma_n(c_N+p_N)\right) \\
%   &=& \left(c_1+\sigma_1(p_1),\ldots,c_1+\sigma_n(p_1)\ldots,c_N+\sigma_1(p_N),\ldots,c_N+\sigma_n(p_N)\right)\\
%   &=& \mathbf{c}\otimes\underbrace{(1,\ldots,1)}_{n-times}+\sigma(\mathbf{p}),
%\end{eqnarray*}
\begin{IEEEeqnarray}{rCl}
  \mathbf{x} &=& \sigma^N(\mathbf{c}+\mathbf{p})\nonumber \\
   &=& \left(\sigma(c_1+p_1),\ldots,\sigma(c_N+p_N)\right) \nonumber\\
   &=& \left(\sigma_1(c_1+p_1),\ldots,\sigma_n(c_1+p_1),\ldots,\sigma_n(c_N+p_N)\right)\nonumber \\
   &=& \left(c_1+\sigma_1(p_1),\ldots,c_1+\sigma_n(p_1),\ldots,c_N+\sigma_n(p_N)\right)\nonumber\\
   &=& \mathbf{c}\otimes\underbrace{(1,\ldots,1)}_{n-times}+\sigma^N(\mathbf{p}), \label{x_general_form}
\end{IEEEeqnarray}
where $\otimes$ is the Kronecker product,  $\mathbf{c}\in \mathcal{C}$ and $\mathbf{p}\in\mathfrak{p}^N$.
%sphere lower bound for infinite
%lattices $\mathcal{S}$ \cite{SLB,tarokh}.
%At a given $i$, $1\leq i\leq N$, a maximum likelihood decoder with perfect CSI makes an error whenever $\|\mathbf{y}_i-\mathbf{H_F}\mathbf{w}\|^2 \leq \|\mathbf{y}_i-\mathbf{H_F}\mathbf{x}_i\|$ for some $\mathbf{w}\in\Lambda$, $\mathbf{w}\neq \mathbf{x}_i$. These inequalities define the so called decision region around $\mathbf{x}$.
%Under ML decoding, the frame error probability is then given by
Define $\mathcal{V}(\mathbf{x},\mathbf{h})$ as the decision region or Voronoi region for a given  lattice  point $\mathbf{x}$ and fading matrix $\mathbf{H_F}=\mathrm{diag}(\mathbf{h})$.
From the geometrical uniformity of lattices we have that $\mathcal{V}(\mathbf{x}, \mathbf{h}) = \mathcal{V}(\mathbf{w}, \mathbf{h})=\mathcal{V}_{\Lambda}(\mathbf{h})$, for all $\mathbf{x},\mathbf{w}\in\Lambda$. Therefore, we may assume the transmission of the all-zero codeword.
If a lattice point $\mathbf{x}\in\Lambda$ is transmitted over a BF channel with additive noise variance $\sigma_{\mathcal{N}}^2$ per dimension, then the \emph{probability of error} $P_e(\Lambda,\sigma_{\mathcal{N}}^2)$ of an ML decoder (or minimum-distance decoder) with perfect CSI for $\Lambda$ is given by \cite[p. 822]{Forney}, \cite[p. 826]{SLB}
%\begin{equation}\label{P_f}
% P_e(\Lambda,\sigma_{\mathcal{N}}^2)=1-\mathbb{E}\left[\left(1-\int_{\mathbf{n}\not\in\mathcal{V}_{\Lambda}(\mathbf{h})}p(\mathbf{n})d\mathbf{n}\right)^N\right],
%\end{equation}
\begin{IEEEeqnarray}{rCl}\label{P_f}
 P_e(\Lambda,\sigma_{\mathcal{N}}^2)&=&\mathbb{E}\left[P_e(\Lambda,\sigma_{\mathcal{N}}^2|\mathbf{h})\right]\nonumber\\
 &=&1-\mathbb{E}\left[\int_{\mathcal{V}_{\Lambda}(\mathbf{h})}g_{\sigma_{\mathcal{N}}^2}(\mathbf{n})d\mathbf{n}\right],
\end{IEEEeqnarray}
where $g_{\sigma_{\mathcal{N}}^2}(\mathbf{n})=\left(2\pi\sigma_{\mathcal{N}}^2\right)^{-nN/2}e^{-\left\|\mathbf{n}\right\|^2/2\sigma_{\mathcal{N}}^2}$ is the probability density function (p.d.f.) of an
$nN$-dimensional zero-mean Gaussian random variable with variance $\sigma_{\mathcal{N}}^2$ per dimension. This expression holds for any lattice point $\mathbf{x}\in\Lambda$.
\textcolor{mycolor4}{
For a fixed lattice $\Lambda$, the decoding error probability $P_e(\Lambda,\sigma_{\mathcal{N}}^2)$  is clearly a function of the SNR $\gamma$. In the rest of this paper, we denote it by $P_e(\gamma)$  in instances where no ambiguity would arise.}
\begin{Definition} The diversity order is defined as the asymptotic (for large SNR) slope of $P_e$ in a log-log scale, that is,
\begin{equation}\label{diversity}
  d\triangleq -\lim_{\gamma \rightarrow \infty}\frac{\log P_e(\gamma)}{\log \gamma}.
\end{equation}
\end{Definition}
}

The diversity order is usually a function of the fading distribution and the signal constellation. It is proved that
the diversity order is the product of the signal space diversity and a parameter of the fading distribution \cite{SLB}. In  Rayleigh fading channels which is the case in this paper, the diversity order $d$ and the signal space diversity coincide and both are denoted by $L$ in the rest of this paper.
\begin{Definition}
Consider a BF channel with $n$ independent fading coefficients per lattice point. The lattice $\Lambda$ is a full-diversity lattice under ML decoding if the diversity order $L$ is equal to the number of fading blocks, that is, $L=n$.
%A constellation $\mathcal{S}\subset \mathbb{R}^n$ has \emph{full-diversity} if the ML decoder is able to decode correctly in presence of $n-1$ \emph{deep fades}
%\footnote{When the transmitter and receiver are surrounded by reflectors, a transmitted signal can traverse in multiple paths and the receiver sees the superposition of multiple copies of the transmitted signal with different attenuations, delays and phase shifts. This can result in either constructive or destructive interference, amplifying or attenuating the signal power of the receiver. Strong destructive interference is frequently referred to as a deep fade and may result in temporary failure of communication due to a severe drop in the channel signal-to-noise ratio.}.
\end{Definition}

\subsection{Good lattices for block-fading channels}\label{Design_sec}
We need an estimate of the error probability of the above system over a BF channel with additive noise with variance $\sigma_{\mathcal{N}}^2$ per dimension  to address the search for good lattices.
In the case of using the lattice $\Lambda$ over this channel, due to the geometrically uniformity of the lattice, we may simply write $P_e(\Lambda)=P_e(\Lambda,\sigma_{\mathcal{N}}^2) = P_e(\Lambda,\sigma_{\mathcal{N}}^2|\mathbf{x})$ for any transmitted point $\mathbf{x}\in \Lambda$. Thus, $\mathbf{x}$ can be considered as the all-zero vector.   By applying the union bound  we obtain an upper bound to the point error probability \cite{viterbo}
\begin{equation}\label{union bound}
 P_e(\Lambda)\leq \sum_{\mathbf{x}\neq \mathbf{w}}P(\mathbf{x}\rightarrow \mathbf{w}),
\end{equation}
where $P(\mathbf{x}\rightarrow \mathbf{w})$ is the pairwise error probability (PEP), the probability that the received point $\mathbf{y}$ is closer to $\mathbf{w}$ than to $\mathbf{x}$ according to the  metric
\begin{equation}\label{metric}
  m(\mathbf{x}|\mathbf{y},\mathbf{h})=\sum_{i=1}^{nN}|y_i-h_ix_i|^2,
\end{equation}
when $\mathbf{x}$ is transmitted. In \cite{viterbo}, using the Chernoff bounding technique, it is shown that \textcolor{mycolor4}{for vanishing noise variance (high SNR)
\begin{equation}\label{viterbo_bound}
  P(\mathbf{x}\rightarrow \mathbf{w})\leq \frac{1}{2}\prod_{x_i\neq w_i}\frac{8\sigma_{\mathcal{N}}^2}{(x_i- w_i)^2}=\frac{(8\sigma_\mathcal{N}^2)^{\ell}}{2d_p^{(\ell)}(\mathbf{x},\mathbf{w})^2},
\end{equation}
}where $\ell=|\left\{1\leq i\leq nN|x_i\neq w_i\right\}|$ and
 $d_p^{(\ell)}(\mathbf{x},\mathbf{w})$ is the  $\ell$-product distance of $\mathbf{x}$
from $\mathbf{w}$ when these two points differ in $\ell$ components
\begin{IEEEeqnarray*}{rCl}
d_p^{(\ell)}(\mathbf{x},\mathbf{w})^2=\prod_{x_i\neq w_i}\left(x_i-w_i\right)^2.
\end{IEEEeqnarray*}
\textcolor{mycolor3}{Let us define $L=\min_{\mathbf{x}\neq \mathbf{w}\in \Lambda}\left\{\ell\right\}$ as the diversity order.
Thus, the point error probability of a lattice is essentially dominated by three factors and to improve the
performance, it is necessary to \cite{viterbo}:
\begin{enumerate}
  %\item minimize the volume of $\Lambda$;
  \item maximize the signal space diversity $L$;
  \item maximize the minimum $L$-product distance
  \begin{equation}\label{d_p_min}
    d_{p,\textrm{min}}^{(L)}=\prod_{x_i\neq y_i}^L |x_i-y_i|,
  \end{equation}
  between any two points $\mathbf{x}$ and $\mathbf{y}$ in lattice;
  \item minimize the product kissing number $\tau_p$ for the $L$-product distance, that is, the total number of points at the
minimum $L$-product distance.
\end{enumerate}
}
To minimize the error probability, one should maximize the diversity
order $L$, that is, have full-diversity $L = n$.
\textcolor{Mycolor}{\begin{Theorem}\cite{viterbo}\label{diverity_order}
Let $(r_1,r_2)$ be the signature of a number field $K$ with the ring of integers $\mathcal{O}_K$. Then, the algebraic lattice of the form $\sigma(\mathcal{O}_K)$ exhibits a diversity $L=r_1+r_2$.
\end{Theorem}
}
% where $\mathcal{O}_K$ is the integers ring of a number field $K$, have diversity order $r_1+r_2$, where $(r_1,r_2)$ is the signature of $K$ \cite{alglattice1}.

\textcolor{Mycolor}{
\begin{Corollary}
Since we have $r_1+2r_2=n=[K:\mathbb{Q}]$ and in totally real number fields $r_2=0$, algebraic lattices obtained from totally real number fields have diversity order $n$, that is, they are full-diversity lattices.
The proposed Construction A in Section~\ref{algebraic_const_A}, which is employed to design the lattices in the rest of this paper, inherits the full-diversity property from the chosen underlying number field \cite[Example 5]{ConstA}.
\end{Corollary}}

%On the other hand, the rank of a lattice determines the number of vectors we get with a given power limit and smaller rank means  less constellation vectors.  Hence, it is preferable to look at full rank lattices.
The three conditions addressed above were introduced first to design good finite lattice constellations for both Rayleigh fading and Gaussian channels \cite{viterbo}. Hence, modifications are required to make some of these conditions applicable in the design of good infinite lattices for fading channels. The following definitions are borrowed from \cite{numberfieldnew}.
\begin{Definition}
Let $\mathbf{v} = (v_1,\ldots,v_n)$ be a vector in $\mathbb{R}^n$. We define the \emph{product norm} of $\mathbf{v}$ as $\mathfrak{N}(\mathbf{v})=\prod_{i=1}^{n}|v_i|$. If $\mathfrak{N}(\mathbf{v})\neq 0$ for all the non zero elements of a lattice $\Lambda$, for example, when $\Lambda$ has full diversity, we can define the \emph{minimum product distance} $d_{p,\mathrm{min}}(\Lambda)$ of $\Lambda$
to be the infimum of the product norms of all non-zero vectors in the lattice.
\end{Definition}

\textcolor{mycolor5}{It should be noted that the definition of minimum $L$-product distance in (\ref{d_p_min}) can be applied for both infinite lattices and finite lattice constellations.  However, finding a finite constellation by maximizing the minimum product norm will not necessarily result in a good finite constellation for fading channels.  When $\mathcal{A}\subset \Lambda$ is a finite lattice constellation with diversity order $L=n$, two cases can be considered: when the all-zero vector is contained in $\mathcal{A}$ or not. When $\mathbf{0}\in\mathcal{A}$, we have
\begin{IEEEeqnarray*}{rCl}
d_{p,\textrm{min}}^{(L)}(\mathcal{A})&=&\min_{\mathbf{x},\mathbf{y}\in\mathcal{A}}\prod_{x_i\neq y_i}^n |x_i-y_i|=\min_{\mathbf{x},\mathbf{y}\in\mathcal{A},\mathbf{x}\neq\mathbf{y}} d_p^{(L)}(\mathbf{x},\mathbf{y})\\
&\leq&\min_{\mathbf{x}\in\mathcal{A}-\{\mathbf{0}\}} d_p^{(L)}(\mathbf{x},\mathbf{0})=\min_{\mathbf{x}\in\mathcal{A}-\{\mathbf{0}\}}\mathfrak{N}(\mathbf{x}).
\end{IEEEeqnarray*}
In this case, the minimum product norm is an upper bound for the minimum $L$-product distance. When $\mathbf{0}\notin\mathcal{A}$, this is not necessarily true. In \figurename~\ref{figsimrot}, we have plotted the minimum product norm and the minimum $L$-product distance of different rotations of 4-QAM constellation in terms of the rotation angle. This figure indicates that maximizing the minimum product norm will not always result in maximizing the minimum $L$-product distance.
}

\textcolor{mycolor5}{For infinite lattices with full diversity, since the all-zero vector is always a lattice vector, due to the linearity of the lattice, one can check that the minimum product norm $d_{p,\mathrm{min}}(\Lambda)$ of the lattice coincides with its minimum $L$-product distance $d_{p,\mathrm{min}}^{(L)}(\Lambda)$. Hence, we can replace $d_p^{(\ell)}(\mathbf{x},\mathbf{w})$ in \eqref{viterbo_bound}, with $\mathfrak{N}(\mathbf{x}-\mathbf{w})$.
\begin{figure}[ht]
\begin{center}
%\vspace{-1cm}
%\includegraphics[width=3.5in]{Performance3.pdf}
\includegraphics[trim={1.4cm 0 1.3cm 1cm},clip,width=4.5in]{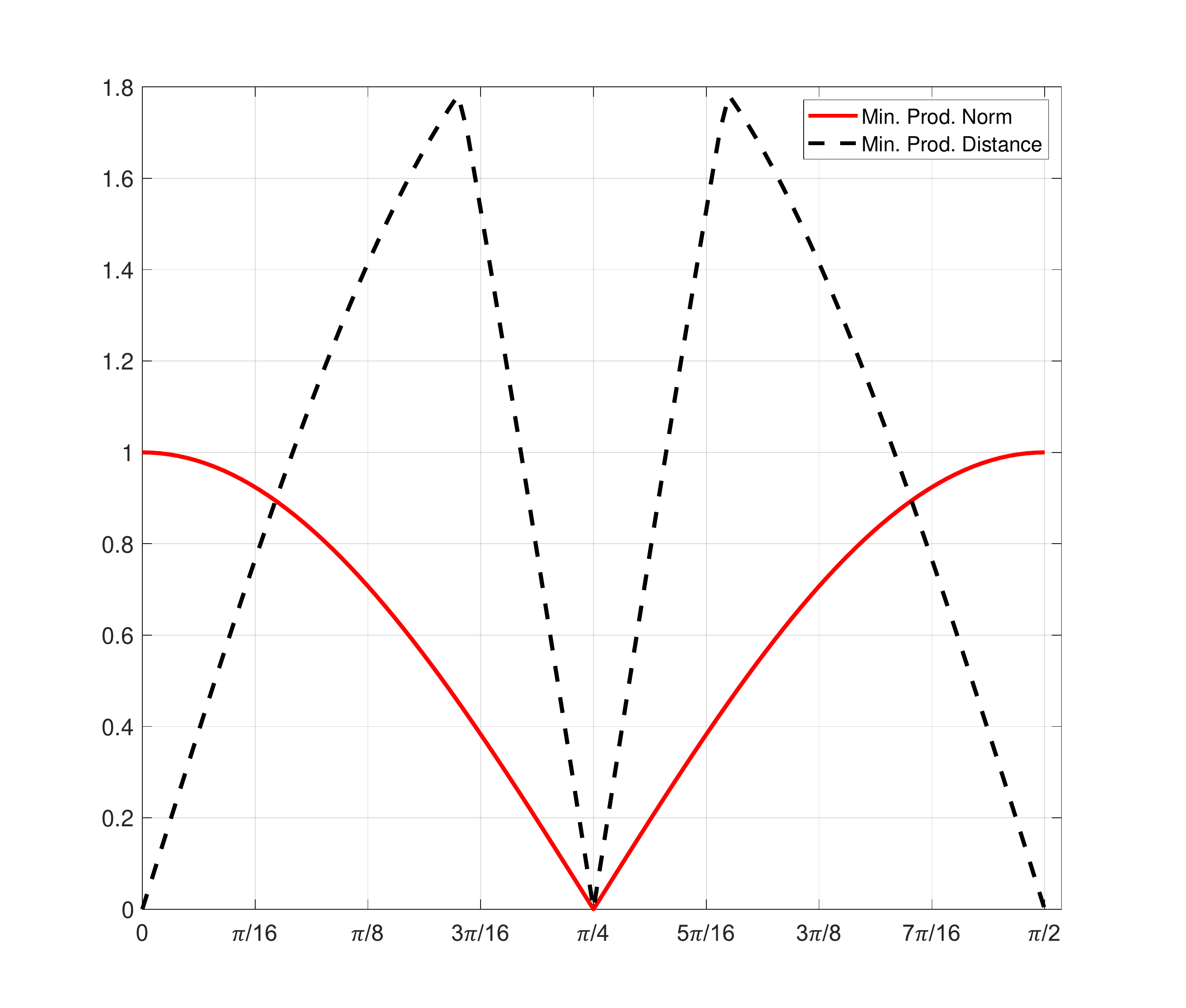}
%\caption{\small{Decoding performance of double-diversity   algebraic LDPC lattices in dimensions
%$n = 200$ and $n=1000$. The plot also shows Poltyrev outage limits.}}
\caption{\small{Comparison of the minimum product norm and the minimum $L$-product distance of different rotations of 4-QAM constellation.}}
\label{figsimrot}
\vspace{-0.5cm}
\end{center}
\end{figure}
}
\begin{Definition}
For a given lattice $\Lambda\subset \mathbb{R}^n$,  the \emph{normalized minimum product distance} is denoted by $Nd_{p,\mathrm{min}}(\Lambda)$ which is obtained by scaling $\Lambda$ to have a
unit size fundamental parallelotope and then taking $d_{p,\mathrm{min}}(\Lambda')$ of the resulting lattice $\Lambda'$. Thus, we have
\begin{IEEEeqnarray}{rCl}\label{Ndpmin}
Nd_{p,\mathrm{min}}(\Lambda)=\frac{d_{p,\mathrm{min}}(\Lambda)}{\mathrm{vol}(\Lambda)}.
\end{IEEEeqnarray}
\end{Definition}
It has been proved that the normalized minimum product distance of the lattices obtained from the ring of integers of number fields depends only on
the discriminant of the field \cite{numberfieldnew}.
\begin{lemma}{\cite[Lemma 3]{numberfieldnew}}\label{lemnum}
Let $K/\mathbb{Q}$ be a totally real number field of degree $n$ and let $\sigma$  be the canonical embedding. Then, $\mathrm{vol}(\sigma(\mathcal{O}_K))=\sqrt{|d_K|}$ and
\begin{IEEEeqnarray}{rCl}\label{Ndpmin2}
Nd_{p,\mathrm{min}}(\sigma(\mathcal{O}_K))=\frac{1}{\sqrt{|d_K|}}.
\end{IEEEeqnarray}
\end{lemma}

\textcolor{mycolor4}{
It is also useful to consider $Nd_{p,\mathrm{min}}^{1/n}$ in order to compare lattices of different dimensions \cite{viterbonew}.
Applying the Chernoff bound on the pairwise error probability of infinite lattices over fading channels  shows that the two relevant design parameters that minimize the PEP are modulation diversity and normalized minimum product distance \cite{viterbonew}. For example, the search for optimal rotated $\mathbb{Z}^n$-lattices in terms of maximal normalized minimum product distance has been done in \cite{viterbonew}.
An algebraic Construction A lattice $\Lambda$ obtained from a number field $K$  is a sub-lattice of $\sigma^N(\mathcal{O}_K^N)$, for some $N$.  According to Lemma~\ref{lemnum}, the normalized minimum product distance of $\Lambda$ is also related to $d_K$.
Hence, in order to find promising algebraic lattices, we need number fields with as small discriminants as possible.
Next, we should select Construction A lattices with the largest normalized minimum product distance. For two full-diversity lattices with the same diversity order and the same minimum product distance, the one with smaller parallelotope or smaller volume, has higher normalized minimum product distance}.
Due to  Theorem~\ref{vol_theorem} and Theorem~\ref{theorem6}, in order to minimize the volume of algebraic lattices it suffices to:
\begin{itemize}
\item minimize the discriminant $d_K$ of the number field $K$,
\item increase the rate of the underlying code $\mathcal{C}$,
\item decrease the alphabet size $p$ of the underlying code $\mathcal{C}$.
\end{itemize}
\textcolor{mycolor3}{The above assertion, in one point of view, indicates the preferability of lower alphabet sizes for underlying code of Construction A lattices. For example, this indicates binary alphabets are preferable for underlying codes of Construction A lattices compared to non-binary alphabets.  This result somehow is confirmed in our simulations (see Section~\ref{Numerical_Results}). In another point of view, this  is in contrast with the known results on AWGN channels in which non-binary Construction A lattices outperform binary ones \cite{Boutros_full_div}. Indeed, binary and non-binary Construction A lattices do not have automatically the same minimum product distance and non-binary Construction A lattices are capable to have larger minimum product distance. In the sequel, we describe an observation which implies  an opposite conclusion about reducing the alphabet size of the underlying code.}

\textcolor{mycolor3}{In our simulations, we observed that decreasing the volume of $\Gamma=\sigma(\mathcal{O}_K)$ or increasing the volume of $\Gamma'=\sigma(\mathfrak{p})$, by choosing an appropriate number filed $K$ and a prime ideal $\mathfrak{p}$ in $\mathcal{O}_K$, improves the error performance of the obtained Construction A lattice $\Lambda$ based on them. We could not prove this observation but we found an explanation for it. Indeed, the reason is related to the error performance of $\Lambda$ over AWGN channels. A necessary but not sufficient condition for a lattice to have good error performance over BF channel is its good error performance over AWGN channel. Construction A lattices are special cases of a larger family of algebraic structures namely \emph{block coset codes} which are proved to be sphere-bound achieving with specific assumptions \cite{Forney}. A block coset code is defined as follows \cite[p. 831]{Forney}.
\begin{Definition}
Let $\Gamma' \subsetneqq \Gamma$ be two nested $n$-dimensional lattices. Let $\mathcal{A}$ be a set of coset
representatives for the cosets of $\Gamma'$ in $\Gamma$ and let $\mathcal{C}$ be a block code
of length $N$  over $\mathcal{A}$, that is, a subset of $\mathcal{A}^N$. Then, a block coset code $\mathcal{L}$ is
\begin{IEEEeqnarray}{rCl}
\mathcal{L}=\left\{\mathbf{x}\in\Gamma^N |\mathbf{x}\equiv \mathbf{c} \bmod{(\Gamma')^N},\,\mbox{for\,some}\,\mathbf{c}\in\mathcal{C} \right\}.
\end{IEEEeqnarray}
If $\mathcal{C}$ is a subgroup of  $\mathcal{A}^N$, then the coset code becomes a lattice.
\end{Definition}
}

Some necessary and  sufficient conditions for a coset code to be \textcolor{mycolor4}{sphere-bound achieving} over AWGN channels are provided in \cite[p. 832]{Forney}. Two of these conditions are expressed as choosing $\mathrm{vol}(\Gamma')$ large enough and $\mathrm{vol}(\Gamma)$ small enough.
\textcolor{mycolor4}{
In this paper, we have considered the coset codes with $\Gamma=\mathcal{O}_K$ and $\Gamma'=\mathfrak{p}$. Applying the provided suggestions in  \cite{Forney} together with our setting verifies our observations.
}This observation motivates the increase of the alphabet size of the underlying code to obtain better performance. Summing up these arguments, no causal inferences can be drawn from the results of this study about the effect of the alphabet size of the underlying codes on the error performance of Construction A lattices over BF channels and we leave it as an open problem.

\begin{remark}
  \textcolor{mycolor3}{In \cite{viterbo}, two disadvantages have been addressed   behind the maximal
diversity and the minimal absolute discriminant design criteria of algebraic lattices.
The main reason for seeking lattices with minimal absolute discriminant is the relation of discriminant and the energy of finite constellations carved from these lattices.
The energy of constellations carved from these lattices is proportional to the volume of lattice and volume is
minimized by selecting the fields with minimum absolute discriminants. The volume can be  reduced further by choosing a complex field, that is, a lattice with $r_2\neq 0$. In this case the volume can be divided by $2^{r_2}$ and the best case in this point of view is working with totally complex fields.
In this sake, the lattices derived from totally real number fields are prone to have bad
performance over a Gaussian channel  mainly due to their high
values of volume. The second disadvantage appears
over the fading channel and is related to the product kissing
number $\tau_p$ which is much higher for real fields lattices than for complex fields lattices \cite{viterbo}.}
\end{remark}
\subsection{Poltyrev outage limit for lattices}
%In the preceding subsections, we introduced the evaluation methods for finite multidimensional constellations, including  lattice constellations.
In order to evaluate infinite lattices over the AWGN channels \cite{QC_LDPC_lattice}, we usually employ Poltyrev limit \cite{polytrev}. Due to this limit, there exists a lattice $\Lambda$, with generator $\mathbf{M}_{\Lambda}$, of high enough dimension $n$ for which the transmission error probability over the AWGN channel decreases  to an arbitrary low value if and only if $\sigma_{\mathcal{N}}^2<\sigma_{max}^2$, where $\sigma_{\mathcal{N}}^2$ is the noise variance per dimension, and  $\sigma_{max}^2$ is the Poltyrev threshold which is given by
\begin{equation}\label{Poltyrev_AWGN}
  \sigma_{max}^2=\frac{\left|\det(\mathbf{M}_{\Lambda})\right|^{\frac{2}{n}}}{2\pi e}.
\end{equation}
Using Poltyrev threshold, a \emph{Poltyrev outage limit} (POL) for lattices over BF channels is proposed in \cite{outage}. It is proved that Poltyrev outage limit has diversity $L$ for a channel with $L$ independent block fadings, that is, Poltyrev outage limit has full-diversity \cite{outage}. Using our notations through this paper, for a fixed instantaneous fading $\mathbf{h}=(h_1,\ldots,h_n)$, Poltyrev threshold becomes \cite{outage}
\begin{equation}\label{Poltyrev_fading}
  \sigma_{max}^2(\mathbf{h})=\frac{\left|\det(\mathbf{M}_{\Lambda})\right|^{\frac{2}{nN}}\prod_{i=1}^n h_i^{\frac{2}{n}}}{2\pi e}.
\end{equation}
The decoding of the lattice with generator $\mathbf{M}_{\Lambda}$ is possible with a vanishing error probability \textcolor{mycolor4}{only if} $\sigma_{\mathcal{N}}^2<\sigma_{max}^2(\mathbf{h})$ \cite{polytrev,outage}.
Thus, for variable fading, an outage event occurs whenever
$\sigma_{\mathcal{N}}^2>\sigma_{max}^2(\mathbf{h})$. The Poltyrev outage limit $P_{out}(\gamma)$ is defined as follows \cite{outage}
\begin{IEEEeqnarray}{rCl}\label{p_out_poly}
  P_{out}(\gamma)&=&\mathrm{Pr}\left(\sigma_{\mathcal{N}}^2>\frac{\left|\det(\mathbf{M}_{\Lambda})\right|^{\frac{2}{nN}}\prod_{i=1}^n h_i^{\frac{2}{n}}}{2\pi e}\right)\nonumber\\
  &=&\mathrm{Pr}\left(\prod_{i=1}^n h_i^2<\frac{(2\pi e)^{n}}{\gamma^n}\right).
\end{IEEEeqnarray}
%where $\left|\det(\mathbf{M}_{\Lambda})\right|=2^{nN+N-k}d_K^{\frac{N}{2}}$ for binary Construction A lattices.
The closed-form expression of $P_{out}(\gamma)$ is not derived in \cite{outage}; however it can be  estimated numerically via Monte Carlo simulation. For a
given lattice, the frame error rate  after lattice decoding over a BF channel, can be compared to $P_{out}(\gamma)$
to measure the gap in SNR and verify the diversity order.
\section{Monogenic Number Fields}\label{monogenic_sec}
In this section, we provide the required algebraic tools for developing Construction A lattices over a wider family of number fields: the monogenic number fields.
\begin{Definition}
Let $K$ be a number field of degree $n$ and $\mathcal{O}_K$ be its ring of integers. If $\mathcal{O}_K$, as a $\mathbb{Z}$-module, has a basis of the form $\left\{1,\alpha,\ldots,\alpha^{n-1}\right\}$, for some $\alpha\in \mathcal{O}_K$, then $\alpha$ is a \emph{power generator}, the basis is a \emph{power basis} and $K$ is a \emph{monogenic number field}.
\end{Definition}

It is a classical problem in algebraic number theory to identify if a number field $K$ is monogenic or not. The quadratic and cyclotomic number fields are monogenic, but in general this is not the case. Dedekind \cite[p. 64]{narkiewicz} was the first to notice this by giving an example of a cubic field generated by a root of $t^3- t^2-2t-8$. The existence of a power generator simplifies the arithmetic in $\mathcal{O}_K$. For instance, if $K$ is monogenic, then the task of factoring
$p\mathcal{O}_K$ into prime ideals over $\mathcal{O}_K$, which is a difficult task in general, reduces to factoring the minimal polynomial of $\alpha$ over $\mathbb{F}_p$, which is significantly easier.

The proposed framework of \cite{ConstA} for developing Construction A lattices assumes that the number field $K$ is a Galois extension of $\mathbb{Q}$. Therefore, our construction method based on monogenic number fields is not a special case of their method since there exist examples of number fields which are monogenic without being Galois extensions. For example let $K = \mathbb{Q}(\alpha)$, where $\alpha^3= 2$ and $\alpha$ is the real cube root of $2$. Then, it is proved \cite[p. 67]{serglang} that $\mathcal{O}_K=\mathbb{Z}[\alpha]$  and $K$ is monogenic. However, it is known that $\mathbb{Q}(\sqrt[3]{2})$ is not a Galois extension.

We start by gathering the proved results about monogenic number fields and then we propose an algorithmic method to develop Construction A over monogenic number fields. We present the results about the number fields with degree less than $4$. More details about monogenic number fields
%and their applications
can be found in \cite{gaal}.
\begin{Theorem}{\cite[p. 76]{serglang}}\label{Th1}
Let $m$ be a non-zero square-free integer and let $K=\mathbb{Q}(\sqrt{m})$.
If $m \equiv 2$ or $3$ $(\bmod \,\,4)$, then $\mathcal{O}_K=\mathbb{Z}[\sqrt{m}]$ and $\left\{1,\sqrt{m}\right\}$ is a basis for $\mathcal{O}_K$ over $\mathbb{Z}$. If $m \equiv  1$ $(\bmod \,\,4)$, then $\mathcal{O}_K=\mathbb{Z}[\frac{1+\sqrt{m}}{2}]$.
\end{Theorem}

Theorem~\ref{Th1} shows that all quadratic fields are monogenic. In the cubic case, however, these studies begin to get more complicated. In fact there are an infinite number of cyclic cubic
fields which have a power basis and also an infinite number which do not, and similarly for quartic fields \cite{Robertson}.
%Generally speaking, one needs to solve a Diophantine equation in order to show that a number
%field is monogenic.

Let $A$ be a Dedekind ring, $K$ its quotient field, $E$ a
finite separable extension of $K$ of degree $n$, and $B$ the integral closure of $A$ in $E$. Let $W = \left\{w_1,\ldots, w_n\right\}$ be any set of $n$ elements of $E$. The discriminant is
\begin{equation}\label{disc_general_def}
  D_{E/K}(W)=\left(\det[\sigma_i(w_j)]_{i,j=1}^n\right)^2,
\end{equation}
where $\sigma_i$'s are $n$ distinct embeddings of $E$ in a given algebraic closure of $K$. If $M$ is a free module of rank $n$ over $A$ (contained in $E$), then we can define the discriminant of $M$ by means of a basis of $M$ over $A$. This notion is well defined up to the square of a unit in $A$.
\begin{Proposition}{\cite[p. 65]{serglang}}\label{prop1}
Let $M_1\subset M_2$ be two free modules of rank $n$ over $A$, contained in $E$. Then $D_{E/K}(M_1)$ divides $D_{E/K}(M_2)$. If $D_{E/K}(M_1)=uD_{E/K}(M_2)$ for some unit $u$ of $A$, then $M_1 = M_2$.
\end{Proposition}
%\begin{Proposition}\cite[p. 15]{neukirch}\label{Th2}
%Let $K$ be a number field of degree $n$ and $\alpha_1,\ldots,\alpha_n\in \mathcal{O}_K$ be linearly independent
%over $\mathbb{Q}$. Let $M$ be the $\mathbb{Z}$-module generated by $\alpha_1,\ldots,\alpha_n$. Then $D_{K/\mathbb{Q}}(M) = [\mathcal{O}_K : M]d_K$. In particular, $D_{K/\mathbb{Q}}(1,\ldots,\alpha^{n-1}) = \textrm{Ind}(\alpha)^2d_K$, for $\alpha\in \mathcal{O}_K$ and $K = \mathbb{Q}(\alpha)$, where $\textrm{Ind}(\alpha) = [\mathcal{O}_K:\mathbb{Z}[\alpha]]$.
%\end{Proposition}
%Let $\mathcal{B} = \left\{1, \beta_1,\ldots ,\beta_n\right\}$ be an integral basis of $\mathcal{O}_K$ and let $\eta=x_1+x_2\beta_2+\cdots+x_n\beta_n\in \mathcal{O}_K$. Then we consider the equation
%\begin{equation}\label{index_form}
%  \textrm{Ind}(\eta)=[\mathcal{O}_K:\mathbb{Z}[\theta]]=I(x_2,\ldots,x_n)=\pm 1
%\end{equation}
%which is solvable if and only if $\left\{1,\theta,\ldots,\theta^{n-1}\right\}$ is a power integral basis of $K$. $I(x_2,\ldots,x_n)$ is called the index form corresponding to the given integral basis.

It is useful to recall the following well-known result.
\begin{lemma}\cite[p. 1-2]{gaal}\label{lemma1}
\textcolor{mycolor3}{Let $K$ be a number field of degree $n$ and $\alpha_1,\ldots,\alpha_n\in \mathcal{O}_K$ be linearly independent elements over $\mathbb{Q}$. Set $Z_K=\mathbb{Z}[\alpha_1, \ldots , \alpha_n]$. Then, we have
\begin{equation*}
D_{K/\mathbb{Q}}(\alpha_1, \ldots , \alpha_n)=J^2\cdot d_K,
\end{equation*}
where $d_K$ is the discriminant of the number field $K$ and $J=[\mathcal{O}_K^{+}:Z_K^{+}]$, in which $\mathcal{O}_K^{+}$ and $Z_K^{+}$ are the additive groups of the modules $\mathcal{O}_K$ and $Z_K$, respectively.}
\end{lemma}

Let $\alpha\in \mathcal{O}_K$ be a primitive element of $K$, that is $K = \mathbb{Q}(\alpha)$. The index of $\alpha$ is
defined by the module index
\begin{equation}\label{index}
  I(\alpha)=[\mathcal{O}_K^{+}:\mathbb{Z}[\alpha]^{+}].
\end{equation}
Obviously, $\alpha$ generates a power integral basis in $K$ if and only if $I(\alpha)=1$. The
minimal index of the field $K$ is defined by
$$\mu(K)=\min_{\alpha} I(\alpha),$$
where the minimum is taken over all primitive integers. The field index of $K$ is
$$m(K)=\min_{\alpha}\gcd I(\alpha),$$
where the greatest common divisor is also taken over all primitive integers of $K$. Monogenic fields have both $\mu(K)= 1$ and $m(K) = 1$, but $m(K) = 1$ is not sufficient for being monogenic.

Let $\left\{1, \omega_2, \ldots , \omega_n\right\}$ be an integral basis of $K$. Let
$$L(\mathbf{x})=x_1+x_2\omega_2+\cdots +x_n\omega_n,$$
with conjugates $L^{(i)}(\mathbf{x})=x_1+x_2\omega_2^{(i)}+\cdots +x_n\omega_n^{(i)}$, where $\omega_j^{(i)}=\sigma_i(\omega_j)$, for $i,j=1,\ldots, n$.
The form $L(\mathbf{x})=L(x_1,\ldots,x_n)$ is  the \emph{fundamental form} and
\begin{equation*}
  D_{K/\mathbb{Q}}\left(L(\mathbf{x})\right)=\prod_{1\leq i<j\leq n}\left(L^{(i)}(\mathbf{x})-L^{(j)}(\mathbf{x})\right)^2
\end{equation*}
is the \emph{fundamental discriminant}.
\begin{lemma}\cite[p. 2]{gaal}
We have
\begin{equation}\label{disc_form}
  D_{K/\mathbb{Q}}\left(L(\mathbf{x})\right)=\left(I(x_2,\ldots,x_n)\right)^2d_K,
\end{equation}
where $d_K$ is the discriminant of the field $K$ and $I(x_2,\ldots ,x_n)$ is a homogeneous form in $n - 1$ variables of degree $n(n- 1)/2$ with integer coefficients. This form $I(x_2,\ldots ,x_n)$ is  the index form corresponding to the integral basis $\left\{1, \omega_2, \ldots , \omega_n\right\}$.
\end{lemma}
\begin{lemma}
For any primitive integer of the form $\alpha=x_1+\omega_2x_2+\cdots +\omega_n x_n\in \mathcal{O}_K$ we have
\begin{equation*}
  I(\alpha)=|I(x_2,\ldots ,x_n)|.
\end{equation*}
\end{lemma}
Indeed, the existence of a power basis is equivalent to the existence of a solution to $I(x_2,\ldots ,x_n) = \pm 1$.
\begin{Theorem}\cite[Theorem 7.1.8]{alaca}\label{disc_squre_free}
Let $K$ be an algebraic number field of degree $n$. Let $\alpha\in \mathcal{O}_K$ be such that $K = \mathbb{Q}(\alpha)$. If $D_{K/\mathbb{Q}}(\alpha)$ is square-free, then $\left\{1,\alpha,\ldots,\alpha^{n-1}\right\}$ is an integral basis for $K$. Indeed, $K$ has a power integral basis.
\end{Theorem}

The computation of the discriminant for some families of polynomials with small degree is a straightforward job. Combining these computations along with the conditions of Theorem~\ref{disc_squre_free} gives some useful results.
\textcolor{mycolor3}{
\begin{Theorem}\cite[Theorems 7.1.10, 7.1.15]{alaca}
%\label{deg3_deg4}
Let $a, b$ be integers such that $x^3 + ax + b$  is irreducible. Let $\theta\in \mathbb{C}$ be a root of $x^3 + ax + b$ so that $K = \mathbb{Q}(\theta)$ is a cubic field and $\theta\in \mathcal{O}_K$. Then $D_{K/\mathbb{Q}}(\theta)=-4a^3 - 27b^2$. If  $D_{K/\mathbb{Q}}(\theta)$ is square-free or $D_{K/\mathbb{Q}}(\theta)=4m$, where $m$ is a square-free integer such that $m\equiv 2$ or $3$ $(\bmod\,\, 4)$, then $\left\{1, \theta, \theta^2\right\}$ is an integral basis for the cubic field $\mathbb{Q}(\theta)$.
 \end{Theorem}
\begin{Theorem}\cite[Theorem 7.1.12]{alaca}
%\label{deg3_deg4}
Let $a, b$ be integers such that $x^4 + ax + b$  is irreducible. Let $\theta\in \mathbb{C}$ be a root of $x^4 + ax + b$ so that $K = \mathbb{Q}(\theta)$ is a quartic field and $\theta\in \mathcal{O}_K$. Then $D_{K/\mathbb{Q}}(\theta)=- 27a^4+256b^3$. If  $D_{K/\mathbb{Q}}(\theta)$ is square-free, then $\left\{1, \theta, \theta^2,\theta^3\right\}$ is an integral basis for the quartic field $\mathbb{Q}(\theta)$.
\end{Theorem}
}
\begin{Theorem}\cite[p. 176]{alaca}\label{Th3}
Let $K = \mathbb{Q}( \sqrt[3]{m})$, with $m \in \mathbb{Z}$ a cube-free number. Assume that $m = hk^2$ with $h, k > 0$ and $hk$ is square-free,  and let $\theta = m^{1/3}$. Then,
\begin{itemize}
  \item for $m^2 \not\equiv 1\pmod{9}$, we have $d_K = -27(hk)^2$, and the numbers $\left\{1,\theta,\theta^2/k\right\}$, form an integral basis of $\mathcal{O}_K$;
  \item for $m^2\equiv\pm 1 \pmod{9}$, we have $d_K = -3(hk)^2$, and the numbers
  $$\left\{1,\theta,\frac{k^2\pm k^2\theta+\theta^2}{3k}\right\},$$
  form an integral basis of $\mathcal{O}_K$.
\end{itemize}
\end{Theorem}
\noindent
This theorem shows that $\mathbb{Q}(\sqrt[3]{p})$ is monogenic for primes $p\equiv \pm 2,\pm5$ $(\bmod\,\, 9)$.

Let $a\in \mathbb{Z}$ be an arbitrary integer  and consider a root $\vartheta$ of the polynomial
\begin{equation}\label{cubic_simplest}
  f(x)=x^3 -ax^2 + (a + 3)x + 1.
\end{equation}
Then, $K = \mathbb{Q}(\vartheta)$ are  the \emph{simplest cubic fields} \cite{cubic}. This cubic equation has discriminant $D = (a^2 + 3a + 9)^2$ and if $a^2 + 3a + 9$ is prime, $D$ is also the discriminant of the field $\mathbb{Q}(\vartheta)$. Accordingly \cite{cubic}, we have $\mathcal{O}_K=\mathbb{Z}[\vartheta]$.
%One may easily verify that the other two roots of (\ref{cubic_simplest}) are
%\begin{equation}\label{roots_simplest_cubic}
%  \vartheta_2=-1/(1+\vartheta),\quad \vartheta_3=-1/(1+\vartheta_2).
%\end{equation}
More information about monogenic number fields with higher degrees can be found in \cite{gaal}. In Section~\ref{new_construction}, we find additional concerns regarding the application of monogenic number fields in this work. These concerns are summarized in this question:  How can we efficiently construct totally real monogenic number fields $K$ of degree $n$ (for arbitrary $n$) with at least one prime ideal $\mathfrak{P}\subset\mathcal{O}_K$ for which $\frac{\mathcal{O}_K}{\mathfrak{P}}\cong \mathbb{F}_2$?
\section{Construction A over Monogenic Number Fields}\label{new_construction}
In this section we give more precise information concerning the splitting of the primes over monogenic number fields that helps us to develop Construction A lattices over monogenic number fields. The construction method is provided for the binary case, but it can be simply modified for the non-binary case.
\begin{Proposition}\cite[p. 27]{serglang}\label{main_theorem}
Let $A$ be a Dedekind ring with quotient field $K$. Let $E$ be a finite separable extension of $K$. Let $B$ be the integral closure of $A$ in $E$ and assume that $B = A[\alpha]$ for some element $\alpha$. Let $f$ be the irreducible
polynomial of $\alpha$ over $K$ and let $\mathfrak{p}$ be a prime of $A$. Consider $\overline{f}$ to be the reduction of $f\,\,(\bmod\,\,\mathfrak{p})$, and let $\overline{f}(x)=\overline{P_1}(x)^{e_1}\cdots \overline{P_r}(x)^{e_r}$
%\begin{equation}\label{poly_decomposition}
% \overline{f}(x)=\overline{P_1}(x)^{e_1}\cdots \overline{P_r}(x)^{e_r},
%\end{equation}
be the factorization of $\overline{f}$ into powers of irreducible factors over  $\overline{A}=A/\mathfrak{p}$. Then
$\mathfrak{p}B=\mathfrak{P}_1^{e_1}\cdots \mathfrak{P}_r^{e_r}$
%, with leading coefficients $1$.
%\begin{equation}\label{prime_decomposition}
%  \mathfrak{p}B=\mathfrak{P}_1^{e_1}\cdots \mathfrak{P}_r^{e_r},
%\end{equation}
is the factorization of $\mathfrak{p}$ in $B$, so that $e_i$ is the ramification index of $\mathfrak{P}_i$ over $\mathfrak{p}$ and
\begin{equation}\label{prime_formula}
  \mathfrak{P}_i=\mathfrak{p}B+P_i(\alpha)B,
\end{equation}
where $P_i\in A[x]$ is a polynomial with leading coefficient $1$ whose reduction $\bmod\, \mathfrak{p}$ is $\overline{P_i}$.
For each $i$, $\mathfrak{P}_i$ has residue class degree $[B/\mathfrak{P}_i:A/\mathfrak{p}]=d_i$, where $d_i=\textrm{deg}(\overline{P_i})$.
\end{Proposition}

\textcolor{mycolor3}{For using Proposition~\ref{main_theorem} in our case, we have} $A=\mathbb{Z}$, $K=\mathbb{Q}$, $E=\mathbb{Q}(\alpha)$, $B=O_E=\mathbb{Z}[\alpha]$ and $\mathfrak{p}=2\mathbb{Z}$. Let $f$ be the minimal polynomial of $\alpha$ over $\mathbb{Q}$ and $\overline{f}=f\,\,(\bmod \,\, 2)$. Write the decomposition of $\overline{f}$ in $\mathbb{F}_2[x]$ as follows
\begin{equation*}
 \overline{f}(x)=\overline{P_1}(x)^{e_1}\cdots \overline{P_r}(x)^{e_r}.
\end{equation*}
Then, we have
\begin{equation*}
  2O_E=\mathfrak{P}_1^{e_1}\cdots \mathfrak{P}_r^{e_r},
\end{equation*}
where $\mathfrak{P}_j=2O_E+P_j(\alpha)O_E$, for $j=1,\ldots,r$. If there exists $\overline{P_i}$ such that $d_i=\textrm{deg}(\overline{P_i})=1$ then $O_E/\mathfrak{P}_i\cong\mathbb{F}_2$. Now, we can define the map $\rho:O_E^N\rightarrow \mathbb{F}_2^N$ as componentwise reduction modulo $\mathfrak{P}_i$ and develop the Construction A lattice $\Gamma_{\mathcal{C}}=\rho^{-1}(\mathcal{C})$ for an $[N,k]$ linear code $\mathcal{C}$.

\textcolor{mycolor5}{
Given the property of the proposed lattices and the considerations of Section~\ref{Design_sec}, the proposed construction is a reasonably good candidate for lattice decoding over fading channels. Summing up all together, gives the following heuristic criterion:
%For a number field $K$, the sufficient conditions to provide a Construction A lattice with good error performance are: 1) being totally real; 2) being monogenic and 3)  having a generator for which the minimal polynomial admits a linear factor after reduction modulo $2$. Considering such number field with the least discriminant in addition to  a high-rate binary code results in the best error performance.
\begin{enumerate}
\item  the number field $K$ should be  totally real;
\item  the number field $K$ should be  monogenic;
\item  the number field $K$ should have a generator for which the minimal polynomial admits a linear factor after reduction modulo $2$;
\item the number field $K$ should have the least discriminant among the totally real monogenic number fields of the same degree.
\end{enumerate}
%The sufficient conditions  for  a number field $K$ to provide a full-diversity binary Construction A lattice
%whose upper bound on the decoding error probability \eqref{viterbo_bound} is minimized are:
%%with error performance close to Poltyrev outage limit  are:
%1) being totally real; 2) being monogenic; 3)  having a generator for which the minimal polynomial admits a linear factor after reduction modulo $2$, and 4) having the least discriminant among the totally real monogenic number fields of the same degree.
%\end{Corollary}
%\begin{IEEEproof}
Among the above conditions, being totally real provides full-diversity and being monogenic is sufficient to have a simple method for decomposing ideals to prime ideals using Proposition~\ref{main_theorem}. For employing the binary codes as underlying code, having a prime ideal $\mathfrak{P}\subset\mathcal{O}_K$ with $\frac{\mathcal{O}_K}{\mathfrak{P}}\cong \mathbb{F}_2$ is necessary. This requirement has  been reduced to the third condition according to the preceding discussion. The last requirement is assumed due to the
intuition provided in Section~\ref{Design_sec} and also the simulation results.
%\end{IEEEproof}
}

As the simplest case, we present our method for BF channels with two fading blocks, that is,  $n=2$.  We require  quadratic fields of the form $K=\mathbb{Q}(\sqrt{m})$, where $m$ is a positive square-free integer; these fields are totally real. Theorem~\ref{Th1} determines the structure of $\mathcal{O}_K$ for these number fields.
\begin{Theorem}\label{quad_theorem}
Let $K=\mathbb{Q}(\sqrt{m})$. Then, $2\mathcal{O}_K$ is totally ramified with $2\mathcal{O}_K\cong \mathfrak{P}^2$
when $m\equiv 2$ $(\bmod\,\,4)$ and $\mathfrak{P}=2\mathbb{Z}[\sqrt{m}]+\sqrt{m}\mathbb{Z}[\sqrt{m}]$, or
$m\equiv 3$ $(\bmod\,\,4)$, $\mathfrak{P}=2\mathbb{Z}[\sqrt{m}]+(\sqrt{m}+1)\mathbb{Z}[\sqrt{m}]$.
%If $m\equiv 2$ or $3$ $(\bmod\,\,4)$ then the ideal $2\mathcal{O}_K$ is totally ramified. Indeed, if $m\equiv 2$ $(\bmod\,\,4)$, $2\mathcal{O}_K\cong \mathfrak{P}^2$, where $\mathfrak{P}=2\mathbb{Z}[\sqrt{m}]+\sqrt{m}\mathbb{Z}[\sqrt{m}]$, if $m\equiv 3$ $(\bmod\,\,4)$, $\mathfrak{P}=2\mathbb{Z}[\sqrt{m}]+(\sqrt{m}+1)\mathbb{Z}[\sqrt{m}]$.
In both of these cases we have $\mathcal{O}_K/\mathfrak{P}\cong\mathbb{F}_2$.
If $m\equiv 1$ $(\bmod\,\,4)$, then $2\mathcal{O}_K$ is not totally ramified, but if $(m-1)/4$ is an even number, then  $2\mathcal{O}_K\cong \mathfrak{P}_1\mathfrak{P}_2$ and $\mathcal{O}_K/\mathfrak{P}_i\cong\mathbb{F}_2$, $i=1,2$, where $\mathfrak{P}_1=2\mathbb{Z}[\alpha]+\alpha\mathbb{Z}[\alpha]$ and $\mathfrak{P}_2=2\mathbb{Z}[\alpha]+(\alpha+1)\mathbb{Z}[\alpha]$, with $\alpha=(1+\sqrt{m})/2$.
\end{Theorem}
\begin{IEEEproof}
All quadratic fields of the form $\mathbb{Q}(\sqrt{m})$, where $m$ is a positive square-free integer, are monogenic and totally real. If $m\equiv 2$ or $3$ $(\bmod\,\,4)$, then $\alpha=\sqrt{m}$ is the generator of the power integral basis with minimal polynomial $f(x)=x^2-m$. In this case, $f$ always has a linear factor after reduction modulo $2$. Indeed, we have $\overline{f}(x)=x^2$ for even $m$'s and $\overline{f}(x)=(x+1)^2$ for odd $m$'s. If $m\equiv 1$ $(\bmod\,\,4)$ then $\alpha=(1+\sqrt{m})/2$ is the generator of power integral basis with minimal polynomial $f(x)=x^2-x-(m-1)/4$. It can be easily seen that in this case, $f$ has a linear factor after reduction modulo $2$  if and only if $(m-1)/4$ is an even number, that is, $m\equiv 1$ $(\bmod\,\,8)$. In this case, $\overline{f}(x)=x(x+1)$. The rest of the proof follows from Proposition~\ref{main_theorem}.
\end{IEEEproof}

In all cases of Theorem~\ref{quad_theorem}, there is at least one prime ideal $\mathfrak{P}_i$ in $\mathcal{O}_K$ such that $\mathcal{O}_K/\mathfrak{P}_i\cong\mathbb{F}_2$.  Define the map $\rho:\mathcal{O}_K^N\rightarrow \mathbb{F}_2^N$ as componentwise reduction modulo $\mathfrak{P}_i$ and implement the Construction A lattice $\Gamma_{\mathcal{C}}=\rho^{-1}(\mathcal{C})$ for an $[N,k]$  binary LDPC code $\mathcal{C}$. Then, $\Lambda=\sigma^N(\Gamma_{\mathcal{C}})$ is an algebraic LDPC lattice of diversity order $2$ in $\mathbb{R}^{2N}$.
\begin{Example}
We have seen that the simplest cubic fields of the form $K=\mathbb{Q}(\vartheta)$, where $\vartheta$ is a root of the polynomial $f(x)=x^3 - ax^2 + (a + 3)x + 1$, are totally real monogenic number fields, when $a^2+3a+9$ is a prime number. Even though this condition holds, these families of number fields are useless for our case since for each $a\in\mathbb{Z}$, $x^3 -ax^2 + (a + 3)x + 1$ $(\bmod \,\,2)$ is one of the polynomials  $x^3+x^2+1$ or $x^3+x+1$ and both of these polynomials are irreducible over $\mathbb{F}_2$.

Another examples are $K=\mathbb{Q}(\theta)$ where $\theta$ has minimal polynomial of the form $x^3+ax+b$. In this case, if $-4a^3-27b^2$  or $(-4a^3-27b^2)/4$ are square free  then $K$ is monogenic. For example put $a=-1$ and $b=-2$. Then $-4a^3-27b^2=-104=-26\cdot 2^2$ which is a square-free integer after dividing by $4$. Hence, $K=\mathbb{Q}(\theta)$ where $f(\theta)=\theta^3-\theta-2=0$ is a monogenic number field \cite[Example 7.1.4]{alaca}. We have
$$\overline{f}(x)=x^3+x=x(x+1)^2.$$
Due to this factorization, each one of the primes $\mathfrak{P}_1=2\mathbb{Z}[\theta]+2\theta\mathbb{Z}[\theta]$ or $\mathfrak{P}_2=2\mathbb{Z}[\theta]+2(\theta+1)\mathbb{Z}[\theta]$  gives us $\mathcal{O}_K/\mathfrak{P_i}\cong \mathbb{F}_2$. It can be easily checked  that $\mathbb{Q}(\theta)$ is not totally real  which is the only problem about these family of cubic polynomials.

Pure cubic fields of the form $\mathbb{Q}(\sqrt[3]{p})$  are monogenic for primes $p\equiv \pm 2,\pm 5$ $(\bmod\,\,9)$. In this case the  factorization of $x^3-p$ always has a linear factor. Unfortunately,  all pure cubic
fields are complex.
$\hfill\square$\end{Example}

In the existing number fields of degree $3$, we did not find any parametric family for which both being totally real and having linear factor after reduction modulo $2$ hold. There are several numerical studies for finding monogenic number fields. An excellent account is provided in the tables of~\cite[Section 11]{gaal} containing  all generators of power integral bases for
$130$ cubic fields with small discriminants (both positive and negative), cyclic quartic,  totally real and totally complex biquadratic number fields up to discriminants $10^6$ and $10^4$, respectively. Furthermore,  the five totally real cyclic sextic fields with smallest discriminants,  the $25$ sextic fields with an imaginary quadratic subfield with smallest absolute value of discriminants and their generators of power integral bases are also given in \cite{gaal}.

We could  generate  many examples of number fields with different degrees of which the aforementioned two conditions are fulfilled. We used SAGE \cite{sage} to generate these examples but most of these results were already included in~\cite{gaal}.

Let us analyze the results of \cite{gaal} about totally real cubic fields. The provided table in \cite[Table 11.1.1]{gaal} contains all power integral bases of totally real cubic fields of discriminants $49 \leq d_K \leq 3137$. The rows contain the following data: $d_K$, $(a_1,a_2,a_3)$, where $d_K$ is the discriminant of the field $K$, generated by a root $\vartheta$ of the polynomial $f(x)=x^3+a_1x^2+a_2x+a_3$, and $(I_0,I_1,I_2,I_3)$ coefficients of the index form equation. In most of these fields $\left\{1, \omega_2=\vartheta,\omega_3=\vartheta^2\right\}$ is an integral basis; if not, then an integral basis is given by $\left\{1, \omega_2, \omega_3\right\}$ with $\omega_2=(u_0+u_1\vartheta+u_2\vartheta^2)/u$, $\omega_3 = (v_0+v_1\vartheta+v_2\vartheta^2)/v$ and the table includes the coefficients $\omega_2 = (u_0, u_1, u_2)/u$, $\omega_3 = (v_0, v_1, v_2)/v$.
Finally, the solutions $(x, y)$, of the index form equation are displayed. All generators of power integral bases of the field $K$ are of the form
$\alpha=a\pm(x\omega_2+y\omega_3)$,
where $a\in \mathbb{Z}$ is arbitrary and $(x, y)$ is a solution of the index form equation. For  $\overline{a_i}\equiv a_i$ $(\bmod\,\, 2)$, $1\leq i\leq 3$, the polynomial $f$ admits a linear factor after reduction modulo $2$, in one of the following cases
\begin{enumerate}
  \item $\overline{a_3}=0$;
  \item $\overline{a_1}\neq 0$ and $\overline{a_2}=\overline{a_3}=0$;
  \item $\overline{a_1}\neq 0$, $\overline{a_2}\neq 0$ and $\overline{a_3}\neq 0$.
  %\item $\overline{a_1}=0$, $\overline{a_2}=0$ and $\overline{a_3}= 0$.
\end{enumerate}
Consequently, for the following values of discriminant in~\cite[Table 11.1.1]{gaal}, we obtain a full-diversity Construction A lattice with binary linear codes as underlying code
$$\begin{array}{l}
  148,229,316,404,469,564,568,621,733,756, \\
  788,837,892,940,1016,1076,1101,1229,1300,1373, \\
  1384,1396,1436,1492,1524,1556,1573,1620,1708,1765, \\
  1901,1940,1944,1957,2021,2024,2101,2213,2296,2300, \\
  2349,2557,2597,2677,2700,2708,2804,2808,2836,2917, \\
  2981,3021,3028,
\end{array}$$
which is $53/93$ or $57\%$ of the cases.
\begin{Example}\label{example_div_3}
Consider the number field $K=\mathbb{Q}(\nu)$, where $\nu$ is the root of the polynomial $f(x)=ax^3+bx^2+cx+d=x^3-x^2-3x+1$. Due to the above discussion, $K$ is monogenic with $d_K=148$ and $\mathcal{O}_K=\mathbb{Z}[\nu]$. Since the discriminant of $f$, which is $\Delta=18abcd-4b^3d+b^2c^2-4ac^3-27a^2d^2=148$, is positive $f$ has $3$ real roots as follows
\begin{IEEEeqnarray*}{rCl}
  x_1 &=& \frac{-1}{3}\left(-1+\zeta^0C+\frac{\Delta_0}{\zeta^0C}\right)=-1.4812, \\
  x_2 &=& \frac{-1}{3}\left(-1+\zeta^1C+\frac{\Delta_0}{\zeta^1C}\right)=2.170086, \\
  x_3 &=& \frac{-1}{3}\left(-1+\zeta^2C+\frac{\Delta_0}{\zeta^2C}\right)=0.311107,
\end{IEEEeqnarray*}
in which $\Delta_0=b^2-3ac$, $\zeta=\frac{-1}{2}+\frac{\sqrt{3}}{2}i$ and
$$C=\sqrt[3]{\frac{\Delta_1\pm \sqrt{\Delta_1^2-4\Delta_0^3}}{2}},\quad\Delta_1=2b^3-9abc+27a^2d.$$
The integral basis of $K$ is generated by $\nu=x_1$ as $\left\{1,\nu,\nu^2\right\}$ and using the embeddings $\sigma_1$ that sends $x_1$ to $x_1$, $\sigma_2$ that sends $x_1$ to $x_3$ and $\sigma_3$ that sends $x_1$ to $x_2$, gives us
\begin{equation*}
  \mathbf{M}=\left[
               \begin{array}{ccc}
                 1 & 1 & 1 \\
                 x_1 & x_3 & x_2 \\
                 x_1^2 & x_3^2 & x_2^2 \\
               \end{array}
             \right],
\end{equation*}
as the generator matrix of the lattice $\sigma(\mathcal{O}_K)$. Decomposing  $\overline{f}(x)\equiv f(x)\pmod{2}=x^3+x^2+x+1$ as $(x+1)^3$ admits the following decomposition
$$2\mathcal{O}_K=\mathfrak{P}^3,\quad \frac{\mathcal{O}_K}{\mathfrak{P}}\cong \mathbb{F}_2,$$
where $\mathfrak{P}=2\mathcal{O}_K+(x_1+1)\mathcal{O}_K$ is a prime ideal of $\mathcal{O}_K$.
It can be checked that $\left\{2,x_1+1,x_1^2-x_1-2\right\}$ is a $\mathbb{Z}$-basis for $\mathfrak{P}$. Thus, the generator matrix of the lattice $\sigma(\mathfrak{P})$ is
\begin{equation*}
  \mathbf{DM}=\left[
               \begin{array}{ccc}
                 2 & 2 & 2 \\
                 x_1+1 & x_3+1 & x_2+1 \\
                 x_1^2-x_1-2 & x_3^2-x_3-2 & x_2^2-x_2-2 \\
               \end{array}
             \right].
\end{equation*}
Now, we consider an $[N,k]$-LDPC code with parity-check matrix $\mathbf{H}_{\mathcal{C}}$ and generator matrix $\mathbf{G}_{\mathcal{C}}=\left[
                            \begin{array}{cc}
                              \mathbf{I}_k & \mathbf{A} \\
                            \end{array}
                          \right]
$ that gives us the parity-check  and generator matrices of the triple diversity   algebraic LDPC lattice $\Lambda=\sigma^N(\Gamma_{\mathcal{C}})$ as  $\mathbf{H}_{\Lambda}$ and $\mathbf{M}_{\Lambda}$ in Theorem~\ref{theorem_parity}, respectively.
$\hfill\square$\end{Example}
\begin{Example}
Next, we analyze the totally real quartic number fields. First examples of such fields are simplest quartic fields which have power integral in only two cases; see \cite{gaal}. These two cases are $K_2=\mathbb{Q}(\vartheta_2)$ and $K_4=\mathbb{Q}(\vartheta_4)$ where $\vartheta_2$ is a root of $f(x)=x^4-2x^3-6x^2+2x+1$ and $\vartheta_4$ is a root of $f(x)=x^4-4x^3-6x^2+4x+1$. The integral bases and solutions of index form equations  with respect to these bases have been presented in \cite{gaal}. Let $\left\{1,\omega_1,\omega_2,\omega_3\right\}$ represent the integral bases of $K_2$ and $K_4$. The generators of the power integral basis of $K_2$ and $K_4$ are of the form $\alpha=a+x_1\omega_1+x_2\omega_2+x_3\omega_3$, where $a\in\mathbb{Z}$ is arbitrary and $(x_1,x_2,x_3)$ is a solution of the corresponding index form equations of $K_2$ and $K_4$. For each $\alpha$ of this form we need to find its minimal polynomial over $\mathbb{Q}$ to check whether its reduction modulo $2$ has linear factors or not.  The minimal polynomials have been computed using SAGE~\cite{sage} and are presented in \tablename~\ref{table1} and \tablename~\ref{table2} for $K_2$ and $K_4$, respectively.
%In the following tables, $(a_0,a_1,a_2,a_3)$ represents the polynomial $x^4+a_0x^3+a_1x^2+a_2x+a_3$.
\begin{table}[h]
\begin{center}
\small
\caption{Minimal polynomials of simplest quartic fields for $a=2$.}\label{table1}
\renewcommand{\arraystretch}{1.3}
\begin{tabular}{|c||l|}
  \hhline{-||-}
  $(x_1,x_2,x_3)$ & Minimal Polynomial \\
  \hhline{=::=}
  $(0,1,0)$ & $t^4-10t^3+25t^2-20t+5$  \\

  $(-1,1,0)$ & $t^4-8t^3+19t^2-12t+1$ \\

  $(6,5,-2)$ & $t^4-22t^3+169t^2-508t+421$ \\

  $(0,4,-1)$ & $t^4-20t^3+115t^2-260t+205$ \\

  $(-12,-4,3)$ & $t^4-4t^3-29t^2-44t-19$ \\

  $(-8,-3,2)$ & $t^4+6t^3+t^2-4t-1$ \\

  $(1,1,0)$& $t^4-12t^3+19t^2-8t+1$ \\

  $(-2,1,0)$ & $t^4-6t^3+t^2+4t+1$ \\

  $(-13,-9,4)$ & $t^4+36t^3+451t^2+2176t+2641$ \\

  $(4,2,-1)$ & $t^4-8t^3+19t^2-12t+1$  \\
  \hhline{-||-}
\end{tabular}
\end{center}
\end{table}
\begin{table}[h]
\begin{center}
\small
\caption{Minimal polynomials of simplest quartic fields for $a=4$.}\label{table2}
\renewcommand{\arraystretch}{1.3}
\begin{tabular}{|c||l|}
   \hhline{-||-}
  $(x_1,x_2,x_3)$ & Minimal Polynomial \\
  \hhline{=::=}
  $(3,2,-1)$ & $t^4-4t^3+2t^2+4t-1$  \\
  $(-2,-2,1)$ & $t^4-8t^2-8t-2$ \\
  $(4,8,-3)$ & $t^4-24t^3+208t^2-760t+958$ \\
  $(-6,-7,3)$ & $t^4+16t^3+88t^2+200t+158$ \\
  $(0,3,-1)$ & $t^4-8t^3+16t^2-8t-2$ \\
  $(1,3,-1)$ & $t^4-12t^3+50t^2-84t+47$ \\
  \hhline{-||-}
\end{tabular}
\end{center}
\end{table}
\begin{table*}[!t]
\centering
\small
\caption{Monogenic totally real bicyclic biquadratic number fields.}
\renewcommand{\arraystretch}{1.5}
\begin{tabular}{|c|c|c|c|c|c|c|}
%  \cline{1-7}
%  & & & & & &\\
 \hline
  $d_K$ & $m$ & $n$ & $l=(m,n)$ & $\alpha$ & Minimal Polynomial $f_\alpha$ & Linear factor in $\overline{f_{\alpha}}$\\
  \hhline{=======}
  $2304$ & $2$ & $3$ & $1$ & $\frac{\sqrt{2}+\sqrt{6}}{2}$ & $t^4-4t^2+1$ & Yes\\
  $7056$ & $7$ & $3$ & $1$ & $\frac{\sqrt{7}+\sqrt{3}}{2}$ & $t^4-5t^2+1$ &No\\
  $24336$ & $39$ & $3$ & $3$ & $-\sqrt{39}+2\frac{\sqrt{39}+\sqrt{3}}{2}+\frac{1+\sqrt{13}}{2}$ & $t^4-2t^3-11t^2+12t-3$ &No\\
  $57600$ & $6$ & $15$ & $3$ & $\frac{\sqrt{6}+\sqrt{10}}{2}$ & $t^4-8t^2+1$ & Yes\\
  $94846$ & $11$ & $7$ & $1$ & $\frac{\sqrt{11}+\sqrt{7}}{2}$ & $t^4-9t^2+1$ & No\\
  $313600$ & $10$ & $35$ & $5$ & $\frac{\sqrt{10}+\sqrt{14}}{2}$ & $t^4-12t^2+1$ &Yes\\
  $435600$ & $15$ & $11$ & $1$ & $\frac{\sqrt{11}+\sqrt{15}}{2}$ & $t^4-13t^2+1$ &No\\
  $659344$ & $203$ & $7$ & $7$ & $-\sqrt{203}+2\frac{\sqrt{203}+\sqrt{7}}{2}+\frac{1+\sqrt{203}}{2}$ & $t^4-2t^3-27t^2+28t-7$ &No\\
  \hline
\end{tabular}\\
\label{table3}
% \vspace*{4pt}
%\hrulefill
\end{table*}
We have that the minimal polynomials of the power generators of $K_2$ are equivalent to $t^4+t^2+1$ modulo $2$ which has no linear factor. For $K_4$, all of them are equivalent to either  $t^4$ or $t^4+1$ which have linear factors. It can be shown that  $d_{K_2}=2000$ and $d_{K_4}=2048$.

Totally real bicyclic biquadratic number fields are other examples. Using the algorithm described in \cite[Section 6.5.2]{gaal}, the minimal index $\mu(K)$ and all elements with minimal index in the $196$ totally real bicyclic biquadratic
number fields $K = \mathbb{Q}(\sqrt{m},\sqrt{n})$ with discriminant smaller than $ 10^6$ have been determined. The results are gathered in \cite[Table 11.2.5]{gaal}. In this table, the solutions of index form equation $I(x_2,x_3,x_4)=\mu(K)$ has been proposed. The cases with $\mu(K)=1$ are the cases that $K$ has power integral basis. In the cases that $K$ has a power integral basis with power generator $\alpha$,  we have computed the minimal polynomial and the results are summarized in \tablename~\ref{table3}.
$\hfill\square$\end{Example}

More quartic fields with certain signatures and Galois
groups are computed and gathered in \cite[Section 11.2.7]{gaal}.
The  tables in \cite[Section 11.2.7]{gaal} contain the following data. In the first column the discriminant of the field $K = \mathbb{Q}(\xi)$, the second column contains the coefficients $(a_1, a_2, a_3, a_4)$ of the minimal polynomial $f_{\xi}(x) = x^4 + a_1x^3 + a_2x^2 + a_3x + a_4$
of $\xi$. In the third column the minimal $m$ for which the index form equation
$I(x_2, x_3, x_4) = \pm m$ has solutions with $|x_2|, |x_3|, |x_4| < 10^{10}$. It is followed by an
integral basis of $K$ in case the integral basis is not the power basis.
Last column contains  the solutions $(x_2, x_3, x_4)$ with absolute values smaller than $ 10^{10}$ of the index
form equation $I(x_2,x_3,x_4)=\pm m$. We have collected the cases that $\mathbb{Q}(\xi)$ has a power integral basis and  $f_{\xi}$ admits a linear factor after reduction modulo $2$. We have presented these cases by their discriminants in the following lists:
\setdefaultleftmargin{0cm}{2cm}{}{}{}{}
\begin{enumerate}[1)]
\item totally real quartic fields with Galois group $A_4$
\begin{equation*}
\begin{array}{l}
  26569,33489,121801,165649,261121,270400,299209, \\
  346921,368449,373321,408321,423801,473344,\\
  502681,529984,582169,660969,877969;
\end{array}
\end{equation*}
\item totally real quartic fields with Galois group $S_4$
\begin{equation*}
\begin{array}{l}
  2777,6224,6809,7537,8468,10273,10889,11324, \\
  11344,11348,13676,13768,14656,15188,15529,15952.
\end{array}
\end{equation*}
\end{enumerate}
\section{Iterative Decoding of Full-diversity Algebraic LDPC Lattices}\label{decod2}
In this section we propose a new decoder for full-diversity  algebraic LDPC lattices, which is based on standard sum-product decoder of binary LDPC codes and sphere decoder \cite{sphere_dec} of low dimensional lattices. We also analyze the decoding complexity of the proposed algorithm.
To simulate the operation of our decoding algorithm, we use Rayleigh BF channel model; see Section \ref{system_model}.
%=============================================================================
%Rayleigh fading is a reasonable model when there are many objects in the environment that scatter the radio signal before it arrives at the receiver. Due to the central limit theorem, if there is sufficiently much scatter, the channel impulse response is modelled as a Gaussian process. If the scatters have no dominant components, then such a process will have zero mean and phase evenly distributed between $0$ and $2\pi$ radians. Thus, the envelope of the channel response is Rayleigh distributed. Often, the gain and phase elements of such channel's distortion are  represented as complex numbers. In this case, Rayleigh fading is exhibited by a complex random variable with real and imaginary parts  modelled by independent and identically distributed zero-mean Gaussian processes.
%With the aid of an in-phase/quadrature component interleaver \cite{ConstA,alglattice1}, it is possible to
%remove the phase of the complex fading coefficients to obtain a real fading which is Rayleigh distributed and guarantee that the fading coefficients are independent from one real symbol to the next.

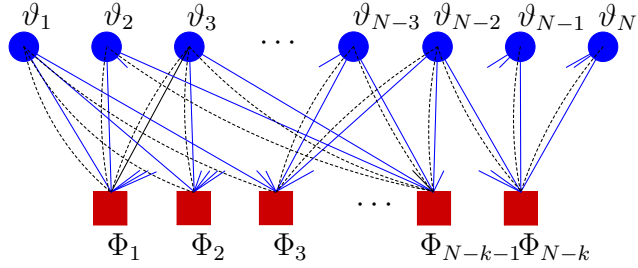
\begin{figure}[bht]
\centering
\definecolor{ccqqqq}{rgb}{0.8,0.,0.}
\definecolor{qqqqff}{rgb}{0.,0.,1.}
\begin{tikzpicture}[line cap=round,line join=round,>=triangle 45,x=0.55cm,y=0.55cm]
\clip(-2.5,-3.2) rectangle (13.3,3.2);
\draw [color=qqqqff,fill=qqqqff,fill opacity=1.0] (0.,2.) circle (0.1958295782955816cm);
\draw [color=qqqqff,fill=qqqqff,fill opacity=1.0] (2.,2.) circle (0.19582957829558148cm);
\draw [color=qqqqff,fill=qqqqff,fill opacity=1.0] (5.968123613353219,2.0036487133059837) circle (0.19582957829558065cm);
\draw [color=qqqqff,fill=qqqqff,fill opacity=1.0] (8.,2.) circle (0.19582957829558237cm);
\draw [color=qqqqff,fill=qqqqff,fill opacity=1.0] (10.,2.) circle (0.19582957829558148cm);
\draw [color=qqqqff,fill=qqqqff,fill opacity=1.0] (12.,2.) circle (0.19582957829558148cm);
\draw [color=qqqqff,fill=qqqqff,fill opacity=1.0] (-2.,2.) circle (0.1958295782955817cm);
\fill[color=ccqqqq,fill=ccqqqq,fill opacity=1.0] (-0.3104072285142831,-1.510438906256035) -- (-0.3104072285142831,-2.3104389062560333) -- (0.4895927714857198,-2.3104389062560333) -- (0.4895927714857198,-1.510438906256035) -- cycle;
\fill[color=ccqqqq,fill=ccqqqq,fill opacity=1.0] (1.7039177652434612,-1.503029812663545) -- (1.7039177652434612,-2.303029812663545) -- (2.503917765243461,-2.303029812663545) -- (2.503917765243461,-1.503029812663545) -- cycle;
\fill[color=ccqqqq,fill=ccqqqq,fill opacity=1.0] (3.6943731093482617,-1.5025459353625241) -- (3.6943731093482617,-2.3025459353625237) -- (4.494373109348263,-2.3025459353625237) -- (4.494373109348263,-1.5025459353625241) -- cycle;
\fill[color=ccqqqq,fill=ccqqqq,fill opacity=1.0] (7.513393340936015,-1.5104477980938014) -- (7.513393340936015,-2.3104477980938007) -- (8.313393340936013,-2.3104477980938007) -- (8.313393340936013,-1.5104477980938014) -- cycle;
\fill[color=ccqqqq,fill=ccqqqq,fill opacity=1.0] (9.614504018844382,-1.5113116207855286) -- (9.614504018844382,-2.3113116207855287) -- (10.414504018844383,-2.3113116207855287) -- (10.414504018844383,-1.5113116207855286) -- cycle;
\draw [color=ccqqqq] (-0.3104072285142831,-1.510438906256035)-- (-0.3104072285142831,-2.3104389062560333);
\draw [color=ccqqqq] (-0.3104072285142831,-2.3104389062560333)-- (0.4895927714857198,-2.3104389062560333);
\draw [color=ccqqqq] (0.4895927714857198,-2.3104389062560333)-- (0.4895927714857198,-1.510438906256035);
\draw [color=ccqqqq] (0.4895927714857198,-1.510438906256035)-- (-0.3104072285142831,-1.510438906256035);
\draw [color=ccqqqq] (1.7039177652434612,-1.503029812663545)-- (1.7039177652434612,-2.303029812663545);
\draw [color=ccqqqq] (1.7039177652434612,-2.303029812663545)-- (2.503917765243461,-2.303029812663545);
\draw [color=ccqqqq] (2.503917765243461,-2.303029812663545)-- (2.503917765243461,-1.503029812663545);
\draw [color=ccqqqq] (2.503917765243461,-1.503029812663545)-- (1.7039177652434612,-1.503029812663545);
\draw [color=ccqqqq] (3.6943731093482617,-1.5025459353625241)-- (3.6943731093482617,-2.3025459353625237);
\draw [color=ccqqqq] (3.6943731093482617,-2.3025459353625237)-- (4.494373109348263,-2.3025459353625237);
\draw [color=ccqqqq] (4.494373109348263,-2.3025459353625237)-- (4.494373109348263,-1.5025459353625241);
\draw [color=ccqqqq] (4.494373109348263,-1.5025459353625241)-- (3.6943731093482617,-1.5025459353625241);
\draw [color=ccqqqq] (7.513393340936015,-1.5104477980938014)-- (7.513393340936015,-2.3104477980938007);
\draw [color=ccqqqq] (7.513393340936015,-2.3104477980938007)-- (8.313393340936013,-2.3104477980938007);
\draw [color=ccqqqq] (8.313393340936013,-2.3104477980938007)-- (8.313393340936013,-1.5104477980938014);
\draw [color=ccqqqq] (8.313393340936013,-1.5104477980938014)-- (7.513393340936015,-1.5104477980938014);
\draw [color=ccqqqq] (9.614504018844382,-1.5113116207855286)-- (9.614504018844382,-2.3113116207855287);
\draw [color=ccqqqq] (9.614504018844382,-2.3113116207855287)-- (10.414504018844383,-2.3113116207855287);
\draw [color=ccqqqq] (10.414504018844383,-2.3113116207855287)-- (10.414504018844383,-1.5113116207855286);
\draw [color=ccqqqq] (10.414504018844383,-1.5113116207855286)-- (9.614504018844382,-1.5113116207855286);
\draw [color=qqqqff] (-2.,2.)-- (0.11278841319085048,-1.510438906256035);
\draw [color=qqqqff] (-2.,2.)-- (2.1167478825169566,-1.503029812663545);
\draw [color=qqqqff] (-2.,2.)-- (4.095386513805523,-1.5025459353625241);
\draw [color=qqqqff] (0.11278841319085048,-1.510438906256035)-- (0.,2.);
\draw (0.11278841319085048,-1.510438906256035)-- (2.,2.);
\draw [color=qqqqff] (2.1167478825169566,-1.503029812663545)-- (2.,2.);
\draw [color=qqqqff] (0.,2.)-- (7.899519721744715,-1.5104477980938014);
\draw [color=qqqqff] (2.,2.)-- (7.899519721744715,-1.5104477980938014);
\draw [color=qqqqff] (5.968123613353219,2.0036487133059837)-- (4.095386513805523,-1.5025459353625241);
\draw [color=qqqqff] (5.968123613353219,2.0036487133059837)-- (7.899519721744715,-1.5104477980938014);
\draw [color=qqqqff] (8.,2.)-- (10.011400871554992,-1.5113116207855286);
\draw [color=qqqqff] (10.,2.)-- (10.011400871554992,-1.5113116207855286);
\draw [color=qqqqff] (12.,2.)-- (10.011400871554992,-1.5113116207855286);
\draw [color=qqqqff] (8.,2.)-- (4.095386513805523,-1.5025459353625241);
\draw [color=qqqqff] (0.11278841319085048,-1.510438906256035)-- (0.7991076116607069,-0.9941380822780541);
\draw [color=qqqqff] (0.11278841319085048,-1.510438906256035)-- (1.,-1.);
\draw [color=qqqqff] (0.11278841319085048,-1.510438906256035)-- (1.1998102649971212,-1.0275299700560887);
\draw [color=qqqqff] (2.1167478825169566,-1.503029812663545)-- (2.490963259081123,-1.005268711537399);
\draw [color=qqqqff] (2.1167478825169566,-1.503029812663545)-- (2.290611932412916,-1.005268711537399);
\draw [color=qqqqff] (2.1167478825169566,-1.503029812663545)-- (2.858274024639503,-1.005268711537399);
\draw [color=qqqqff] (2.1167478825169566,-1.503029812663545)-- (2.6801839564899854,-1.005268711537399);
\draw [color=qqqqff] (4.095386513805523,-1.5025459353625241)-- (3.8544653433508667,-1.0108340261670716);
\draw [color=qqqqff] (4.095386513805523,-1.5025459353625241)-- (3.6318527581639697,-0.9997033969077269);
\draw [color=qqqqff] (4.095386513805523,-1.5025459353625241)-- (5.,-1.);
\draw (7.899519721744715,-1.5104477980938014)-- (7.400399641521534,-0.9949329469790268);
\draw [color=qqqqff] (7.899519721744715,-1.5104477980938014)-- (8.14816505113125,-1.0053185776680507);
\draw [color=qqqqff] (10.011400871554992,-1.5113116207855286)-- (9.290584426923871,-0.9949329469790268);
\draw [color=qqqqff] (10.011400871554992,-1.5113116207855286)-- (9.,-1.);
\draw [color=qqqqff] (10.,2.)-- (9.5,1.5);
\draw [color=qqqqff] (10.,2.)-- (9.208111902683846,1.5026964103991582);
\draw [color=qqqqff] (12.,2.)-- (11.301105197665144,1.5017383379988685);
\draw [color=qqqqff] (12.,2.)-- (11.112275548773802,1.5017383379988685);
\draw [color=qqqqff] (5.968123613353219,2.0036487133059837)-- (5.1382864465709135,1.5149480789360534);
\draw [color=qqqqff] (8.,2.)-- (7.899519721744715,-1.5104477980938014);
\draw (3.425402198207871,2.532672014252459) node[anchor=north west] {$\cdots$};
\draw (5.757036404563646,-1.3784563318927696) node[anchor=north west] {$\cdots$};
\draw [shift={(4.124599955571245,3.2766717586023306)},dash pattern=on 1pt off 1pt]  plot[domain=3.347099681359604:4.014876804903414,variable=\t]({1.*6.256246110487987*cos(\t r)+0.*6.256246110487987*sin(\t r)},{0.*6.256246110487987*cos(\t r)+1.*6.256246110487987*sin(\t r)});
\draw [shift={(4.42831078837296,5.337978284072746)},dash pattern=on 1pt off 1pt]  plot[domain=3.6205309447518905:4.386535873380058,variable=\t]({1.*7.243291973740491*cos(\t r)+0.*7.243291973740491*sin(\t r)},{0.*7.243291973740491*cos(\t r)+1.*7.243291973740491*sin(\t r)});
\draw [shift={(9.830998451556667,0.410112344773896)},dash pattern=on 1pt off 1pt]  plot[domain=2.9812589334493156:3.336702528985785,variable=\t]({1.*9.958728498696503*cos(\t r)+0.*9.958728498696503*sin(\t r)},{0.*9.958728498696503*cos(\t r)+1.*9.958728498696503*sin(\t r)});
\draw [shift={(11.134597424970197,0.5478873059648064)},dash pattern=on 1pt off 1pt]  plot[domain=2.9839433651101266:3.365903332315412,variable=\t]({1.*9.249297345877162*cos(\t r)+0.*9.249297345877162*sin(\t r)},{0.*9.249297345877162*cos(\t r)+1.*9.249297345877162*sin(\t r)});
\draw [shift={(9.312529092245304,14.53386859280516)},dash pattern=on 1pt off 1pt]  plot[domain=3.978163042885976:4.397857743331845,variable=\t]({1.*16.88405094651766*cos(\t r)+0.*16.88405094651766*sin(\t r)},{0.*16.88405094651766*cos(\t r)+1.*16.88405094651766*sin(\t r)});
\draw [shift={(11.48988739397793,17.276877721549788)},dash pattern=on 1pt off 1pt]  plot[domain=4.067539052318516:4.523559938129143,variable=\t]({1.*19.115452002123234*cos(\t r)+0.*19.115452002123234*sin(\t r)},{0.*19.115452002123234*cos(\t r)+1.*19.115452002123234*sin(\t r)});
\draw [shift={(10.602859478519747,9.802321202627384)},dash pattern=on 1pt off 1pt]  plot[domain=3.878231600333007:4.477824205365199,variable=\t]({1.*11.614017709480466*cos(\t r)+0.*11.614017709480466*sin(\t r)},{0.*11.614017709480466*cos(\t r)+1.*11.614017709480466*sin(\t r)});
\draw [color=qqqqff] (0.,2.)-- (0.5,1.5);
\draw [shift={(9.414802827072858,-4.224054951974129)},dash pattern=on 1pt off 1pt]  plot[domain=2.4432792197153597:2.8577461830518023,variable=\t]({1.*9.680814067502862*cos(\t r)+0.*9.680814067502862*sin(\t r)},{0.*9.680814067502862*cos(\t r)+1.*9.680814067502862*sin(\t r)});
\draw [shift={(11.982747083609802,-3.4010713159230543)},dash pattern=on 1pt off 1pt]  plot[domain=2.409553467973505:2.9053813851370984,variable=\t]({1.*8.086203935303072*cos(\t r)+0.*8.086203935303072*sin(\t r)},{0.*8.086203935303072*cos(\t r)+1.*8.086203935303072*sin(\t r)});
\draw [shift={(15.691129913738598,-10.472250617386686)},dash pattern=on 1pt off 1pt]  plot[domain=2.1233753735879874:2.483198970966547,variable=\t]({1.*14.653003610622202*cos(\t r)+0.*14.653003610622202*sin(\t r)},{0.*14.653003610622202*cos(\t r)+1.*14.653003610622202*sin(\t r)});
\draw [shift={(19.144522588106415,0.25736424307646744)},dash pattern=on 1pt off 1pt]  plot[domain=2.953284414293158:3.3328799987347333,variable=\t]({1.*9.309085494595958*cos(\t r)+0.*9.309085494595958*sin(\t r)},{0.*9.309085494595958*cos(\t r)+1.*9.309085494595958*sin(\t r)});
\draw [shift={(17.60594384281732,-0.04019476215796479)},dash pattern=on 1pt off 1pt]  plot[domain=2.9323136709165296:3.2919220554649073,variable=\t]({1.*9.820211391762237*cos(\t r)+0.*9.820211391762237*sin(\t r)},{0.*9.820211391762237*cos(\t r)+1.*9.820211391762237*sin(\t r)});
\draw [shift={(18.585169375661387,6.598172273142701)},dash pattern=on 1pt off 1pt]  plot[domain=3.4908190755674484:3.790722747189913,variable=\t]({1.*13.427564578510623*cos(\t r)+0.*13.427564578510623*sin(\t r)},{0.*13.427564578510623*cos(\t r)+1.*13.427564578510623*sin(\t r)});
\draw [shift={(17.580925338081716,5.204226720345692)},dash pattern=on 1pt off 1pt]  plot[domain=3.464337192078189:3.86728012680366,variable=\t]({1.*10.102534296367104*cos(\t r)+0.*10.102534296367104*sin(\t r)},{0.*10.102534296367104*cos(\t r)+1.*10.102534296367104*sin(\t r)});
\draw [shift={(19.806946995881276,-4.734232092754369)},dash pattern=on 1pt off 1pt]  plot[domain=2.4298319354548292:2.8237300410752284,variable=\t]({1.*10.310106850832467*cos(\t r)+0.*10.310106850832467*sin(\t r)},{0.*10.310106850832467*cos(\t r)+1.*10.310106850832467*sin(\t r)});
\draw (-2.315933643248822,3.3349547519232754) node[anchor=north west] {$\vartheta_1$};
\draw (-0.3352981346240238,3.3349547519232754) node[anchor=north west] {$\vartheta_2$};
\draw (1.670408709552987,3.309883416371062) node[anchor=north west] {$\vartheta_3$};
\draw (5.681822397907008,3.3349547519232754) node[anchor=north west] {$\vartheta_{N-3}$};
\draw (7.637386570979594,3.3349547519232754) node[anchor=north west] {$\vartheta_{N-2}$};
\draw (9.668164750708817,3.309883416371062) node[anchor=north west] {$\vartheta_{N-1}$};
\draw (11.623728923781401,3.309883416371062) node[anchor=north west] {$\vartheta_{N}$};
\draw (-0.2600841279673859,-2.2559530762202247) node[anchor=north west] {$\Phi_1$};
\draw (1.7706940517618375,-2.2559530762202247) node[anchor=north west] {$\Phi_2$};
\draw (3.7513295603866355,-2.2559530762202247) node[anchor=north west] {$\Phi_3$};
%\draw (7.562172564322956,-2.2559530762202247) node[anchor=north west] {$\Phi_{N-k+1}$};
\draw (7.302172564322956,-2.2559530762202247) node[anchor=north west] {$\Phi_{N-k-1}$};
\draw (9.668164750708817,-2.2559530762202247) node[anchor=north west] {$\Phi_{N-k}$};
\end{tikzpicture}
\caption{Tanner graph for a full-diversity   algebraic LDPC lattice with regular $(3,6)$ LDPC code as underlying code.}\label{taner_graph}
\end{figure}

Let $\mathbf{y}$ be the received  vector from Rayleigh BF channel with $n$ fading blocks and coherence time $N$ which is given in (\ref{AWGN_output1}).
%\begin{equation}\label{AWGN_output1}
%\mathbf{y}^{t}=(\mathbf{I}_N\otimes\mathbf{H_F})\mathbf{x}^t+\mathbf{n}^t,
%\end{equation}
%where $\mathbf{H_F}=\textrm{diag}(|h_1|,\ldots,|h_n|)$ and the fading coefficients $h_i$ are complex Gaussian random
%variables with variance $\sigma_b^2$, so that $|h_i|$ are Rayleigh distributed with parameter $\sigma_b^2$, for all $i = 1,\ldots , n$, and $\mathbf{n}=(\nu_1,\ldots,\nu_{nN})$, where $\nu_i\sim \mathcal{N}(0,\sigma^2)$ is the Gaussian noise, for $i=1,\ldots,nN$.
%=============================================================================================================================
%Even in absence of fading effect, minimum component-wise distance between two points of the above lattice is $2$ which is not enough for error correction purposes.
In the sequel, we propose two different decoders for full-diversity algebraic LDPC lattices. The first one is described in this section which contains iterative and non-iterative phases.
In the case of using iterative phase of our decoding algorithm, in order to employ the standard sum-product decoder of binary LDPC codes,  we use the scaled and translated version of $\sigma^N(\Gamma_C)$ \cite[\S 20.5]{2}, \cite{sloane}. Hence, instead of $\mathbf{x}$, we use $\mathbf{x}'=2\mathbf{x}-(1,\ldots,1)$ as transmitted vector. In this case, the received vector is
\begin{IEEEeqnarray}{rCl}\label{fading_output}
\mathbf{y}'^{t}&=&(\mathbf{I}_N\otimes\mathbf{H_F})\mathbf{x}'^{t}+\mathbf{n}^t \\
&=&2(\mathbf{I}_N\otimes\mathbf{H_F})\mathbf{x}^{t}-\left(\underbrace{(1,\ldots,1)}_N \otimes (|h_1|,\ldots,|h_n|) \right)^{t}\hspace{-0.15cm}+\mathbf{n}^{t}. \nonumber
\end{IEEEeqnarray}
\textcolor{mycolor4}{The decoding of $\mathbf{x}$ entails obtaining the components $\mathbf{p}$ and $\mathbf{c}$ in \eqref{x_general_form} from $\mathbf{y}'$.} First, we decode $\mathbf{p}$ and then we find $\mathbf{c}$. It is interesting to simulate iterative decoding of full-diversity   algebraic LDPC lattices for $n=2$, where the underlying code $\mathcal{C}$ is the $(3,6)$ ensemble (generalizations to other degree distributions and rates are treated similarly). In order to simulate iterative decoding of full-diversity  algebraic LDPC lattices, the definition of  \emph{Tanner graph} is needed. \textcolor{mycolor3}{The original Tanner graph of algebraic LDPC lattices can be defined using the parity check matrix of Theorem~\ref{theorem_parity}. Moreover, we associate another Tanner graph to these lattices which is presented in \figurename~\ref{taner_graph} for a $(3,6)$ ensemble full-diversity algebraic LDPC lattice. We describe this second Tanner graph in the sequel.}
%------------------------------------------------
%------------------------------------------------
%========================================
%========================================

In the Tanner graph of \figurename~\ref{taner_graph}, the transmitted information symbols are split into two classes: $N$ symbols are transmitted on $h_1$, while $N$ symbols are transmitted on $h_2$. Thus, there are two types of edges in \figurename~\ref{taner_graph}. Solid-line edges connect a variable node to a check node, both affected by $h_1$, and dashed-line edges connect a variable node to a check node, both affected by $h_2$.
The Tanner graph of the underlying code and the Tanner graph corresponding to the parity check matrix obtained using Theorem~\ref{theorem_parity}
are related as follows. Let us denote the Tanner graph of the underlying code by $G_1$ and the Tanner graph of the lattice (Theorem~\ref{theorem_parity})  by $G_2$. Then, $G_2$ is a disjoint union of $n$ copies of $G_1$, that is, $G_2=G_1\cup G_1\cup \cdots \cup G_1$.
\textcolor{mycolor3}{Due to the structure of the parity-check matrix of full-diversity   algebraic LDPC lattice in Theorem~\ref{theorem_parity}, in the original Tanner graph of this lattice, there is no edge between the affected variable nodes by $h_1$ and the affected check nodes by $h_2$, conversely, there is no edge between the affected variable nodes by $h_2$ and the affected check nodes by $h_1$.}
This indicates that the decoding problem using the Tanner graph $G_2$ can be partitioned into $n$ equivalent decoding instances using $G_1$. Thus, each variable node has $n$ representations, and all are connected to each other which results in the second Tanner graph of \figurename~\ref{taner_graph}. This graph is a multigraph and is used only to indicate that among the $nN$ variable nodes, there are only $N$ variable nodes with independent values and the rest are dependent to these $N$ nodes. For check nodes, the situation is similar and there are only $k$ check nodes with independent values.

%==============================================================================
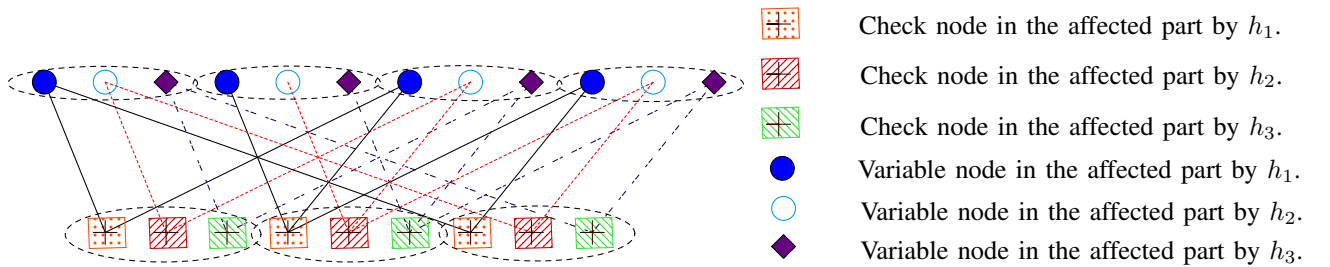
\begin{figure*}
  \centering
  \hrulefill
\definecolor{ffcctt}{rgb}{.2,0.9,0.2}
\definecolor{ccqqqq}{rgb}{0.8,0.,0.}
\definecolor{ffxfqq}{rgb}{1.,0.2980392156862745,0.}
\definecolor{xfqqff}{rgb}{0.1,0.,0.3}
\definecolor{ffqqqq}{rgb}{1.,0.,0.}
\definecolor{ttqqqq}{rgb}{0.2,0.,0.}
\definecolor{wwqqcc}{rgb}{0.4,0.,0.5}
\definecolor{qqzzff}{rgb}{0.,0.6,1.}
\definecolor{qqqqff}{rgb}{0.,0.,1.}
\begin{tikzpicture}[line cap=round,line join=round,>=triangle 45,x=0.8095238095238095cm,y=0.6666666666666666cm]
\clip(-3.,0.) rectangle (20.,6.);
\fill[color=ffxfqq,fill=ffxfqq,pattern=dots,pattern color=ffxfqq] (-1.28,1.28) -- (-1.26,0.68) -- (-0.66,0.7) -- (-0.68,1.3) -- cycle;
\fill[color=ffxfqq,fill=ffxfqq,pattern=dots,pattern color=ffxfqq] (1.7,1.28) -- (1.72,0.68) -- (2.32,0.7) -- (2.3,1.3) -- cycle;
\fill[color=ffxfqq,fill=ffxfqq,pattern=dots,pattern color=ffxfqq] (4.72,1.28) -- (4.74,0.68) -- (5.34,0.7) -- (5.32,1.3) -- cycle;
\fill[color=ccqqqq,fill=ccqqqq,pattern=north east lines,pattern color=ccqqqq] (-0.28,1.28) -- (-0.26,0.68) -- (0.34,0.7) -- (0.32,1.3) -- cycle;
\fill[color=ccqqqq,fill=ccqqqq,pattern=north east lines,pattern color=ccqqqq] (2.7,1.28) -- (2.72,0.68) -- (3.32,0.7) -- (3.3,1.3) -- cycle;
\fill[color=ccqqqq,fill=ccqqqq,pattern=north east lines,pattern color=ccqqqq] (5.72,1.28) -- (5.74,0.68) -- (6.34,0.7) -- (6.32,1.3) -- cycle;
\fill[color=ffcctt,fill=ffcctt,pattern=north west lines,pattern color=ffcctt] (0.7,1.28) -- (0.72,0.68) -- (1.32,0.7) -- (1.3,1.3) -- cycle;
\fill[color=ffcctt,fill=ffcctt,pattern=north west lines,pattern color=ffcctt] (3.7,1.28) -- (3.72,0.68) -- (4.32,0.7) -- (4.3,1.3) -- cycle;
\fill[color=ffcctt,fill=ffcctt,pattern=north west lines,pattern color=ffcctt] (6.72,1.28) -- (6.74,0.68) -- (7.34,0.7) -- (7.32,1.3) -- cycle;
\fill[color=ffxfqq,fill=ffxfqq,pattern=dots,pattern color=ffxfqq] (9.78,5.46) -- (9.8,4.86) -- (10.4,4.88) -- (10.38,5.48) -- cycle;
\fill[color=ccqqqq,fill=ccqqqq,pattern=north east lines,pattern color=ccqqqq] (9.78,4.46) -- (9.8,3.86) -- (10.4,3.88) -- (10.38,4.48) -- cycle;
\fill[color=ffcctt,fill=ffcctt,pattern=north west lines,pattern color=ffcctt] (9.78,3.48) -- (9.8,2.88) -- (10.4,2.9) -- (10.38,3.5) -- cycle;
\draw (-2.,4.)-- (-1.,1.);
\draw (4.,4.)-- (-1.,1.);
\draw [dash pattern=on 1pt off 1pt,color=ffqqqq] (-1.,4.)-- (0.,1.);
\draw [dash pattern=on 1pt off 1pt,color=ffqqqq] (5.,4.)-- (0.,1.);
\draw [dash pattern=on 1pt off 1pt on 2pt off 4pt,color=xfqqff] (0.,4.)-- (1.,1.);
\draw [dash pattern=on 1pt off 1pt on 2pt off 4pt,color=xfqqff] (6.,4.)-- (1.,1.);
\draw (1.,4.)-- (2.,1.);
\draw (4.,4.)-- (2.,1.);
\draw (7.,4.)-- (2.,1.);
\draw [dash pattern=on 1pt off 1pt,color=ffqqqq] (2.,4.)-- (3.,1.);
\draw [dash pattern=on 1pt off 1pt,color=ffqqqq] (5.,4.)-- (3.,1.);
\draw [dash pattern=on 1pt off 1pt,color=ffqqqq] (8.,4.)-- (3.,1.);
\draw [dash pattern=on 1pt off 1pt on 2pt off 4pt,color=xfqqff] (3.,4.)-- (4.,1.);
\draw [dash pattern=on 1pt off 1pt on 2pt off 4pt,color=xfqqff] (6.,4.)-- (4.,1.);
\draw [dash pattern=on 1pt off 1pt on 2pt off 4pt,color=xfqqff] (9.,4.)-- (4.,1.);
\draw (-2.,4.)-- (5.,1.);
\draw [dash pattern=on 1pt off 1pt,color=ffqqqq] (-1.,4.)-- (6.,1.);
\draw [dash pattern=on 1pt off 1pt on 2pt off 4pt,color=xfqqff] (0.,4.)-- (7.,1.);
\draw (5.,1.)-- (7.,4.);
\draw [dash pattern=on 1pt off 1pt,color=ffqqqq] (6.,1.)-- (8.,4.);
\draw [dash pattern=on 1pt off 1pt on 2pt off 4pt,color=xfqqff] (7.,1.)-- (9.,4.);
\draw [rotate around={0.:(-1.04,3.98)},dash pattern=on 2pt off 2pt] (-1.04,3.98) ellipse (1.2599998247177517cm and 0.22330847424983846cm);
\draw [rotate around={0.3819662047290258:(1.98,3.99)},dash pattern=on 2pt off 2pt] (1.98,3.99) ellipse (1.243392232342656cm and 0.22015964066186258cm);
\draw [rotate around={-0.7690246825780411:(4.97,3.98)},dash pattern=on 2pt off 2pt] (4.97,3.98) ellipse (1.2388085738182524cm and 0.2321843088289628cm);
\draw [rotate around={0.:(7.95,3.96)},dash pattern=on 2pt off 2pt] (7.95,3.96) ellipse (1.238436944394749cm and 0.23122060405169553cm);
\draw [color=ffxfqq] (-1.28,1.28)-- (-1.26,0.68);
\draw [color=ffxfqq] (-1.26,0.68)-- (-0.66,0.7);
\draw [color=ffxfqq] (-0.66,0.7)-- (-0.68,1.3);
\draw [color=ffxfqq] (-0.68,1.3)-- (-1.28,1.28);
\draw [color=ffxfqq] (1.7,1.28)-- (1.72,0.68);
\draw [color=ffxfqq] (1.72,0.68)-- (2.32,0.7);
\draw [color=ffxfqq] (2.32,0.7)-- (2.3,1.3);
\draw [color=ffxfqq] (2.3,1.3)-- (1.7,1.28);
\draw [color=ffxfqq] (4.72,1.28)-- (4.74,0.68);
\draw [color=ffxfqq] (4.74,0.68)-- (5.34,0.7);
\draw [color=ffxfqq] (5.34,0.7)-- (5.32,1.3);
\draw [color=ffxfqq] (5.32,1.3)-- (4.72,1.28);
\draw [color=ccqqqq] (-0.28,1.28)-- (-0.26,0.68);
\draw [color=ccqqqq] (-0.26,0.68)-- (0.34,0.7);
\draw [color=ccqqqq] (0.34,0.7)-- (0.32,1.3);
\draw [color=ccqqqq] (0.32,1.3)-- (-0.28,1.28);
\draw [color=ccqqqq] (2.7,1.28)-- (2.72,0.68);
\draw [color=ccqqqq] (2.72,0.68)-- (3.32,0.7);
\draw [color=ccqqqq] (3.32,0.7)-- (3.3,1.3);
\draw [color=ccqqqq] (3.3,1.3)-- (2.7,1.28);
\draw [color=ccqqqq] (5.72,1.28)-- (5.74,0.68);
\draw [color=ccqqqq] (5.74,0.68)-- (6.34,0.7);
\draw [color=ccqqqq] (6.34,0.7)-- (6.32,1.3);
\draw [color=ccqqqq] (6.32,1.3)-- (5.72,1.28);
\draw [color=ffcctt] (0.7,1.28)-- (0.72,0.68);
\draw [color=ffcctt] (0.72,0.68)-- (1.32,0.7);
\draw [color=ffcctt] (1.32,0.7)-- (1.3,1.3);
\draw [color=ffcctt] (1.3,1.3)-- (0.7,1.28);
\draw [color=ffcctt] (3.7,1.28)-- (3.72,0.68);
\draw [color=ffcctt] (3.72,0.68)-- (4.32,0.7);
\draw [color=ffcctt] (4.32,0.7)-- (4.3,1.3);
\draw [color=ffcctt] (4.3,1.3)-- (3.7,1.28);
\draw [color=ffcctt] (6.72,1.28)-- (6.74,0.68);
\draw [color=ffcctt] (6.74,0.68)-- (7.34,0.7);
\draw [color=ffcctt] (7.34,0.7)-- (7.32,1.3);
\draw [color=ffcctt] (7.32,1.3)-- (6.72,1.28);
\draw [color=ffxfqq] (9.78,5.46)-- (9.8,4.86);
\draw [color=ffxfqq] (9.8,4.86)-- (10.4,4.88);
\draw [color=ffxfqq] (10.4,4.88)-- (10.38,5.48);
\draw [color=ffxfqq] (10.38,5.48)-- (9.78,5.46);
\draw [color=ccqqqq] (9.78,4.46)-- (9.8,3.86);
\draw [color=ccqqqq] (9.8,3.86)-- (10.4,3.88);
\draw [color=ccqqqq] (10.4,3.88)-- (10.38,4.48);
\draw [color=ccqqqq] (10.38,4.48)-- (9.78,4.46);
\draw [color=ffcctt] (9.78,3.48)-- (9.8,2.88);
\draw [color=ffcctt] (9.8,2.88)-- (10.4,2.9);
\draw [color=ffcctt] (10.4,2.9)-- (10.38,3.5);
\draw [color=ffcctt] (10.38,3.5)-- (9.78,3.48);
\draw (11.18,2.68) node[anchor=north west] {\small Variable node in the affected part by $h_1$.};
\draw (11.22,1.84) node[anchor=north west] {\small Variable node in the affected part by $h_2$.};
\draw (11.22,1.06) node[anchor=north west] {\small Variable node in the affected part by $h_3$.};
\draw (11.2,5.54) node[anchor=north west] {\small Check node in the affected part by $h_1$.};
\draw (11.22,4.56) node[anchor=north west] {\small Check node in the affected part by $h_2$.};
\draw (11.22,3.54) node[anchor=north west] {\small Check node in the affected part by $h_3$.};
\draw [rotate around={0.:(3.01,0.96)},dash pattern=on 2pt off 2pt] (3.01,0.96) ellipse (1.3004908371026282cm and 0.365577705731889cm);
\draw [rotate around={0.:(-0.05,0.96)},dash pattern=on 2pt off 2pt] (-0.05,0.96) ellipse (1.300490837102624cm and 0.36557770573188786cm);
\draw [rotate around={0.:(6.09,0.98)},dash pattern=on 2pt off 2pt] (6.09,0.98) ellipse (1.3004908371026325cm and 0.3655777057318902cm);
\begin{scriptsize}
\draw [fill=qqqqff] (-2.,4.) circle (4.5pt);
\draw [color=qqzzff] (-1.,4.) circle (4.5pt);
\draw [fill=wwqqcc] (0.,4.) ++(-4.5pt,0 pt) -- ++(4.5pt,4.5pt)--++(4.5pt,-4.5pt)--++(-4.5pt,-4.5pt)--++(-4.5pt,4.5pt);
\draw [fill=qqqqff] (1.,4.) circle (4.5pt);
\draw [color=qqzzff] (2.,4.) circle (4.5pt);
\draw [fill=wwqqcc] (3.,4.) ++(-4.5pt,0 pt) -- ++(4.5pt,4.5pt)--++(4.5pt,-4.5pt)--++(-4.5pt,-4.5pt)--++(-4.5pt,4.5pt);
\draw [fill=qqqqff] (4.,4.) circle (4.5pt);
\draw [color=qqzzff] (5.,4.) circle (4.5pt);
\draw [fill=wwqqcc] (6.,4.) ++(-4.5pt,0 pt) -- ++(4.5pt,4.5pt)--++(4.5pt,-4.5pt)--++(-4.5pt,-4.5pt)--++(-4.5pt,4.5pt);
\draw [fill=qqqqff] (7.,4.) circle (4.5pt);
\draw [color=qqzzff] (8.,4.) circle (4.5pt);
\draw [fill=wwqqcc] (9.,4.) ++(-4.5pt,0 pt) -- ++(4.5pt,4.5pt)--++(4.5pt,-4.5pt)--++(-4.5pt,-4.5pt)--++(-4.5pt,4.5pt);
\draw [color=ttqqqq] (-1.,1.)-- ++(-4.5pt,0 pt) -- ++(9.0pt,0 pt) ++(-4.5pt,-4.5pt) -- ++(0 pt,9.0pt);
\draw [color=ttqqqq] (0.,1.)-- ++(-4.5pt,0 pt) -- ++(9.0pt,0 pt) ++(-4.5pt,-4.5pt) -- ++(0 pt,9.0pt);
\draw [color=ttqqqq] (1.,1.)-- ++(-4.5pt,0 pt) -- ++(9.0pt,0 pt) ++(-4.5pt,-4.5pt) -- ++(0 pt,9.0pt);
\draw [color=ttqqqq] (2.,1.)-- ++(-4.5pt,0 pt) -- ++(9.0pt,0 pt) ++(-4.5pt,-4.5pt) -- ++(0 pt,9.0pt);
\draw [color=ttqqqq] (3.,1.)-- ++(-4.5pt,0 pt) -- ++(9.0pt,0 pt) ++(-4.5pt,-4.5pt) -- ++(0 pt,9.0pt);
\draw [color=ttqqqq] (4.,1.)-- ++(-4.5pt,0 pt) -- ++(9.0pt,0 pt) ++(-4.5pt,-4.5pt) -- ++(0 pt,9.0pt);
\draw [color=ttqqqq] (5.,1.)-- ++(-4.5pt,0 pt) -- ++(9.0pt,0 pt) ++(-4.5pt,-4.5pt) -- ++(0 pt,9.0pt);
\draw [color=ttqqqq] (6.,1.)-- ++(-4.5pt,0 pt) -- ++(9.0pt,0 pt) ++(-4.5pt,-4.5pt) -- ++(0 pt,9.0pt);
\draw [color=ttqqqq] (7.,1.)-- ++(-4.5pt,0 pt) -- ++(9.0pt,0 pt) ++(-4.5pt,-4.5pt) -- ++(0 pt,9.0pt);
\draw [color=ttqqqq] (10.06,5.16)-- ++(-4.5pt,0 pt) -- ++(9.0pt,0 pt) ++(-4.5pt,-4.5pt) -- ++(0 pt,9.0pt);
\draw [color=ttqqqq] (10.06,4.16)-- ++(-4.5pt,0 pt) -- ++(9.0pt,0 pt) ++(-4.5pt,-4.5pt) -- ++(0 pt,9.0pt);
\draw [color=ttqqqq] (10.06,3.16)-- ++(-4.5pt,0 pt) -- ++(9.0pt,0 pt) ++(-4.5pt,-4.5pt) -- ++(0 pt,9.0pt);
\draw [fill=qqqqff] (10.12,2.28) circle (4.5pt);
\draw [color=qqzzff] (10.14,1.48) circle (4.5pt);
\draw [fill=wwqqcc] (10.14,0.72) ++(-4.5pt,0 pt) -- ++(4.5pt,4.5pt)--++(4.5pt,-4.5pt)--++(-4.5pt,-4.5pt)--++(-4.5pt,4.5pt);
\end{scriptsize}
\end{tikzpicture}
  \caption{Notation and diagram for the Tanner graph of a full-diversity algebraic LDPC lattice for a BF channel with $3$ fading blocks.}\label{Ex_graph}
   \hrulefill
\end{figure*}
\textcolor{Mycolor}{For each variable node $\vartheta_i$, $i=1,\ldots,N$, and check node $\Phi_j$, $j=1,\ldots,N-k$, we denote by $\varepsilon_{i,j}$ and $\varepsilon_{i,j}'$  the edges that connect $\vartheta_i$ to $\Phi_j$ in the affected part by $h_1$ and $h_2$, respectively. Indeed, $\varepsilon_{i,j}$ is one of the solid-line edges while $\varepsilon_{i,j}'$ is one of the dashed-line edges. Only one of these two edges with smaller fading effect, is chosen for decoding. This guarantees full-diversity under iterative message passing decoding \cite{rootLDPC}.}

%====================================================================================================
\begin{Example}
Let $\mathcal{C}$ be a binary code with parity-check matrix $\mathbf{H}_{\mathcal{C}}$ as follows
\begin{equation}\label{H_C}
  \mathbf{H}_{\mathcal{C}}=\left[
                             \begin{array}{cccc}
                               1 & 0 & 1 & 0 \\
                               0 & 1 & 1 & 1\\
                               1 & 0 & 0 & 1\\
                             \end{array}
                           \right].
\end{equation}
A full-diversity  algebraic Construction A lattice $\Lambda$ with diversity order $3$ based on $\mathcal{C}$ has the following parity-check matrix
 \begin{equation}\label{H_{land}}
   \mathbf{H}_{\Lambda}=\left[
                          \begin{array}{cccccccccccc}
                            1 & 0 & 0 & 0 & 0 & 0 & 1 & 0 & 0 & 0 & 0 & 0 \\
                            0 & 1 & 0 & 0 & 0 & 0 & 0 & 1 & 0 & 0 & 0 & 0 \\
                            0 & 0 & 1 & 0 & 0 & 0 & 0 & 0 & 1 & 0 & 0 & 0 \\
                            0 & 0 & 0 & 1 & 0 & 0 & 1 & 0 & 0 & 1 & 0 & 0 \\
                            0 & 0 & 0 & 0 & 1 & 0 & 0 & 1 & 0 & 0 & 1 & 0 \\
                            0 & 0 & 0 & 0 & 0 & 1 & 0 & 0 & 1 & 0 & 0 & 1 \\
                            1 & 0 & 0 & 0 & 0 & 0 & 0 & 0 & 0 & 1 & 0 & 0 \\
                            0 & 1 & 0 & 0 & 0 & 0 & 0 & 0 & 0 & 0 & 1 & 0 \\
                            0 & 0 & 1 & 0 & 0 & 0 & 0 & 0 & 0 & 0 & 0 & 1 \\
                          \end{array}
                        \right].
\end{equation}
\textcolor{mycolor3}{The parity-check matrix $\mathbf{H}_{\mathcal{C}}$ of the underlying code of $\Lambda$ is not sparse enough to call $\mathcal{C}$ an LDPC code; however, $\mathbf{H}_{\Lambda}$ is sparse enough and we can consider $\Lambda$ as an algebraic LDPC lattice.}  The Tanner graph of this lattice is presented in \figurename~\ref{Ex_graph}. For decoding, we use the Tanner graph in \figurename~\ref{Ex_graph2} in which the solid line edges, corresponding to the edges with lower fading effect or higher value of fading gain $h_i$, are used in iterative decoding. This Tanner graph is obtained by merging similar nodes in \figurename~\ref{Ex_graph} which are grouped by dashed-line ellipses.   If we apply the Tanner graph of \figurename~\ref{Ex_graph} for our iterative decoding, the generated messages during the message passing iterations do not necessarily  preserve full-diversity \cite{rootLDPC}.
\begin{figure}[ht]
  \centering
\definecolor{xfqqff}{rgb}{0.4980392156862745,0.,1.}
\definecolor{ffqqqq}{rgb}{1.,0.,0.}
\definecolor{ffxfqq}{rgb}{1.,0.4980392156862745,0.}
\definecolor{ttqqqq}{rgb}{0.2,0.,0.}
\definecolor{qqqqff}{rgb}{0.,0.,1.}
\begin{tikzpicture}[line cap=round,line join=round,>=triangle 45,x=0.8cm,y=0.7777777777777778cm]
\clip(-1.,-0.5) rectangle (9.,4.);
\fill[color=ffxfqq,fill=ffxfqq,pattern=dots,pattern color=ffxfqq] (0.16,0.64) -- (0.18,0.04) -- (0.78,0.06) -- (0.76,0.66) -- cycle;
\fill[color=ffxfqq,fill=ffxfqq,pattern=dots,pattern color=ffxfqq] (3.14,0.64) -- (3.16,0.04) -- (3.76,0.06) -- (3.74,0.66) -- cycle;
\fill[color=ffxfqq,fill=ffxfqq,pattern=dots,pattern color=ffxfqq] (6.16,0.64) -- (6.18,0.04) -- (6.78,0.06) -- (6.76,0.66) -- cycle;
\draw (-0.56,3.36)-- (0.44,0.36);
\draw (5.44,3.36)-- (0.44,0.36);
\draw (2.44,3.36)-- (3.44,0.36);
\draw (5.44,3.36)-- (3.44,0.36);
\draw (8.44,3.36)-- (3.44,0.36);
\draw (-0.56,3.36)-- (6.44,0.36);
\draw (6.44,0.36)-- (8.44,3.36);
\draw [color=ffxfqq] (0.16,0.64)-- (0.18,0.04);
\draw [color=ffxfqq] (0.18,0.04)-- (0.78,0.06);
\draw [color=ffxfqq] (0.78,0.06)-- (0.76,0.66);
\draw [color=ffxfqq] (0.76,0.66)-- (0.16,0.64);
\draw [color=ffxfqq] (3.14,0.64)-- (3.16,0.04);
\draw [color=ffxfqq] (3.16,0.04)-- (3.76,0.06);
\draw [color=ffxfqq] (3.76,0.06)-- (3.74,0.66);
\draw [color=ffxfqq] (3.74,0.66)-- (3.14,0.64);
\draw [color=ffxfqq] (6.16,0.64)-- (6.18,0.04);
\draw [color=ffxfqq] (6.18,0.04)-- (6.78,0.06);
\draw [color=ffxfqq] (6.78,0.06)-- (6.76,0.66);
\draw [color=ffxfqq] (6.76,0.66)-- (6.16,0.64);
\draw [shift={(6.6966666666666725,4.112222222222225)},dash pattern=on 1pt off 1pt,color=ffqqqq]  plot[domain=3.244883215773977:3.6818032001988934,variable=\t]({1.*7.295549971230143*cos(\t r)+0.*7.295549971230143*sin(\t r)},{0.*7.295549971230143*cos(\t r)+1.*7.295549971230143*sin(\t r)});
\draw [shift={(-5.6027272727272734,0.012424242424242404)},dash pattern=on 1pt off 1pt on 1pt off 4pt,color=xfqqff]  plot[domain=0.05745637380226927:0.5860447349910151,variable=\t]({1.*6.052715241923772*cos(\t r)+0.*6.052715241923772*sin(\t r)},{0.*6.052715241923772*cos(\t r)+1.*6.052715241923772*sin(\t r)});
\draw [shift={(8.890645161290331,3.8435483870967775)},dash pattern=on 1pt off 1pt,color=ffqqqq]  plot[domain=3.216413962157864:3.7102724538150067,variable=\t]({1.*6.468743466821224*cos(\t r)+0.*6.468743466821224*sin(\t r)},{0.*6.468743466821224*cos(\t r)+1.*6.468743466821224*sin(\t r)});
\draw [shift={(-2.7018749999999945,-0.02062499999999823)},dash pattern=on 1pt off 1pt on 1pt off 4pt,color=xfqqff]  plot[domain=0.06189296441496237:0.5816081443783221,variable=\t]({1.*6.153657766422335*cos(\t r)+0.*6.153657766422335*sin(\t r)},{0.*6.153657766422335*cos(\t r)+1.*6.153657766422335*sin(\t r)});
\draw [shift={(16.518461538461512,-20.770769230769186)},dash pattern=on 1pt off 1pt,color=ffqqqq]  plot[domain=2.0011928631657856:2.221238790965176,variable=\t]({1.*26.55233198285619*cos(\t r)+0.*26.55233198285619*sin(\t r)},{0.*26.55233198285619*cos(\t r)+1.*26.55233198285619*sin(\t r)});
\draw [shift={(-7.908085106382984,19.94014184397164)},dash pattern=on 1pt off 1pt on 1pt off 4pt,color=xfqqff]  plot[domain=5.115406494030344:5.390210467280204,variable=\t]({1.*21.28549927940293*cos(\t r)+0.*21.28549927940293*sin(\t r)},{0.*21.28549927940293*cos(\t r)+1.*21.28549927940293*sin(\t r)});
\draw [shift={(11.321142857142844,-2.727428571428564)},dash pattern=on 1pt off 1pt,color=ffqqqq]  plot[domain=2.338960579503949:2.768219520580502,variable=\t]({1.*8.46431497041356*cos(\t r)+0.*8.46431497041356*sin(\t r)},{0.*8.46431497041356*cos(\t r)+1.*8.46431497041356*sin(\t r)});
\draw [shift={(-2.2475,6.318333333333343)},dash pattern=on 1pt off 1pt on 1pt off 4pt,color=xfqqff]  plot[domain=5.474535520971676:5.915829886292362,variable=\t]({1.*8.237074259778858*cos(\t r)+0.*8.237074259778858*sin(\t r)},{0.*8.237074259778858*cos(\t r)+1.*8.237074259778858*sin(\t r)});
\draw [shift={(14.98457142857143,-13.214285714285715)},dash pattern=on 1pt off 1pt,color=ffqqqq]  plot[domain=1.9468663871309164:2.2755652670000455,variable=\t]({1.*17.819606115805396*cos(\t r)+0.*17.819606115805396*sin(\t r)},{0.*17.819606115805396*cos(\t r)+1.*17.819606115805396*sin(\t r)});
\draw [shift={(-4.279354838709662,18.892258064516103)},dash pattern=on 1pt off 1pt on 1pt off 4pt,color=xfqqff]  plot[domain=5.107068858505279:5.398548102805268,variable=\t]({1.*20.07568250634913*cos(\t r)+0.*20.07568250634913*sin(\t r)},{0.*20.07568250634913*cos(\t r)+1.*20.07568250634913*sin(\t r)});
\draw [shift={(14.29971428571428,-2.7131428571428504)},dash pattern=on 1pt off 1pt,color=ffqqqq]  plot[domain=2.338310607519766:2.7688694925646864,variable=\t]({1.*8.43915372969762*cos(\t r)+0.*8.43915372969762*sin(\t r)},{0.*8.43915372969762*cos(\t r)+1.*8.43915372969762*sin(\t r)});
\draw [shift={(0.7666666666666575,6.308888888888893)},dash pattern=on 1pt off 1pt on 1pt off 4pt,color=xfqqff]  plot[domain=5.474082215875044:5.916283191388994,variable=\t]({1.*8.220461673376803*cos(\t r)+0.*8.220461673376803*sin(\t r)},{0.*8.220461673376803*cos(\t r)+1.*8.220461673376803*sin(\t r)});
\draw [shift={(15.788571428571437,31.84)},dash pattern=on 1pt off 1pt,color=ffqqqq]  plot[domain=4.1912787833663145:4.423715604832898,variable=\t]({1.*32.83879089971346*cos(\t r)+0.*32.83879089971346*sin(\t r)},{0.*32.83879089971346*cos(\t r)+1.*32.83879089971346*sin(\t r)});
\draw [shift={(-9.046,-26.107333333333337)},dash pattern=on 1pt off 1pt on 1pt off 4pt,color=xfqqff]  plot[domain=1.0414057910866443:1.290403289932982,variable=\t]({1.*30.66489735475692*cos(\t r)+0.*30.66489735475692*sin(\t r)},{0.*30.66489735475692*cos(\t r)+1.*30.66489735475692*sin(\t r)});
\begin{scriptsize}
\draw [fill=qqqqff] (-0.56,3.36) circle (4.5pt);
\draw [fill=qqqqff] (2.44,3.36) circle (4.5pt);
\draw [fill=qqqqff] (5.44,3.36) circle (4.5pt);
\draw [fill=qqqqff] (8.44,3.36) circle (4.5pt);
\draw [color=ttqqqq] (0.44,0.36)-- ++(-4.5pt,0 pt) -- ++(9.0pt,0 pt) ++(-4.5pt,-4.5pt) -- ++(0 pt,9.0pt);
\draw [color=ttqqqq] (3.44,0.36)-- ++(-4.5pt,0 pt) -- ++(9.0pt,0 pt) ++(-4.5pt,-4.5pt) -- ++(0 pt,9.0pt);
\draw [color=ttqqqq] (6.44,0.36)-- ++(-4.5pt,0 pt) -- ++(9.0pt,0 pt) ++(-4.5pt,-4.5pt) -- ++(0 pt,9.0pt);
\end{scriptsize}
\end{tikzpicture}
  \caption{Tanner graph of a full-diversity   algebraic LDPC lattice after choosing the edges with the least fading effect. }\label{Ex_graph2}
\end{figure}
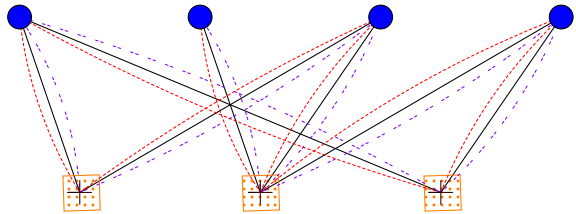
%In \figurename~\ref{Ex_graph2}, the specified groups of nodes in  \figurename~\ref{Ex_graph} inside the dashed-line ellipses are merged.
$\hfill\square$\end{Example}

Define $\hat{\mathbf{p}}$, the estimation of $\mathbf{p}$, as follows
\begin{equation}\label{p_hat}
\hat{\mathbf{p}}=Q_{\Lambda_P'}\left(\mathbf{y}'^t\right),
\end{equation}
where $\Lambda_P'$ is the lattice with the following generator matrix $\mathbf{P}'$ and  $Q_{\Lambda_P'}(\mathbf{y}'^t)$ is a lattice quantizer returning $\hat{\mathbf{z}}\mathbf{P}'$, where  $\hat{\mathbf{z}}=\textrm{argmin}_{\mathbf{z}\in\mathbb{Z}^{nN}}\|\mathbf{y}'^t-\mathbf{P}'\mathbf{z}^t\|^2$ with
$$\mathbf{P}'=2(\mathbf{I}_N\otimes\mathbf{H_F}\mathbf{P}^t),$$
in which $\mathbf{P}$ is the generator matrix of $\mathfrak{P}$ in $\mathbb{R}^n$. This decoding step seems to be a hard problem due to the high dimension of $\Lambda_P'$ which is $nN$. Here, we present a method which makes the complexity of this step affordable.   We use the following property of the Kronecker product in simplifying matrix equations. Consider three matrices $\mathbf{A}$, $\mathbf{B}$ and $\mathbf{X}$ such that $\mathbf{C}=\mathbf{AXB}$. Then \cite{horn1994topics}
\begin{equation}\label{vec_identity}
 ( \mathbf{B}^t\otimes \mathbf{A})\textrm{vec}(\mathbf{X})=\textrm{vec}(\mathbf{C}),
\end{equation}
where $\textrm{vec}(\mathbf{X})$ denotes the vectorization of the matrix $\mathbf{X}$ formed by stacking the columns of $X$ into a single column vector. For each $\mathbf{z}=(z_{1},\ldots,z_{nN})\in\mathbb{Z}^{nN}$, we consider
\begin{eqnarray*}
  \mathbf{Z} &=& \left[
                   \begin{array}{cccc}
                     z_1 & z_{n+1} & \cdots & z_{(n-1)N+1} \\
                     z_2 & z_{n+2} & \cdots & z_{(n-1)N+2} \\
                     \vdots & \vdots & \ddots & \vdots \\
                     z_n & z_{2n} & \cdots & z_{nN} \\
                   \end{array}
                 \right].
\end{eqnarray*}
It is clear that $\textrm{vec}(\mathbf{Z})=\mathbf{z}^t$. By using (\ref{vec_identity}), we have
\begin{eqnarray*}
  \mathbf{P}'\mathbf{z}^t&=&2(\textrm{vec}\left(\mathbf{H_F}\mathbf{P}^t\mathbf{Z}\right))\\
  &=&\left(2\mathbf{z}_1\mathbf{P}\mathbf{H_F},\ldots,2\mathbf{z}_N\mathbf{P}\mathbf{H_F}\right)^t,
\end{eqnarray*}
where $\mathbf{z}_i^t$ is the $i$th column of $\mathbf{Z}$, for $i=1,\ldots,N$.
In a similar manner we can write
 \begin{eqnarray*}
  (\mathbf{I}_N\otimes\mathbf{H_F})\mathbf{x}^t&=&\left(\mathbf{x}_1\mathbf{H_F},\ldots,\mathbf{x}_N\mathbf{H_F}\right)^t,
\end{eqnarray*}
where $\mathbf{x}_i=\mathbf{x}((i-1)\cdot n+1:i\cdot n)$, for $i=1,\ldots,N$. Consequently, we have
%\begin{eqnarray*}
%% \nonumber to remove numbering (before each equation)
%  \textrm{argmin}_{\mathbf{z}\in\mathbb{Z}^{nN}}\|\mathbf{y}'^t-\mathbf{P}'\mathbf{z}^t\|^2 &=& \sum_{i=1}^N\textrm{argmin}_{\mathbf{z}_i\in\mathbb{Z}^{nN}}\| \mathbf{y}_i'^t-2\mathbf{HP}\mathbf{z}_i\|^2.\\
%\end{eqnarray*}
\begin{IEEEeqnarray}{rCl}\label{ML_split}
  \|\mathbf{y}'^t-\mathbf{P}'\mathbf{z}^t\|^2 &=& \sum_{i=1}^N\| \mathbf{y}_i'^t-2\mathbf{H_F}\mathbf{P}^t\mathbf{z}_i^t\|^2,
\end{IEEEeqnarray}
where
\begin{IEEEeqnarray*}{rCl}
  \mathbf{y}_i'&=&2\mathbf{x}_i\mathbf{H_F}-(|h_1|,\ldots,|h_n|)+\mathbf{n}_i\\
  &=&2\mathbf{y}\left((i-1)\cdot n+1:i\cdot n\right)-(|h_1|,\ldots,|h_n|),
\end{IEEEeqnarray*}
and $\mathbf{z}_i=\mathbf{z}\left((i-1)\cdot n+1:i\cdot n\right)$. Indeed, it is enough to find $\textrm{argmin}_{\mathbf{z}_i\in\mathbb{Z}^{n}}\| \mathbf{y}_i'^t-2\mathbf{H_F}\mathbf{P}^t\mathbf{z}_i^t\|^2$, for $i=1,\ldots,N$, which are $N$ instances of maximum likelihood (ML) decoding in dimension $n$. Since $n$ is the number of fading blocks, $n$ is small in comparison to the dimension of lattice $\Lambda=\sigma^N(\Gamma_C)$.
For computing the ML solutions, less complex methods exist; one of the most prominent ones being sphere decoding which is based on searching for the closest lattice point
within a given hyper-sphere \cite{sphere_dec}. In small dimensions, typically less than $100$, sphere decoding
is feasible after computing the Gram matrix \cite{sphere_dec}.
%=====================
Using the preceding discussion,  the steps for estimating $\hat{\mathbf{p}}$ is presented in Algorithm~\ref{decoding_step1}. The inputs of this algorithm are the matrices $\mathbf{P}$ and $\mathbf{H_F}$ and the received vector $\mathbf{y}'$ in Equation~(\ref{fading_output}).

\begin{algorithm}
%\scriptsize
 \begin{algorithmic}[1]
 \Procedure{Low-Dim-ML}{$\mathbf{y}',\mathbf{P},\mbox{diag}(|h_1|,\ldots,|h_n|)$}
% \State $\hat{\mathbf{z}}\gets [\,\,]$
% \State $\hat{\mathbf{y}}\gets [\,\,]$
% \State $\hat{\mathbf{h}}\gets [\,\,]$
% \State $\hat{\mathbf{p}}\gets [\,\,]$
 \State $\hat{\mathbf{y}}\gets \mathbf{0}_{1\times N}$
  \State $\hat{\mathbf{p}}\gets \mathbf{0}_{1\times nN}$
% \State $\hat{\mathbf{y}}\gets [\,\,]$
% \State $\hat{\mathbf{h}}\gets [\,\,]$
% \State $\hat{\mathbf{p}}\gets [\,\,]$
 \For{$i=1:N$}
 \State $\mathbf{y}_i'\gets \mathbf{y}'((i-1)\cdot n+1:i\cdot n)$
 \State $\hat{\mathbf{p}}_i\gets \hat{\mathbf{p}}((i-1)\cdot n+1:i\cdot n)$
 % \State $\hat{\mathbf{y}}_i\gets \hat{\mathbf{y}}(n(i-1)+1:ni)$
 \State $\mathbf{y}_i^{+}\gets \mathbf{y}_i'-(|h_1|,\ldots,|h_n|)$
  \State $\mathbf{y}_i^{-}\gets \mathbf{y}_i'+(|h_1|,\ldots,|h_n|)$
 %\State  $i_0 \gets \underset{{1\leq i\leq n}}{\textrm{arg}\max}\left(|h_1|,\ldots,|h_n|\right)$

  \State $\hat{\mathbf{z}}_i^{+}\gets \underset{{\mathbf{z}_i\in\mathbb{Z}^{n}}}{\textrm{arg}\min}\|
                \mathbf{y}_i^{+}-2\mathbf{z}_i\mathbf{P}\mathbf{H_F}\|^2$
    \State $\hat{\mathbf{z}}_i^{-}\gets \underset{{\mathbf{z}_i\in\mathbb{Z}^{n}}}{\textrm{arg}\min}\|
                \mathbf{y}_i^{-}-2\mathbf{z}_i\mathbf{P}\mathbf{H_F}\|^2$
  \State $\hat{\mathbf{p}}_i^{+}\gets 2\mathbf{z}_i^{+}\mathbf{P}\mathbf{H_F}$
 \State $\hat{\mathbf{p}}_i^{-}\gets 2\mathbf{z}_i^{-}\mathbf{P}\mathbf{H_F}$
%\State $\mathbf{y}_i^{ ''t}\gets \mathbf{R}^{(1 \leftrightarrow i_0)}\mathbf{H_F}^{-1}\mathbf{y}_i^{'t}$
% \State $\hat{\mathbf{z}}_i^t\gets \underset{{\mathbf{z}_i\in\mathbb{Z}^{n}}}{\textrm{arg}\min}\|
 %               \mathbf{y}_i^{ ''t}-2\mathbf{R}^{(1 \leftrightarrow i_0)}\mathbf{P}\mathbf{z}_i\|^2$

 %\State $\hat{\mathbf{z}}_i^t\gets \textrm{argmin}_{\mathbf{z}_i\in\mathbb{Z}^{n}}\| \mathbf{y}_i'^t-2\mathbf{HP}\mathbf{z}_i\|^2$
 %\State $\mathbf{f}=\left(f_1,\ldots,f_n\right)\gets \mathbf{y}'_i-2\hat{\mathbf{z}}_i\mathbf{H_FP}$
 \State $i_m \gets \underset{{1\leq i\leq n} }{\textrm{arg}\max}\left(|h_1|,\ldots,|h_n|\right)$
 \If{$\|\mathbf{y}_i^{+}-\hat{\mathbf{p}}_i^{+}\|\leq \| \mathbf{y}_i^{-}-\hat{\mathbf{p}}_i^{-}\|$}
 \State $\hat{\mathbf{z}}_i\gets \hat{\mathbf{z}}_i^{+}$
 \Else
 \State $\hat{\mathbf{z}}_i\gets \hat{\mathbf{z}}_i^{-}$
 \EndIf
 \State $\hat{\mathbf{p}}_i\gets \hat{\mathbf{z}}_i\mathbf{P}$
 \State $\hat{\mathbf{y}}(i)\gets \mathbf{y}_i'(i_m)-2\mathbf{h}(i_m)\hat{\mathbf{p}}_i(i_m)$

% \State $\hat{\mathbf{h}}(i)\gets \mathbf{H_F}(i_m,i_m)$
% \State $\hat{\mathbf{z}}(i)\gets \hat{\mathbf{z}}_i$
% \State $\hat{\mathbf{p}}(i)\gets \hat{\mathbf{p}}_i(i_m)$
 \EndFor
 \State \textbf{return} $\hat{\mathbf{y}},\hat{\mathbf{p}}$.
 %\Else
  %\State $\hat{c}_{int}\gets [\hat{c}_1,\ldots ,\hat{c}_{n-m} ] \times G_{\mathcal{C}}$
  %\State $z\gets \lfloor \frac{y}{4}-\frac{\hat{c}_{int}}{4}\rceil$
   %\State \textbf{return} $r=\hat{c}_{int}+4z$ \Comment{The decoded value is $r$.  }
  %\EndIf
 \EndProcedure
 \end{algorithmic}
 \caption{\small{First step of decoding for full-diversity   algebraic LDPC lattices}}
 \label{decoding_step1}
\end{algorithm}

%\normalsize
%Now define $a_i=y_i-4\hat{z}_i$, for $1\leq i\leq n$, and $S=\left\{1 \leq i\leq n\,\,|\,\, a_i>1\right\}$.
%Put
%\begin{equation}\label{input_decoder}
%\hat{a}_i=\left\{ \begin{array}{l}
%                           2-a_i, \quad i\in S,\\
%                            a_i, \quad \textrm{otherwise}.
%                         \end{array}
%\right.
%\end{equation}
%Define the $i^{th}$ LLR value as follows
%\begin{equation}\label{LLR2}
%\gamma^i =\left(\frac{\hat{a}_i+1}{\hat{a}_i-1}\right)^2.
%\end{equation}
After finding $\hat{\mathbf{p}}$, the estimation of $\mathbf{p}$, we need to find $\mathbf{c}$. After choosing the appropriate edges and discarding the remaining edges, we reach to an identical Tanner graph of the underlying code and we employ the standard sum-product algorithm of binary LDPC codes \cite{sara}.
%The sum-product algorithm is a soft decision message-passing algorithm. For the sum-product decoder, the extrinsic information passed between nodes is also given as probabilities rather than hard decisions. The aim of sum-product decoding is  to compute the a posteriori probability  for each codeword bit, $p_i = p\left\{c_i = 1|s = 0\right\}$ and to select the decoded value for each bit as the value with the maximum APP (a posteriori probability).
The sum-product algorithm iteratively computes an approximation of the MAP (maximum a posteriori probability) value for each code bit. The inputs are the log likelihood ratios (LLR) for the a priori message probabilities from each channel.
In the sequel, we introduce our method to estimate the vector of log likelihood ratios $\boldsymbol{\Upsilon}=(\Upsilon_1,\ldots, \Upsilon_N)$  for full-diversity algebraic LDPC lattices in presence of perfect CSI.
% Then $\boldsymbol{\gamma}$ will be passed to the  SPA decoder of the binary LDPC codes.
We define the  vector of log likelihood ratios as
\begin{IEEEeqnarray}{rCl}\label{LLR}
\boldsymbol{\Upsilon}=\frac{2\max\left\{|h_1|,\ldots,|h_n|\right\}\cdot\hat{\mathbf{y}}}{ \sigma_{\mathcal{N}}^2}.
\end{IEEEeqnarray}
%where $\circ $ is the Hadamard product or entrywise product.
Then, we input $\boldsymbol{\Upsilon}$ to the sum-product decoder of LDPC codes that gives us $\hat{\mathbf{c}}$. We convert $\hat{\mathbf{c}}$ to $\pm 1$ notation and we denote the obtained vector by $\hat{\mathbf{c}}'$.
The final decoded vector is
\begin{equation*}
\hat{\mathbf{x}}'=\hat{\mathbf{c}}'\otimes\overbrace{(1,\ldots,1)}^{n}+2\hat{\mathbf{p}}.
\end{equation*}
Decoding error happens when $\hat{\mathbf{c}} \neq \mathbf{c}$ or $\hat{\mathbf{p}} \neq \mathbf{p}$.
\subsection{Decoding analysis}\label{decoding_analysis}
In \cite{myISIT}, a decoder has been proposed for full-diversity algebraic LDPC lattices which provides diversity $n-1$ for an algebraic LDPC lattice  with diversity $n$.
The results of \cite{myISIT} are provided for diversity order $2$, but they can be generalized for diversity order $n$.
In this section, we give an improvement of this result. We also employ the  notation introduced in the previous section.

\textcolor{mycolor4}{The analysis of the iterative decoding performance of LDPC and root-LDPC codes over BF channels has been provided in \cite{rootLDPC1,rootLDPC2}. In the rest of this section, we make a connection between the error performance of full-diversity algebraic LDPC lattices and the one of their  underlying codes over a BF channel with one fading block. In binary coding over a BF channel with one fading block, the input-output channel model is $y_i=hc_i'+n_i$, where $c_i'$ is the $i$th component of the transmitted binary codeword and $c_i'\in\{-1,+1\}$ for $i=1,\ldots,N$. Here, the employed error-correcting code is an instance from an LDPC ensemble defined by a Tanner graph and its degree distribution \cite{rootLDPC1}. The coding rate is denoted by $R=k/N$. The fading coefficient $h$ is Rayleigh distributed, that is, $h^2$ is $\chi^2$-distributed with degree 2 and normalized moment $\mathbb{E}[h^2]=1$, and the noise $n_i$ is Gaussian distributed $\mathcal{N}(0,\sigma_{\mathcal{N}}^2)$. We also define the SNR as  $\gamma_{\mathcal{C}}=1/\sigma_{\mathcal{N}}^2$.
}

\textcolor{mycolor4}{For efficient LDPC coding on BF channels, the main objective is at rendering a
frame error rate $P_{\mathcal{C}}$ of the LDPC code as close as possible to the information theoretical limit $P_{out,\mathcal{C}}(\gamma_{\mathcal{C}})$ which is defined next. The instantaneous capacity (that is, conditioned on
the fading instance) of the  channel model described above is \cite{rootLDPC1,rootLDPC2}
\begin{IEEEeqnarray}{rCl}\label{pout0}
C(\gamma_{\mathcal{C}}|h)=1-\mathbb{E}_{X}\left[\log_2\left(1+e^{-2h^2X} \right)\right],
\end{IEEEeqnarray}
where $X\sim \mathcal{N}(\gamma_{\mathcal{C}},\gamma_{\mathcal{C}})$. An outage event occurs each time $C(\gamma_{\mathcal{C}}|h)<R$. The outage probability limit is defined as $P_{out,\mathcal{C}}(\gamma_{\mathcal{C}})=\mathrm{Pr}\left(C(\gamma_{\mathcal{C}}|h)<R\right)$ \cite{rootLDPC1,rootLDPC2}. Unfortunately, $P_{out,\mathcal{C}}(\gamma_{\mathcal{C}})$ has no simple closed form expression. However, by performing the density evolution techniques, some numerical methods are provided to calculate the outage probability  for a given code ensemble \cite{rootLDPC2}.  In order to simplify the expression of $P_{out,\mathcal{C}}(\gamma_{\mathcal{C}})$, define
\begin{IEEEeqnarray}{rCl}\label{poutnew}
\mathfrak{g}(h,\gamma_{\mathcal{C}})&\triangleq&\mathbb{E}_{X}\left[\log_2\left(1+e^{-2h^2X} \right)\right]\\
&=&\frac{1}{\sqrt{2\pi\gamma_{\mathcal{C}}}}\int_{-\infty}^{+\infty}\log_2\left(1+e^{-2h^2x}\right)e^{-(x-\gamma_{\mathcal{C}})^2/\gamma_{\mathcal{C}}}dx. \nonumber
\end{IEEEeqnarray}
A good approximation to \eqref{poutnew} is proposed in \cite{rootLDPC2} as
\begin{IEEEeqnarray}{rCl}
\mathfrak{g}(h,\gamma_{\mathcal{C}})\approx \log_2\left(1+e^{-h^2\gamma_{\mathcal{C}}}\right).
\end{IEEEeqnarray}
Under the approximation above, the condition for an outage becomes
\begin{IEEEeqnarray*}{rCl}
1-\log_2\left(1+e^{-h^2\gamma_{\mathcal{C}}}\right)<R,
\end{IEEEeqnarray*}
which is equivalent to $h^2<\frac{-\ln\left(2^{1-R}-1\right)}{\gamma_{\mathcal{C}}}$. Under the assumption of Rayleigh fading, $h^2$ has an exponential density, and hence we may use the approximation $\mathrm{Pr}(h^2<x)\approx x$ valid for small $x$ \cite[p. 170]{rootLDPC2}. Hence, we compute the outage probability using this approximation as follows:
\begin{IEEEeqnarray}{rCl}\label{pnew2}
P_{out,\mathcal{C}}(\gamma_{\mathcal{C}})&\approx& \mathrm{Pr}\left(1-\log_2\left(1+e^{-h^2\gamma_{\mathcal{C}}}\right)<R\right)\nonumber\\
&=&\mathrm{Pr}\left(h^2<\frac{-\ln\left(2^{1-R}-1\right)}{\gamma_{\mathcal{C}}} \right)\nonumber\\
&\approx& \frac{-\ln\left(2^{1-R}-1\right)}{\gamma_{\mathcal{C}}}.
\end{IEEEeqnarray}
In our application,  underlying codes with high rates are desirable. When $R$ approaches 1, the numerator of \eqref{pnew2} approaches $+\infty$. In practical values of $R$ which are less than $0.99$, the numerator of \eqref{pnew2} is less than $4.97$ and $P_{out,\mathcal{C}}(\gamma_{\mathcal{C}})$ is upper bounded by $\frac{4.97}{\gamma_{\mathcal{C}}}$.
In the sequel, we assume  that the iterative performance of the underlying code of our lattices at high
SNRs is the same as the one of the outage boundary, that is $\frac{1}{\gamma_{\mathcal{C}}}$. Before explaining the main result of this section, we recall a classical result from statistics \cite[p. 75]{stat1page75}, \cite[47]{stat2page47}.
\begin{lemma}\label{lemmstat}
Let $X_1,X_2,\ldots,X_s$ be a sequence  of i.i.d. random variables  with cumulative distribution function (CDF) $\mathcal{F}_X$. Define the random variable $Y=\max\left\{X_1,X_2,\ldots,X_s\right\}$. Then, the CDF of $Y$ is
\begin{IEEEeqnarray}{rCl}
\mathcal{F}_{Y}(x)=\mathrm{Pr}(Y\leq x)=\left(\mathcal{F}_{X}(x)\right)^s.
\end{IEEEeqnarray}
\end{lemma}
}

\begin{Theorem}\label{div_theorem}
\textcolor{mycolor5}{Let $P_{\mathcal{C}}$ denote the  frame error probability  of the  code $\mathcal{C}$ using the iterative decoding of LDPC codes over a one-block fading channel.  Moreover, assume that $P_{\mathcal{C}}$ is equivalent to the outage probability over a BF channel with one fading block. Then, the algebraic LDPC lattice based on the underlying code $\mathcal{C}$ and with diversity $n$  achieves diversity $n$ over a BF channel with $n$ fading blocks using the  decoder proposed in Section~\ref{decod2}.}
\end{Theorem}
\begin{IEEEproof}
\textcolor{mycolor3}{Before going through the details of the proof, we explain three notations. We use $\gamma_{\mathcal{C}}=1/\sigma_{\mathcal{N}}^2$ to denote the SNR in a scenario in which the underlying code $\mathcal{C}$ has been employed for communication over a BF channel with one fading block. In this case, the error probability is dominated by $\frac{1}{\gamma_{\mathcal{C}}}$. We also use the following notations
\begin{IEEEeqnarray*}{rCl}
\gamma_{\Lambda}&=&\frac{\mathrm{vol}(\Lambda)^{2/nN}}{\sigma_{\mathcal{N}}^2}=\frac{\left(d_K^{N/2}2^{N-k}\right)^{2/nN}}{\sigma_{\mathcal{N}}^2},\\
\gamma_{\mathfrak{P}}&=&\frac{\mathrm{vol}\left(\sigma(\mathfrak{P})\right)^{2/n}}{\sigma_{\mathcal{N}}^2}=\frac{\left(2\sqrt{d_K}\right)^{2/n}}{\sigma_{\mathcal{N}}^2},
\end{IEEEeqnarray*}
as the SNR in scenarios in which $\Lambda$ and $\sigma(\mathfrak{P})$ have been employed for communication over a BF channel with $n$ fading blocks, respectively. When both cases achieve full diversity, their error probabilities are dominated by $1/\gamma_{\Lambda}^n$ and $1/\gamma_{\mathfrak{P}}^n$, respectively. All these three definitions are connected to each other. Indeed, we have $\gamma_{\Lambda}=\mathrm{vol}(\Lambda)^{2/nN}\gamma_{\mathcal{C}}$ and $\gamma_{\Lambda}=2^{-2k/nN}\gamma_{\mathfrak{P}}$ which implies $O(1/\gamma_{\mathfrak{P}})=O(1/\gamma_{\mathcal{C}})=O(1/\gamma_{\Lambda})$. In high  SNRs, that is, when $\sigma_{\mathcal{N}}^2\rightarrow 0$, there is no significant difference between $\gamma_{\mathcal{C}}$, $\gamma_{\Lambda}$ and $\gamma_{\mathfrak{P}}$. Hence, without loss of generality, all of them will be denoted by $\gamma$ in the rest of proof.}

In the first part of our decoding algorithm, we have $2N$ instances of optimal decoding, for the lattice generated by $\mathbf{P}$, over an $n$-block-fading channel. First, we assume that the transmitted codeword $\mathbf{c}$ in (\ref{x_general_form}) is the all-zero codeword. In the absence of codeword $\mathbf{c}$, using Equation~(\ref{ML_split}),  our decoding problem is equivalent to $N$ instances of optimal decoding over an $n$-block-fading channel with an additive noise with variance $\sigma_{\mathcal{N}}^2$. The lattice generated by $\mathbf{P}$ comes from a totally real algebraic number field and it has diversity order $n$. Thus, at high SNRs, that is, when $\sigma_{\mathcal{N}}^2\rightarrow 0$, optimal decoding of this lattice admits diversity order $n$. Now, we consider the general case that $\mathbf{c}=(c_1,\ldots,c_N)$ is not the all-zero codeword. In this case, the purpose of the instance $i$ of our optimal decoding, for $i=1,\ldots,N$, is to obtain  $\mathbf{p}_i=(\sigma_1(p_i),\ldots,\sigma_n(p_i))=(p_{i,1},\ldots,p_{i,n})\in\mathbf{P}$ from the received vector of the form
\begin{IEEEeqnarray*}{rCl}
\mathbf{y}_i'=\left(|h_1|(2p_{i,1}+c_i')+e_{i,1},\ldots,|h_n|(2p_{i,n}+c_i')+e_{i,n}\right),
\end{IEEEeqnarray*}
in which $c_i'=2c_i-1$ and $e_{i,j}\sim\mathcal{N}(0,\sigma_{\mathcal{N}}^2)$, for $j=1,\ldots,n$. We consider $e_{i,j}'=|h_j|c_i'+e_{i,j}$ as the effective noise that is not necessarily small  in high SNRs and we reach to an error floor in the performance curve. Without loss of generality, assume  $c_i=1$.
Then, $c_i'=1$ and we have
\begin{IEEEeqnarray*}{rCl}
\mathbf{y}_i^{+}&=& \mathbf{y}_i'-(|h_1|,\ldots,|h_n|)\\
&=&\left(2|h_1|p_{i,1}+e_{i,1},\ldots,2|h_n|p_{i,n}+e_{i,n}\right),\\
\mathbf{y}_i^{-}&=& \mathbf{y}_i'+(|h_1|,\ldots,|h_n|)\\
&=&\left(2|h_1|(p_{i,1}+1)+e_{i,1},\ldots,2|h_n|(p_{i,n}+1)+e_{i,n}\right).
\end{IEEEeqnarray*}
For each $\mathbf{z}_i\in\mathbb{Z}^n$, $\|\mathbf{y}_i^{+}-2\mathbf{z}_i\mathbf{P}\mathbf{H_F}\|^2$ is smaller than $\|\mathbf{y}_i^{-}-2\mathbf{z}_i\mathbf{P}\mathbf{H_F}\|^2$ which implies $\|\mathbf{y}_i^{+}-\hat{\mathbf{p}}_i^{+}\|\leq \| \mathbf{y}_i^{-}-\hat{\mathbf{p}}_i^{-}\|$ and $\hat{\mathbf{z}}_i= \hat{\mathbf{z}}_i^{+}$. In this case, $\mathbf{y}_i^{+}$ is the correct input for the ML decoder in which the effect of the non-zero  value $c_i'$ is removed. Thus, we have an optimal decoding over an $n$-block-fading channel with an additive noise with variance $\sigma_{\mathcal{N}}^2$ and when $\sigma_{\mathcal{N}}^2\rightarrow 0$, optimal decoding of the lattice generated by $\mathbf{P}$  admits diversity order $n$. If $c_i=0$ or equivalently $c_i'=-1$, we have
\begin{IEEEeqnarray*}{rCl}
\mathbf{y}_i^{+}&=& \mathbf{y}_i'-(|h_1|,\ldots,|h_n|)\\
&=&\left(2|h_1|(p_{i,1}-1)+e_{i,1},\ldots,2|h_n|(p_{i,n}-1)+e_{i,n}\right),\\
\mathbf{y}_i^{-}&=& \mathbf{y}_i'+(|h_1|,\ldots,|h_n|)\\
&=&\left(2|h_1|p_{i,1}+e_{i,1},\ldots,2|h_n|p_{i,n}+e_{i,n}\right).
\end{IEEEeqnarray*}
Thus, $\|\mathbf{y}_i^{-}-\hat{\mathbf{p}}_i^{-}\|<\| \mathbf{y}_i^{+}-\hat{\mathbf{p}}_i^{+}\|$ which implies $\hat{\mathbf{z}}_i= \hat{\mathbf{z}}_i^{-}$. In this case, $\mathbf{y}_i^{-}$ is the correct input for the ML decoder in which the effect of the non-zero  value $c_i'$ is removed. Hence, for $i=1,\ldots,n$, we obtain $\hat{\mathbf{p}}_i$ the estimation of $\mathbf{p}_i$ with diversity $n$, that is, $\mathrm{Pr}\left\{\hat{\mathbf{p}}_i\neq \mathbf{p}_i \right\}\approx \gamma^{-n}$ asymptotically.
After $N$ steps, we obtain $\hat{\mathbf{p}}$ the estimation of $\mathbf{p}$ and
$$\mathrm{Pr}\left\{\hat{\mathbf{p}}\neq \mathbf{p} \right\}=\sum_{i=1}^{N}\mathrm{Pr}\left\{\hat{\mathbf{p}}_i\neq \mathbf{p}_i \right\}\approx N\gamma^{-n},$$
which admits diversity $n$, too.
Now, assume $\mathbf{p}$ is estimated correctly.
 Without loss of generality let $h_1>0$ be the maximum of $\left\{|h_1|,\ldots,|h_n|\right\}$.
%  and let us consider $n-1$ deep fades as $h_2=h_3=\cdots=h_{n}=0$.
In this case, we have
\begin{IEEEeqnarray*}{rCl}
\frac{\hat{\mathbf{y}}}{h_1}&=&\frac{\left(h_1(2p_{1,1}+c_1')+e_{1,1},\ldots,h_1(2p_{N,1}+c_N')+e_{N,1}\right)}{h_1}\\
&&-\frac{\left(2h_1\hat{p}_{1,1},\ldots,2h_1\hat{p}_{N,1}\right)}{h_1}\\
&=& \mathbf{c}'+(e_{1,1}',\ldots ,e_{N,1}'),
\end{IEEEeqnarray*}
where $e_{i,1}'\sim \mathcal{N}(0,\sigma_{\mathcal{N}}^{'2})$ for $i=1,\ldots,N$, is the Gaussian noise with $\sigma_{\mathcal{N}}^{'2}=\sigma_{\mathcal{N}}^2/h_1^2$.
This is exactly the setting in which a codeword of the LDPC code $\mathcal{C}$ has been transmitted over a BF channel with one fading block  using BPSK modulation. Thus, the LLR for a specific SNR and symbol $c_i'$ can be estimated as follows
\begin{IEEEeqnarray}{rCl}\label{LLR2}
\boldsymbol{\Upsilon}(i)&=&\log\frac{\mathrm{Pr}\left\{\mathbf{y}(i)/h_1 | c_i'=+1,h_1 \right\}}{\mathrm{Pr}\left\{\mathbf{y}(i)/h_1 | c_i'=-1,h_1 \right\}}\nonumber\\
&=&\frac{2\hat{\mathbf{y}}(i)/h_1}{\sigma_{\mathcal{N}}^{'2}}=\frac{2\hat{\mathbf{y}}(i)h_1}{\sigma_{\mathcal{N}}^{2}},
\end{IEEEeqnarray}
which is the same as Equation~(\ref{LLR}) if $h_1=\max \left\{|h_1|,\ldots,|h_n|\right\}$. Let $\boldsymbol{\Upsilon}^i$ denote the LLR vector obtained by replacing $h_1$ with $h_i$ in \eqref{LLR2} and $\hat{\mathbf{c}}_i'$ denote the estimation of the codeword $\mathbf{c}'$ by giving $\boldsymbol{\Upsilon}^i$ as the input of the sum-product decoder and $P_{\mathcal{C}}^i$ be the frame error rate of this estimation, that is, $P_{\mathcal{C}}^i=\mathrm{Pr}\left\{\hat{\mathbf{c}}_i'\neq\mathbf{c}' \right\}$. Since obtaining $\hat{\mathbf{c}}_i'$ is equivalent to retrieving a codeword transmitted over a BF channel with one fading block, $P_{\mathcal{C}}^i$ is upper bounded by $\gamma^{-1}$. If for $i=1,\ldots,n$, $\hat{\mathbf{c}}_i'\neq \mathbf{c}'$, an error happens in the estimation of $\mathbf{c}'$.
\textcolor{mycolor3}{Indeed, using the received  vector $\mathbf{y}$ of length $nN$,  $n$ erroneous replicas of $\mathbf{c}'$ can be found each of which is attenuated by one of $h_j$'s, for $j=1,\ldots,n$. Hence, for each transmitted codeword $\mathbf{c}'$, $n$ different decodings can be done via $\mathbf{y}$  and equivalently $n$ different estimations can be obtained from $\mathbf{y}$. Each of these instances is equivalent to retrieving $\mathbf{c}'$ from a vector of the form $\mathbf{y}_j''=h_j\mathbf{c}'+\mathbf{e}_j''$, where $\mathbf{e}_j''$ is the Gaussian noise with zero mean and variance $\sigma_{\mathcal{N}}^2$ per dimension. The larger the coefficient $h_j$, the better the approximation of $\mathbf{c}'$. Therefore, if for the largest value of $h_1,\ldots,h_n$, the decoder returns a wrong estimation, it would return wrong estimations for other $h_j$'s too and all $n$ instances of decoding would be failed. Hence, $n$ instances of wrong decoding  is equivalent to the case in which error happens in the estimation of $\hat{\mathbf{c}}_1'$ because $h_1=\max \left\{|h_1|,\ldots,|h_n|\right\}$ and $\boldsymbol{\Upsilon}^1$ is the best approximation of LLR among all $\boldsymbol{\Upsilon}^i$'s.  Let us define $n$ events corresponding to the mistake in each one of these $n$ decoding instances with outputs $\hat{\mathbf{c}}_j'$'s.  Since these events are independent, we have the final erroneous decoding $\hat{\mathbf{c}}'\neq\mathbf{c}'$ if and only if $\hat{\mathbf{c}}_j'\neq\mathbf{c}'$, for $j=1,\ldots,n$. Consequently, we have
\begin{IEEEeqnarray*}{rCl}
\mathrm{Pr}\left\{\hat{\mathbf{c}}'\neq\mathbf{c}' \right\}&\equiv&\mathrm{Pr}\left(\hat{\mathbf{c}}_1'\neq\mathbf{c}'\right)\&\cdots \&\mathrm{Pr}\left(\hat{\mathbf{c}}_n'\neq\mathbf{c}' \right)\\
&=& P_{\mathcal{C}}^1\times \cdots \times P_{\mathcal{C}}^n \approx\gamma^{-n}.
\end{IEEEeqnarray*}
}
\textcolor{mycolor4}{
The above result can also be obtained using the definition of outage probability in \eqref{pnew2} and Lemma~\ref{lemmstat}. For a fixed high SNR $\gamma$ and a fading coefficient $h_i$, $C(\gamma|h_i)=1-\log_2\left(1+e^{-h_i^2\gamma}\right)$  is a random variable depending only on the fading coefficient $h_i$. Let us denote the random variable corresponding to the $i$th fading coefficient by $H_i$ and  the random variable $C(\gamma|H_i)$ by $\mathcal{H}_i$. For a fixed value of $\gamma$, since $C(\gamma|h_i)$ is an increasing function in terms of $h_i$, and $h_1=\max\{h_1,\ldots,h_n\}$, we can assume $\mathcal{H}_1=\max\{\mathcal{H}_1,\ldots,\mathcal{H}_n\}$. According to \eqref{pnew2}, the outage probability corresponding to $\mathcal{H}_1$ in SNR $\gamma$  is $P_{out,\mathcal{C}}(\gamma)=\mathrm{Pr}(\mathcal{H}_1<R)$ which can be computed using Lemma~\ref{lemmstat} as $[\mathrm{Pr}(\mathcal{H}<R)]^n$, where $\mathcal{H}$ denotes the common distribution of all $\mathcal{H}_i$'s. Due to our assumption
that the iterative performance of $\mathcal{C}$ at high SNRs is the same as the one of the outage probability, $\mathrm{Pr}(\mathcal{H}<R)=\gamma^{-1}$ and $P_{out,\mathcal{C}}(\gamma)=[\mathrm{Pr}(\mathcal{H}<R)]^n=\gamma^{-n}$. Hence, at high SNRs, the frame error rate of $\mathcal{C}$ also behaves like $\gamma^{-n}$.
}
Thus we have
%It is also clear that $\mathrm{Pr}\left(\hat{\mathbf{c}}'\neq\mathbf{c}' ,\hat{\mathbf{p}}\neq\mathbf{p} \right)\leq %\mathrm{Pr}\left(\hat{\mathbf{p}}\neq\mathbf{p} \right)\approx N\gamma^{-n}$. Hence
\begin{IEEEeqnarray*}{rCl}
\mathrm{Pr}\left\{\hat{\mathbf{x}}\neq\mathbf{x} \right\}\leq \mathrm{Pr}\left\{\hat{\mathbf{c}}'\neq\mathbf{c}'  \right\}+\mathrm{Pr}\left\{\hat{\mathbf{p}}\neq\mathbf{p} \right\}
\approx (N+1)\gamma^{-n},
%\mathrm{Pr}\left(\hat{\mathbf{c}}'\neq\mathbf{c}'  \right)&=&
%\mathrm{Pr}\left(\hat{\mathbf{c}}'\neq\mathbf{c}' ,\hat{\mathbf{p}}\neq\mathbf{p} \right)+\mathrm{Pr}\left(\hat{\mathbf{c}}'\neq\mathbf{c}' ,\hat{\mathbf{p}}=\mathbf{p} \right)\\
%&\leq& \frac{N+1}{\gamma^n }.
\end{IEEEeqnarray*}
which indicates diversity $n$ of algebraic LDPC lattices using the proposed decoder in Section~\ref{decod2}.
\end{IEEEproof}
\begin{remark}
In Theorem~\ref{div_theorem}, we have considered a sufficient condition about the frame error of the underlying code of algebraic LDPC lattices to achieve full-diversity over BF channels. However, this assumption is not a necessary condition to achieve full-diversity. In the next section we modify the proposed algorithm in this section by removing its iterative phase which enables full-diversity decoding of general Construction A lattices without any assumption about their underlying code. We believe that the second part of the proof of Theorem~\ref{div_theorem} can be provided by using diversity population evolution (DPE) and Density Evolution (DE) techniques similar to the proofs of \cite{punekar} and \cite{rootLDPC}. However, going through the details of these techniques pulls us away from our main goal.
\end{remark}

In order to discus the decoding complexity of the proposed algorithm, let us consider the complexity of the used optimal decoder in dimension $n$ as $f(n)$, which is cubic in high SNRs for heuristic methods and exponential in worst-case complexity \cite{hassibi}.
Since our decoding involves $2N$ uses of an optimal decoder in dimension $n$,  the complexity of our decoding method is $O(2N\cdot f(n))+O(N\cdot d\cdot t)$ in which $t$ is the maximum number of iterations in the iterative decoding and $d$ is the average column degree of $\mathbf{H}_{\mathcal{C}}$. This complexity is dominated by $O(N\cdot d\cdot t)$ as $N$ is much greater than $n$.
\section{Decoding of General Full-diversity Construction A Lattices}\label{non_binary_decoding_sec}
In this section, we remove the iterative phase of the  algorithm  proposed in Section~\ref{decod2} which enables full-diversity decoding of general Construction A lattices without any assumption about their underlying code.Indeed, using the proposed algorithm,  all generalized Construction A lattices with any binary or non-binary underlying code can be decoded with full diversity and linear complexity in the dimension of the lattice.

Let $p$ be a prime number and $\mathcal{C}\subset \mathbb{F}_p^N$ be an arbitrary  linear $[N,k]$ code and $\mathcal{O}_K$ be the integers ring of a totally real number field $K$ of degree $n$. Let $\mathfrak{p}$ be a prime ideal of $\mathcal{O}_K$ such that $\mathcal{O}_K/\mathfrak{P}\cong \mathbb{F}_p$. Also, consider $\sigma_1,\ldots,\sigma_n$ to be $n$ real embeddings of $K$. Every lattice vector $\mathbf{x}$ in $\sigma^N(\Gamma_{\mathcal{C}})=\sigma^N(\rho^{-1}(\mathcal{C}))\subset\mathbb{R}^{nN}$ has the same  form given in (\ref{x_general_form}).

Let $\mathbf{y}$ be the received  vector from Rayleigh BF channel with $n$ fading blocks and coherence time $N$ which is given in (\ref{AWGN_output1}).
In this decoding procedure, we do not need the iterative phase of previous decoding based on standard sum-product decoder of binary LDPC codes. Hence, we do not scale or translate  $\sigma^N(\Gamma_C)$ and $\mathbf{x}$ is the transmitted vector. In this case, the received vector is
\begin{IEEEeqnarray}{rCl}\label{fading_output_2}
\mathbf{y}^{t}=(\mathbf{I}_N\otimes\mathbf{H_F})\mathbf{x}^{t}+\mathbf{n}^t.
\end{IEEEeqnarray}
Unlike the previous method, we decode $\mathbf{p}$ and $\mathbf{c}$ in a single phase. The steps of decoding $\hat{\mathbf{c}}$ and $\hat{\mathbf{p}}$ is provided in Algorithm~\ref{decoding_alg2}.  The final decoded lattice vector is
\begin{equation*}
\hat{\mathbf{x}}=\hat{\mathbf{c}}\otimes\overbrace{(1,\ldots,1)}^{n}+\hat{\mathbf{p}}.
\end{equation*}
\textcolor{mycolor3}{
In order to give some insight about this decoding method, we provide the following toy example.
\begin{Example}
 Consider the cyclotomic field $K=\mathbb{Q}(\xi_3)$, where $\xi_3=e^{2\pi i/3}$, and $\mathcal{O}_{K}=\mathbb{Z}[\xi_3]$ as its ring of integers. We have $3\mathcal{O}_{K}=\mathfrak{P}^2$ and $\mathcal{O}_{K}/\mathfrak{P}\cong \mathbb{F}_3$, where $\mathfrak{P}=(1-\xi_3)$ is a prime ideal of  $\mathcal{O}_{K}$. Let $\mathbf{G}_{\mathcal{C}}=\left[
                                \begin{array}{ccc}
                                  1 & 0 & 1 \\
                                  0& 1 & 2 \\
                                \end{array}
                              \right]$ be the generator matrix of the $3$-ary underlying code $\mathcal{C}$ of $\Lambda=\sigma^3(\rho^{-1}(\mathcal{C}))$. Consider $\mathbf{c}=(c_1,c_2,c_3)=(2,1,1)$  and $\mathbf{p}=(p_1,p_2,p_3)=(2+\xi_3,1-\xi_3,1+2\xi_3)$ as randomly chosen elements in $\mathcal{C}$ and $\mathfrak{P}^3$, respectively. It should be noted that every member of $\mathfrak{P}$ is of the form $(a+b\xi_3)(1-\xi_3)$, for $a,b\in\mathbb{Z}$, which can be simplified to $a(1-\xi_3)+b(1+2\xi_3)$ using the fact that $\xi_3^2+\xi_3+1=0$. Hence, $\left\{1-\xi_3,1+2\xi_3\right\}$ is a $\mathbb{Z}$-basis of $\mathfrak{P}$. Using the fact that the identity map $\sigma_1$ and  $\sigma_2$, that maps $\xi_3$ to $\bar{\xi_3}$, are two embeddings of $K$, the transmitted vector $\mathbf{x}$ with components  $\mathbf{p}$ and $\mathbf{c}$ is of the following form
 \begin{IEEEeqnarray*}{rCl}
 \mathbf{x}&=&\sigma^3(\mathbf{c}+\mathbf{p})=\left(\sigma(c_1+p_1),\sigma(c_2+p_2),\sigma(c_3+p_3)\right)\\
 &=&\left(c_1+p_1,\overline{c_1+p_1},c_2+p_2,\overline{c_2+p_2},c_3+p_3,\overline{c_3+p_3}\right)\\
 &=&\left(c_1+p_1,c_1+\bar{p_1},c_2+p_2,c_2+\bar{p_2},c_3+p_3,c_3+\bar{p_3}\right)\\
  &=& \mathbf{c}\otimes(1,1)+\sigma^3(\mathbf{p}).
 \end{IEEEeqnarray*}
Let $\mathbf{h}=(h_1,h_2)$ be a realization of the fading coefficients of the BF channel with two fading blocks and $\mathbf{n}=(n_1,\ldots,n_6)$ be the additive Gaussian noise with zero mean and variance $\sigma_{\mathcal{N}}^2$ per dimension. Then, the received vector has the following form
\begin{IEEEeqnarray*}{rCl}
\mathbf{y}= \mathbf{c}\otimes(h_1,h_2)+\sigma^3(\mathbf{p})\mathrm{diag}(h_1,h_2,h_1,h_2,h_1,h_2).
\end{IEEEeqnarray*}
Using the above representation, we can split the decoding of $\mathbf{y}$ into three separate phases each of which are equivalent to obtaining $c_i$ and $p_i$ from the following subvector
\begin{IEEEeqnarray*}{rCl}
\mathbf{y}_i=c_i(h_1,h_2)+(p_i,\bar{p_i})\left[
                                           \begin{array}{cc}
                                             h_1 & 0 \\
                                             0 & h_2 \\
                                           \end{array}
                                         \right]
+\mathbf{n}_i,
\end{IEEEeqnarray*}
where $\mathbf{n}_i=\mathbf{n}(2i-1:2i,)$ and $i=1,2,3$. Using the $\mathbb{Z}$-basis of $\mathfrak{P}$, the generator matrix of $\sigma(\mathfrak{P})$ is
\begin{IEEEeqnarray*}{rCl}
\mathbf{P}&=&\left[
              \begin{array}{cc}
                \Re \sigma_1(1-\xi_3) & \Im \sigma_2(1-\xi_3)  \\
                \Re \sigma_1(1+2\xi_3) & \Im \sigma_2(1+2\xi_3) \\
              \end{array}
            \right]
=  \left[
              \begin{array}{cc}
                \frac{3}{2} & -\frac{\sqrt{3}}{2}  \\
                0 & -\sqrt{3} \\
              \end{array}
            \right].
\end{IEEEeqnarray*}
We employ exhaustive search to find $c_i$. If we guess the value of $c_i\in\mathbb{F}_3$ correctly and subtract $c_i(h_1,h_2)$ from $\mathbf{y}_i$ and denote the obtained vector by $\mathbf{y}_i'$, then the decoding problem is reduced to finding $\mathbf{z}_0\in \mathbb{Z}^2$ such that $\left\|\mathbf{y}_i'-\mathbf{z}_0\mathbf{P}\mathrm{diag}(\mathbf{h})\right\|^2$ is minimized. In high SNRs, the additive noise variance approaches zero and the last statement is approximately the decoding problem in the case of using the lattice $\sigma(\mathfrak{P})$ in a fading channel with slightly lower SNR compared to the SNR of the channel in which $\Lambda$ has been empolyed. Indeed, this approximation is due  to the difference between the definition of SNR  for $\sigma(\mathfrak{P})$ and $\Lambda$ which is related to their different volumes.  The error probability of this scenario is related to the diversity order of $\sigma(\mathfrak{P})$.  Since the signature of $K$ is $(r_1,r_2)=(0,1)$, the diversity order of $\sigma(\mathfrak{P})$ is $r_1+r_2$ which is one. If we denote $\mathrm{diag}(\mathbf{h})$ by $\mathbf{H_F}$ and we guess $\hat{c}_i\neq c_i$ as the value of $c_i$, then we are encountered with an additive noise of the form $\mathbf{n}_i'=\mathbf{n}_i+(c_i-\hat{c}_i)(h_1,h_2)$ in our decoding. The vector $\mathbf{n}_i'$  is still a Gaussian vector but except for the deep fades, that is, when $h_1=h_2=0$, its components have different nonzero variances.   Therefore,  we find a vector $\hat{\mathbf{z}}_i\in\mathbb{Z}^2$ using our decoding that minimizes $\left\|\mathbf{y}_i-\hat{c}_i(h_1,h_2)-\mathbf{z}\mathbf{P}\mathbf{H_F}\right\|^2$ but $\hat{\mathbf{z}}_i\mathbf{P}\neq\mathbf{z}_i\mathbf{P}=(p_i,\bar{p_i})$ with high probability. Then, for high SNRs, the components of $\mathbf{n}_i$ are small and we have
\begin{IEEEeqnarray*}{lll}
\left\|\mathbf{y}_i-\hat{c}_i\mathbf{h}-\hat{\mathbf{z}}_i\mathbf{P}\mathbf{H_F}\right\|^2&=& \left\|(c_i-\hat{c}_i)\mathbf{h}+(\mathbf{z}_i-\hat{\mathbf{z}}_i)\mathbf{P}\mathbf{H_F}+\mathbf{n}_i\right\|^2\\
&>&\left\|\mathbf{y}_i-c_i\mathbf{h}-\mathbf{z}_i\mathbf{P}\mathbf{H_F}\right\|^2
=\left\|\mathbf{n}_i\right\|^2.
\end{IEEEeqnarray*}
This comparison  indicates situations in which a wrong decision has been taken regarding the value of $c_i$. Hence, for $i=1,2,3$, if we guess the value of $c_i$ correctly and choose $\mathbf{z}\in\mathbb{Z}^2$ that makes $\left\|\mathbf{y}_i-c_i(h_1,h_2)-\mathbf{z}\mathbf{P}\mathbf{H_F}\right\|^2$ closer to its minimum value, that is $\left\|\mathbf{n}_i\right\|^2$, we have reached to an estimation of the transmitted point.
\end{Example}
}
\begin{algorithm}
%\scriptsize
 \begin{algorithmic}[1]
 \Procedure{DEC}{$\mathbf{y},\mathbf{P},\mathbf{H_F}=\mbox{diag}(|h_1|,\ldots,|h_n|),p$}
% \State $\hat{\mathbf{z}}\gets [\,\,]$
% \State $\hat{\mathbf{y}}\gets [\,\,]$
% \State $\hat{\mathbf{h}}\gets [\,\,]$
% \State $\hat{\mathbf{p}}\gets [\,\,]$
 %\State $\hat{\mathbf{y}}\gets \mathbf{0}_{1\times N}$
 \State $\hat{\mathbf{c}}\gets \mathbf{0}_{1\times N}$
  \State $\hat{\mathbf{p}}\gets \mathbf{0}_{1\times nN}$
% \State $\hat{\mathbf{y}}\gets [\,\,]$
% \State $\hat{\mathbf{h}}\gets [\,\,]$
% \State $\hat{\mathbf{p}}\gets [\,\,]$
 \For{$i=1:N$}
 \State $\mathbf{y}_i\gets \mathbf{y}((i-1)\cdot n+1:i\cdot n)$
 \State $\hat{\mathbf{p}}_i\gets \hat{\mathbf{p}}((i-1)\cdot n+1:i\cdot n)$
 \State $\textrm{Threshold} \gets +\infty$
 % \State $\hat{\mathbf{y}}_i\gets \hat{\mathbf{y}}(n(i-1)+1:ni)$
 \For{$j=0:p-1$}
 \State $\mathbf{y}_i^{c_i=j}\gets \mathbf{y}_i-j\cdot(|h_1|,\ldots,|h_n|)$
 \State $\hat{\mathbf{z}}_i^{c_i=j}\gets \underset{{\mathbf{z}_i\in\mathbb{Z}^{n}}}{\textrm{arg}\min}\|
                \mathbf{y}_i^{c_i=j}-\mathbf{z}_i\mathbf{P}\mathbf{H_F}\|^2$
   \State $\hat{\mathbf{p}}_i^{c_i=j}\gets \mathbf{z}_i^{c_i=j}\mathbf{P}\mathbf{H_F}$
\If{$\|\mathbf{y}_i-\hat{\mathbf{p}}_i^{c_i=j}\|< \textrm{Threshold}$}
 \State $\hat{\mathbf{z}}_i\gets \hat{\mathbf{z}}_i^{c_i=j}$
 \State $\hat{c}_i \gets j$
 \State $\textrm{Threshold}\gets \|\mathbf{y}_i-\hat{\mathbf{p}}_i^{c_i=j}\| $
% \Else
% \State $\hat{\mathbf{z}}_i\gets \hat{\mathbf{z}}_i^{-}$
 \EndIf

  \EndFor
 % \State $\mathbf{y}_i^{-}\gets \mathbf{y}_i'+(|h_1|,\ldots,|h_n|)$
 %\State  $i_0 \gets \underset{{1\leq i\leq n}}{\textrm{arg}\max}\left(|h_1|,\ldots,|h_n|\right)$

  %\State $\hat{\mathbf{z}}_i^{+}\gets \underset{{\mathbf{z}_i\in\mathbb{Z}^{n}}}{\textrm{arg}\min}\|
%                \mathbf{y}_i^{+}-2\mathbf{z}_i\mathbf{P}\mathbf{H_F}\|^2$
%    \State $\hat{\mathbf{z}}_i^{-}\gets \underset{{\mathbf{z}_i\in\mathbb{Z}^{n}}}{\textrm{arg}\min}\|
%                \mathbf{y}_i^{-}-2\mathbf{z}_i\mathbf{P}\mathbf{H_F}\|^2$
%  \State $\hat{\mathbf{p}}_i^{+}\gets 2\mathbf{z}_i^{+}\mathbf{P}\mathbf{H_F}$
% \State $\hat{\mathbf{p}}_i^{-}\gets 2\mathbf{z}_i^{-}\mathbf{P}\mathbf{H_F}$
%\State $\mathbf{y}_i^{ ''t}\gets \mathbf{R}^{(1 \leftrightarrow i_0)}\mathbf{H_F}^{-1}\mathbf{y}_i^{'t}$
% \State $\hat{\mathbf{z}}_i^t\gets \underset{{\mathbf{z}_i\in\mathbb{Z}^{n}}}{\textrm{arg}\min}\|
 %               \mathbf{y}_i^{ ''t}-2\mathbf{R}^{(1 \leftrightarrow i_0)}\mathbf{P}\mathbf{z}_i\|^2$

 %\State $\hat{\mathbf{z}}_i^t\gets \textrm{argmin}_{\mathbf{z}_i\in\mathbb{Z}^{n}}\| \mathbf{y}_i'^t-2\mathbf{HP}\mathbf{z}_i\|^2$
 %\State $\mathbf{f}=\left(f_1,\ldots,f_n\right)\gets \mathbf{y}'_i-2\hat{\mathbf{z}}_i\mathbf{H_FP}$
 %\State $i_m \gets \underset{{1\leq i\leq n} }{\textrm{arg}\max}\left(|h_1|,\ldots,|h_n|\right)$

 \State $\hat{\mathbf{p}}_i\gets \hat{\mathbf{z}}_i\mathbf{P}$
% \State $\hat{\mathbf{y}}(i)\gets \mathbf{y}_i'(i_m)-2\mathbf{h}(i_m)\hat{\mathbf{p}}_i(i_m)$

% \State $\hat{\mathbf{h}}(i)\gets \mathbf{H_F}(i_m,i_m)$
% \State $\hat{\mathbf{z}}(i)\gets \hat{\mathbf{z}}_i$
% \State $\hat{\mathbf{p}}(i)\gets \hat{\mathbf{p}}_i(i_m)$
 \EndFor
 \State \textbf{return} $\hat{\mathbf{c}},\hat{\mathbf{p}}$.
 %\Else
  %\State $\hat{c}_{int}\gets [\hat{c}_1,\ldots ,\hat{c}_{n-m} ] \times G_{\mathcal{C}}$
  %\State $z\gets \lfloor \frac{y}{4}-\frac{\hat{c}_{int}}{4}\rceil$
   %\State \textbf{return} $r=\hat{c}_{int}+4z$ \Comment{The decoded value is $r$.  }
  %\EndIf
 \EndProcedure
 \end{algorithmic}
 \caption{\small{Decoding of general full-diversity algebraic Construction A lattices}}
 \label{decoding_alg2}
\end{algorithm}
Using the notation of Section~\ref{decoding_analysis}, let us consider the complexity of the used optimal decoder in dimension $n$ as $f(n)$. Since our decoding involves $pN$ uses of an optimal decoder in dimension $n$,  the complexity of our decoding method is $O(pN\cdot f(n))$. This complexity is almost linear in terms of $N$ since $N$ is much greater than $n$ and $p$. In order to make a comparison, consider full-rate uncoded transmission with $\log_2 (M)$ bit/s/Hz. The optimal decoding in this case entails $M^{nN}$ searches. Using our proposed algorithm requires only $pN\cdot M^n$ searches which indicates a remarkable reduction in the complexity. For $M=4$, $n=3$, $p=2$ and $N=100$ which are typical values in our simulations, the number of trials is $200\times 2^6\approx 2^{14}$ for our decoder  versus $2^{600}$ for ML decoder. This results in $2^{586}$ times faster decoding compared to ML decoding.
\begin{Theorem}\label{div_theorem2}
 Let $\mathcal{C}\subset\mathbb{F}_p^N$ be the underlying code  of a generalized Construction A lattice $\Lambda$ with diversity $n$. Then, $\Lambda$   achieves full diversity over a BF channel with $n$ fading blocks using the  decoder proposed in Algorithm~\ref{decoding_alg2}.
\end{Theorem}
\begin{IEEEproof}
For $\mathbf{P}'=\mathbf{I}_N\otimes\mathbf{P}\mathbf{H_F}$ the optimal decoding of $\Lambda$ means finding $\underset{{\mathbf{z}\in\mathbb{Z}^{nN}}}{\textrm{arg}\min} \|\mathbf{y}^t-(\mathbf{z}\mathbf{M}_{\Lambda})^t\|^2$ which is equivalent to solving the following problem
\begin{IEEEeqnarray*}{rCl}
(\hat{\mathbf{z}},\hat{\mathbf{c}})=\underset{\mathbf{z}\in\mathbb{Z}^{nN},\mathbf{c}\in\mathbb{F}_p^N}{\textrm{arg}\min} \|\mathbf{y}^t-(\mathbf{c}\otimes(|h_1|,\ldots ,|h_n|))^t-(\mathbf{z}\mathbf{P}')^t\|^2.\,\,\,\,\,\,
\end{IEEEeqnarray*}
Next, the decoded lattice vector is $\hat{\mathbf{x}}=\hat{\mathbf{c}}\otimes \mathbf{1}_n+\hat{\mathbf{z}}(\mathbf{I}_N\otimes\mathbf{P})$, in which $\mathbf{1}_n$ denotes the all-one vector of length $n$.
By splitting $\mathbf{z}$ and $\mathbf{y}$ to $N$ vectors $\mathbf{z}_1,\ldots,\mathbf{z}_N$ and $\mathbf{y}_1,\ldots,\mathbf{y}_N$ each of length $n$, $\hat{\mathbf{x}}$ can  be written as
\begin{IEEEeqnarray}{rCl}\label{ML_split22}
% \|\mathbf{y}^t-(\mathbf{z}\mathbf{M}_{\Lambda})^t\|^2&=&
%  \underset{\mathbf{z}\in\mathbb{Z}^{nN},\mathbf{c}\in\mathbb{F}_p^N}{\textrm{arg}\min} \|\mathbf{y}^t-(\mathbf{z}\mathbf{P}')^t\|^2 \\
\bigoplus_{i=1}^N  \underset{\mathbf{z}_i\in\mathbb{Z}^{n},c_i\in\mathbb{F}_p}{\textrm{arg}\min} \| \mathbf{y}_i^t-c_i(|h_1|,\ldots ,|h_n|)^t-\mathbf{H_F}\mathbf{P}^t\mathbf{z}_i^t\|^2,\,\,\,\,\,\,
\end{IEEEeqnarray}
where $\oplus$ denotes the concatenation of $N$ vectors  given afterwards as $\hat{\mathbf{x}}_i=\hat{c}_i\mathbf{1}_n+\hat{\mathbf{z}}_i\mathbf{P}$, in which
\begin{IEEEeqnarray*}{rCl}
(\hat{\mathbf{z}}_i,\hat{c}_i)=\underset{\mathbf{z}_i\in\mathbb{Z}^{n},c_i\in\mathbb{F}_p}{\textrm{arg}\min} \| \mathbf{y}_i^t-c_i(|h_1|,\ldots ,|h_n|)^t-\mathbf{H_F}\mathbf{P}^t\mathbf{z}_i^t\|^2.
\end{IEEEeqnarray*}
 Instead of finding the minimum of $ \|\mathbf{y}^t-(\mathbf{z}\mathbf{M}_{\Lambda})^t\|^2$ for $\mathbf{z}\in\mathbb{Z}^{nN}$, which is an ML decoding in dimension $nN$, our algorithm solves the minor minimization problems in~(\ref{ML_split22}) which are $pN$ instances of ML decoding in dimension $n$. Since all these ML decoding instances provide diversity $n$, their point error probability is upper bounded by $\gamma^{-n}$ and error happens in their concatenation if error happens in at least one of them. Hence, the point error probability of our decoder is upper bounded by $N\gamma^{-n}$ which admits diversity $n$.
\end{IEEEproof}
\textcolor{mycolor3}{
\begin{remark}
According to our simulation results in Section~\ref{Numerical_Results}, the iterative and non-iterative decoding algorithms have comparable error performance. The complexity of the algorithms are also comparable. Moreover, the non-iterative algorithm works for non-binary and non-LDPC codes which are lacked for iterative algorithm. The question arises spontaneously: in this context, are there reasons to prefer the iterative algorithm? The answer to this question maybe found in the future. Indeed, the proposed decoding algorithms in this paper can be easily generalized for coset codes based on arbitrary lattices $\Gamma'\subset\Gamma$. In this case, the role of sphere decoder would be played by the decoder of $\Gamma'$. Hence, if we could find an appropriate sub-lattice $\Gamma'$ with iterative decoding, it seems that a family of coset codes with fully iterative decoding algorithm over BF channels can be obtained.  The fully iterative decoding algorithm  of such coset codes can outperform our non-iterative algorithm in terms of complexity.
\end{remark}
}
\section{Numerical Results}\label{Numerical_Results}
\begin{figure}[ht]
\begin{center}
%\vspace{-1cm}
\includegraphics[trim={1.4cm 0 1.3cm 1cm},clip,width=5in]{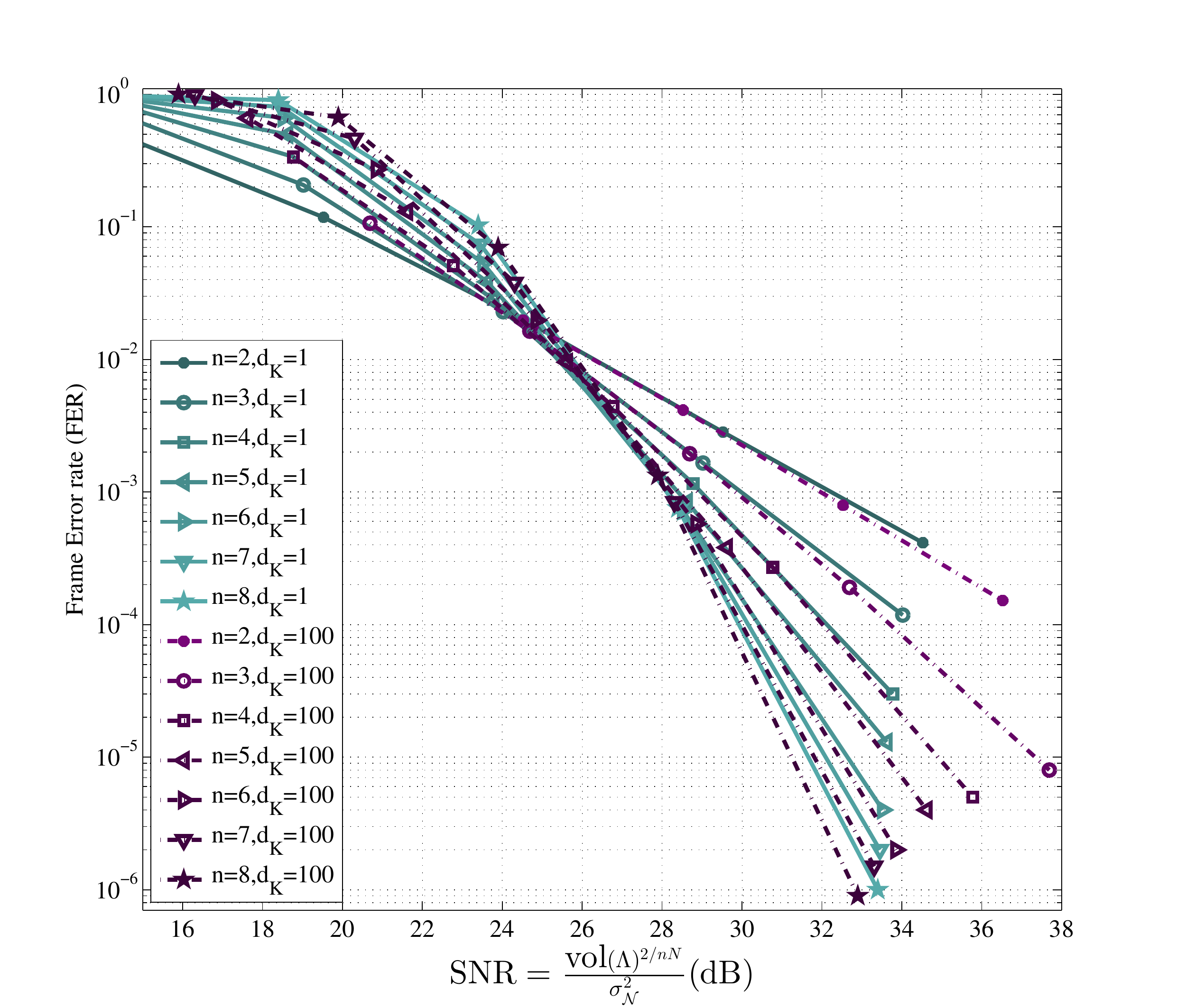}
%\caption{\small{Decoding performance of double-diversity   algebraic LDPC lattices in dimensions
%$n = 200$ and $n=1000$. The plot also shows Poltyrev outage limits.}}
\caption{\small{Poltyrev outage limit for   algebraic LDPC lattices with $[N,k]=[100,50]$ and different diversity orders.}}
\label{figsim0}
\vspace{-0.5cm}
\end{center}
\end{figure}

In this section, we present numerical results of simulating full-diversity Construction A lattices for BF channels.
%We illustrate the results on SER curves for following scenario.
%Assume that $50\%$ of the information integers are zero for resolution and vestigial information.
In the binary cases in which iterative decoding has been used, randomly generated MacKay LDPC codes \cite{mackey} with parity-check matrices of size $45\times 50$, $50\times 100$, $90\times 100$,  and $250\times 500$  are used in our simulations. Frame error rate (FER) performance of   all lattices are plotted versus SNR $\gamma=\mathrm{vol}(\Lambda)^{2/nN}/\sigma_{\mathcal{N}}^2$. We have compared the obtained results with the proposed Poltyrev outage limit (POL) in \cite{outage}. This POL is related to the fading distribution and determinant of the lattice which itself is related to $d_K$ and the rate of its underlying code. The Poltyrev outage limit of full-diversity   algebraic LDPC lattices with different parameters and diversity orders are plotted in  \figurename~\ref{figsim0}.

\begin{figure}[ht]
\begin{center}
%\vspace{-1cm}
%\includegraphics[width=3.5in]{Performance2.pdf}
\includegraphics[trim={1.4cm 0 1.3cm 1cm},clip,width=5in]{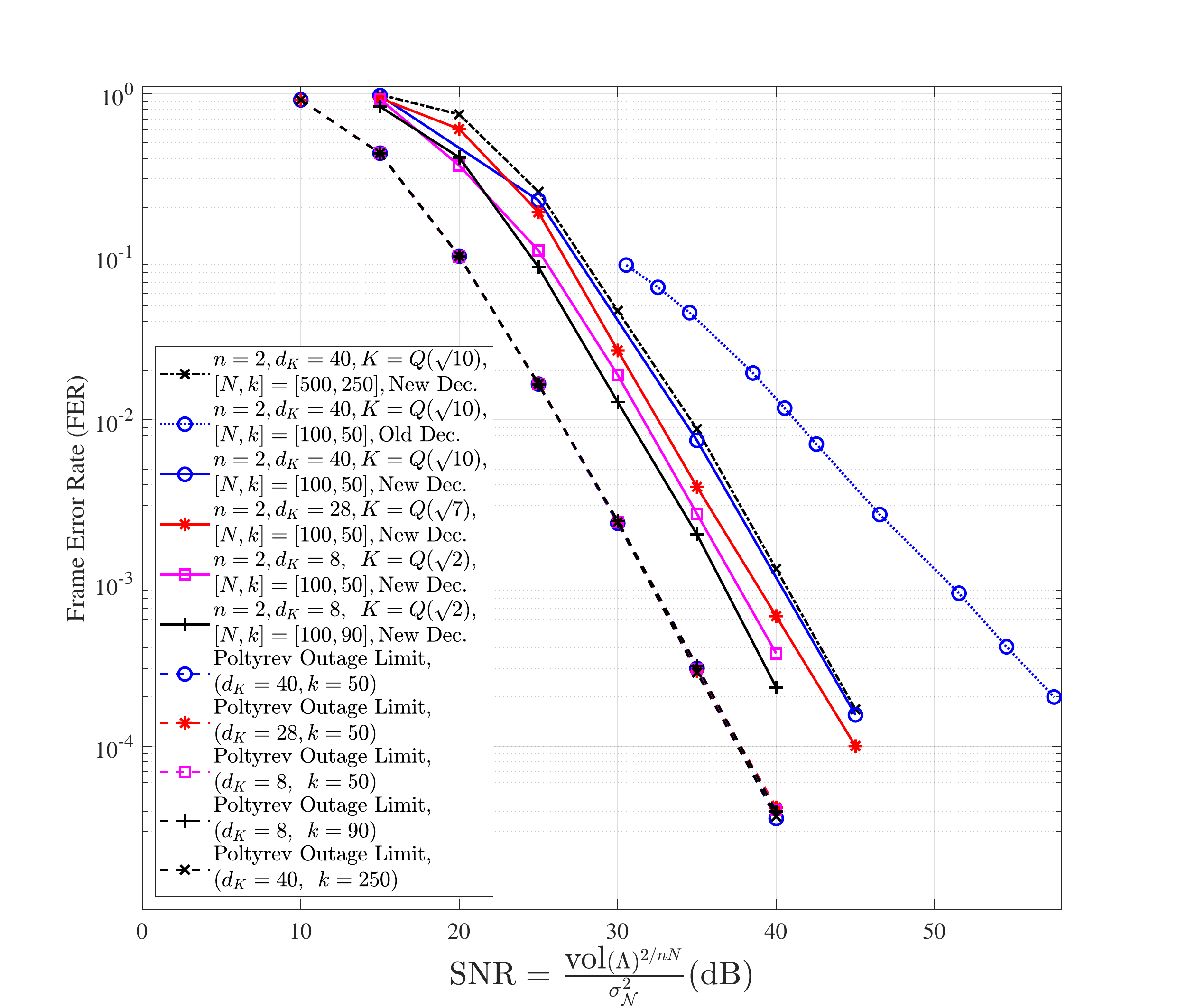}
%\caption{\small{Decoding performance of double-diversity   algebraic LDPC lattices in dimensions
%$n = 200$ and $n=1000$. The plot also shows Poltyrev outage limits.}}
\caption{\small{Decoding of double-diversity   algebraic LDPC lattices and comparison with the proposed decoding algorithm in \cite{myISIT}.}}
\label{figsim1}
\vspace{-0.5cm}
\end{center}
\end{figure}

In~\figurename~\ref{figsim1}, decoding of double-diversity algebraic LDPC lattices and comparison with the proposed decoding algorithm in \cite{myISIT} are presented.
In simulations we have used the construction of Theorem~\ref{main_theorem} with $m=10,7,2$ and the  decoding algorithm proposed in  Section~\ref{decod2}.
For $m=10$, $m\equiv 2 \pmod 4$, and $d_K=4m=40$. Here, $K=\mathbb{Q}(\sqrt{10})$, $\mathcal{O}_K=\mathbb{Z}[\sqrt{10}]$ and the prime ideal is $\mathfrak{P}=2\mathcal{O}_K+\sqrt{10}\mathcal{O}_K$. In this case, the integral basis of $\mathfrak{P}$ is $\left\{2,\sqrt{10}\right\}$ and the matrix $\mathbf{P}$  in Algorithm~\ref{decoding_step1} which was denoted by $\mathbf{DM}$ in~(\ref{Gamma_C_gen}) is
\begin{IEEEeqnarray}{rCl}\label{P_mat_Ex1}
\mathbf{P}=\left[
             \begin{array}{cc}
               2 & 2 \\
               \sqrt{m} & -\sqrt{m} \\
             \end{array}
           \right].
\end{IEEEeqnarray}
For $m=7$, $m\equiv 3 \pmod 4$, and $d_K=4m=28$. Here, $K=\mathbb{Q}(\sqrt{7})$, $\mathcal{O}_K=\mathbb{Z}[\sqrt{7}]$ and the prime ideal is $\mathfrak{P}=2\mathcal{O}_K+(\sqrt{7}+1)\mathcal{O}_K$. In this case, the integral basis of $\mathfrak{P}$ is $\left\{2,1+\sqrt{7}\right\}$ because  each element of $\mathfrak{P}$ has the form $x=2(a+b\sqrt{7})+(c+d\sqrt{7})(1+\sqrt{7})$, for $a,b,c,d\in\mathbb{Z}$. It can be checked that $x$ can also be written as follows:
$$x=(c+d+2b)(1+\sqrt{7})+(6d+2a-2b).$$
Hence, $x$ can be generated by $\left\{2,1+\sqrt{7}\right\}$ as a $\mathbb{Z}$-basis.  The matrix $\mathbf{P}$ in this case is
\begin{IEEEeqnarray}{rCl}\label{P_mat_Ex2}
\mathbf{P}=\left[
             \begin{array}{cc}
               2 & 2 \\
               1+\sqrt{m} & 1-\sqrt{m} \\
             \end{array}
           \right].
\end{IEEEeqnarray}
\textcolor{mycolor3}{For $m=2$, $m\equiv 2 \pmod 4$, and $d_K=4m=8$. Here, $K=\mathbb{Q}(\sqrt{2})$, $\mathcal{O}_K=\mathbb{Z}[\sqrt{2}]$ and the desired prime ideal is $\mathfrak{P}=2\mathcal{O}_K+\sqrt{2}\mathcal{O}_K$. In this case, the integral basis of $\mathfrak{P}$ is $\left\{2,\sqrt{2}\right\}$ because  each element $x$ of $\mathfrak{P}$ has the form $x=2(a+b\sqrt{2})+(c+d\sqrt{2})\sqrt{2}$, for $a,b,c,d\in\mathbb{Z}$, and it can also be written as follows:
$$x=2(d+a)+(2b+c)\sqrt{2}.$$
The matrix $\mathbf{P}$ in this case is of form given in~(\ref{P_mat_Ex1}).}

In~\figurename~\ref{figsim1}, at FER of $10^{-4}$, the double-diversity algebraic LDPC lattice  based on $\mathbb{Q}(\sqrt{10})$ with $[N,k]=[100,50]$ performs 8.45dB away from its corresponding POL. Using the decoder proposed in \cite{myISIT}, this lattice performs 21.6dB away from its corresponding POL. This indicates 13.15dB improvement compared to the previous decoder of full-diversity algebraic LDPC lattices in \cite{myISIT}. The double-diversity algebraic LDPC lattice  based on $\mathbb{Q}(\sqrt{7})$ with $[N,k]=[100,50]$ performs 7.25dB away from its corresponding POL which outperforms the one based on $\mathbb{Q}(\sqrt{10})$ by $1.2$dB. This better performance is caused by lower discriminant of  $\mathbb{Q}(\sqrt{7})$ compared to $\mathbb{Q}(\sqrt{10})$ which is in accordance with our expectations (see Section~\ref{Design_sec}). Among all quadratic number fields, $\mathbb{Q}(\sqrt{5})$ has the least positive discriminant.
\textcolor{mycolor3}{Unfortunately, the minimal polynomial $x^2-x-1$ of  $(1+\sqrt{5})/2$, which is the generator of the integers ring of $\mathbb{Q}(\sqrt{5})$, has no linear factor after reduction modulo $2$. Hence, it is not possible to employ this number field to obtain any full-diversity binary Construction A lattice. After $\mathbb{Q}(\sqrt{5})$, $\mathbb{Q}(\sqrt{2})$ has the least positive discriminant which is $8$. We see in~\figurename~\ref{figsim1} that the algebraic LDPC lattice  based on $\mathbb{Q}(\sqrt{2})$ with $[N,k]=[100,50]$ performs 5.5dB away from its corresponding POL. According to our provided design paradigms in Section~\ref{Design_sec}, we can further  improve the performance by increasing the rate of the underlying code. In~\figurename~\ref{figsim1}, the double-diversity algebraic LDPC lattice  based on $\mathbb{Q}(\sqrt{2})$ with $[N,k]=[100,90]$ performs 4.2dB away from its corresponding POL.} In all simulations, the full-diversity property of the iterative decoder proposed in this paper has been verified. Another result in~\figurename~\ref{figsim1} is the FER of  an algebraic LDPC lattice  based on $\mathbb{Q}(\sqrt{10})$ with $[N,k]=[500,250]$ which performs 8.6dB away from its corresponding POL. Hence, by increasing the dimension from 200 to 1000, 0.15dB loss in the performance happens which is quite natural in BF channels.
%The results for dimension $200$ are provided in presence and absence of multiplying by $\mathbf{R}$ in Algorithm~\ref{decoding_step1}.
%\vspace{-1cm}

\begin{figure}[ht]
\begin{center}
%\vspace{-1cm}
%\includegraphics[width=3.5in]{Performance3.pdf}
\includegraphics[trim={1.4cm 0 1.3cm 1cm},clip,width=5in]{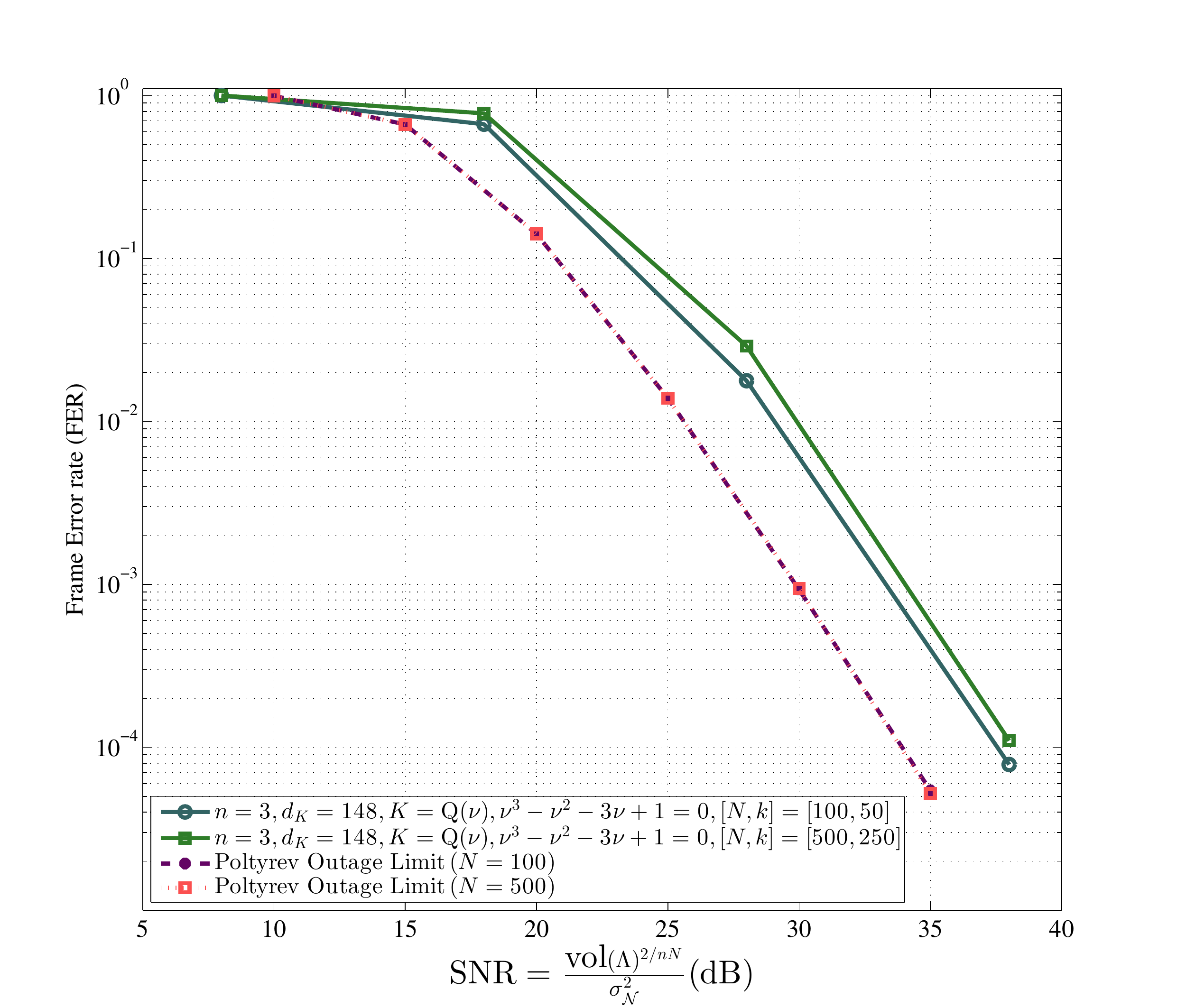}
%\caption{\small{Decoding performance of double-diversity   algebraic LDPC lattices in dimensions
%$n = 200$ and $n=1000$. The plot also shows Poltyrev outage limits.}}
\caption{\small{Decoding of triple-diversity   algebraic LDPC lattices.}}
\label{figsim2}
\vspace{-0.5cm}
\end{center}
\end{figure}

In \figurename~\ref{figsim2} we  present the FER performance of triple-diversity   algebraic LDPC lattices, obtained from Example~\ref{example_div_3} by employing $[100,50]$ and $[500,250]$ binary LDPC codes as underlying code. The POL with diversity order $3$ is plotted for comparison. Due to the results of \figurename~\ref{figsim2}, triple diversity   algebraic LDPC lattices indicate diversity order $3$ under the proposed iterative decoding algorithm in Section~\ref{decod2}, that confirms the proven result in Section~\ref{decoding_analysis}. For  $[N,k]=[100,50]$, the triple-diversity   algebraic LDPC lattice performs 3.65dB away from its corresponding POL. This error performance can be improved by considering underlying codes with higher rates and number fields with lower discriminant. In dimension 1500, that is, for  $[N,k]=[500,250]$, the triple-diversity   algebraic LDPC lattice performs 4.3dB away from its corresponding POL.

\begin{figure}[ht]
\begin{center}
%\vspace{-1cm}
%\includegraphics[width=3.5in]{Performance3.pdf}
\includegraphics[trim={1.4cm 0 1.3cm 1cm},clip,width=5in]{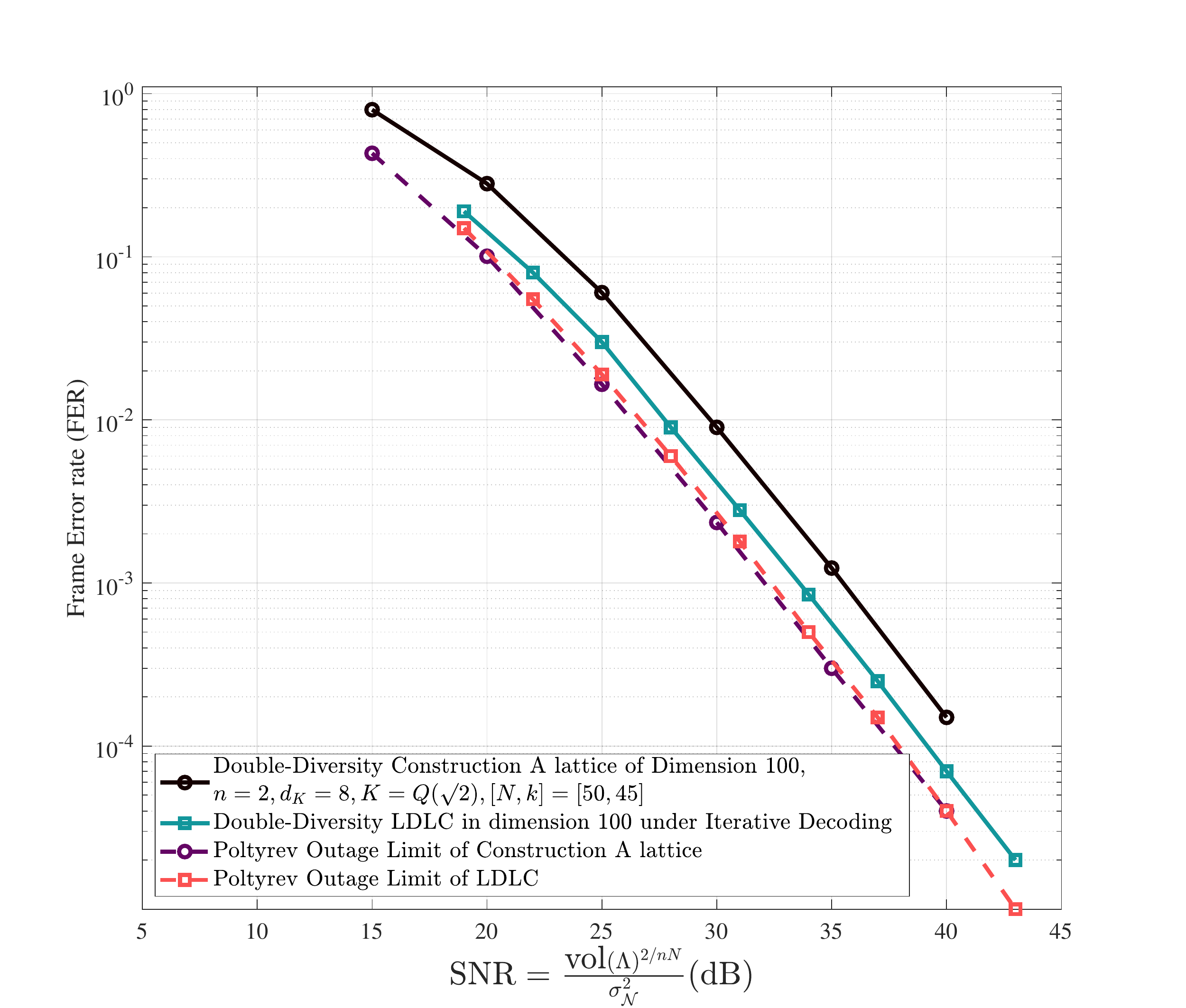}
%\caption{\small{Decoding performance of double-diversity   algebraic LDPC lattices in dimensions
%$n = 200$ and $n=1000$. The plot also shows Poltyrev outage limits.}}
\caption{\small{Comparison between error performance of a Construction A lattice under iterative decoding algorithm and an LDLC of dimension $100$ with diversity order $2$. }}
\label{figsim3}
\vspace{-0.5cm}
\end{center}
\end{figure}

In~\figurename~\ref{figsim3}, the comparison between the FER of a double-diversity algebraic LDPC lattice under iterative decoding and the FER of a low-density lattice code (LDLC) of dimension $100$ with diversity order $2$ is provided \cite{punekar}. \textcolor{mycolor3}{The full-diversity algebraic LDPC lattice is based on $\mathbb{Q}(\sqrt{2})$ and its underlying code is a binary $[50,45]$ LDPC code.  According to the results of \figurename~\ref{figsim3}, at FER of $10^{-4}$, LDLC performs 1.35dB away from its corresponding POL and LDPC performs 3.4dB away from its corresponding POL.}

\begin{figure}[ht]
\begin{center}
%\vspace{-1cm}
%\includegraphics[width=3.5in]{Performance3.pdf}
\includegraphics[trim={1.4cm 0 1.3cm 1cm},clip,width=5in]{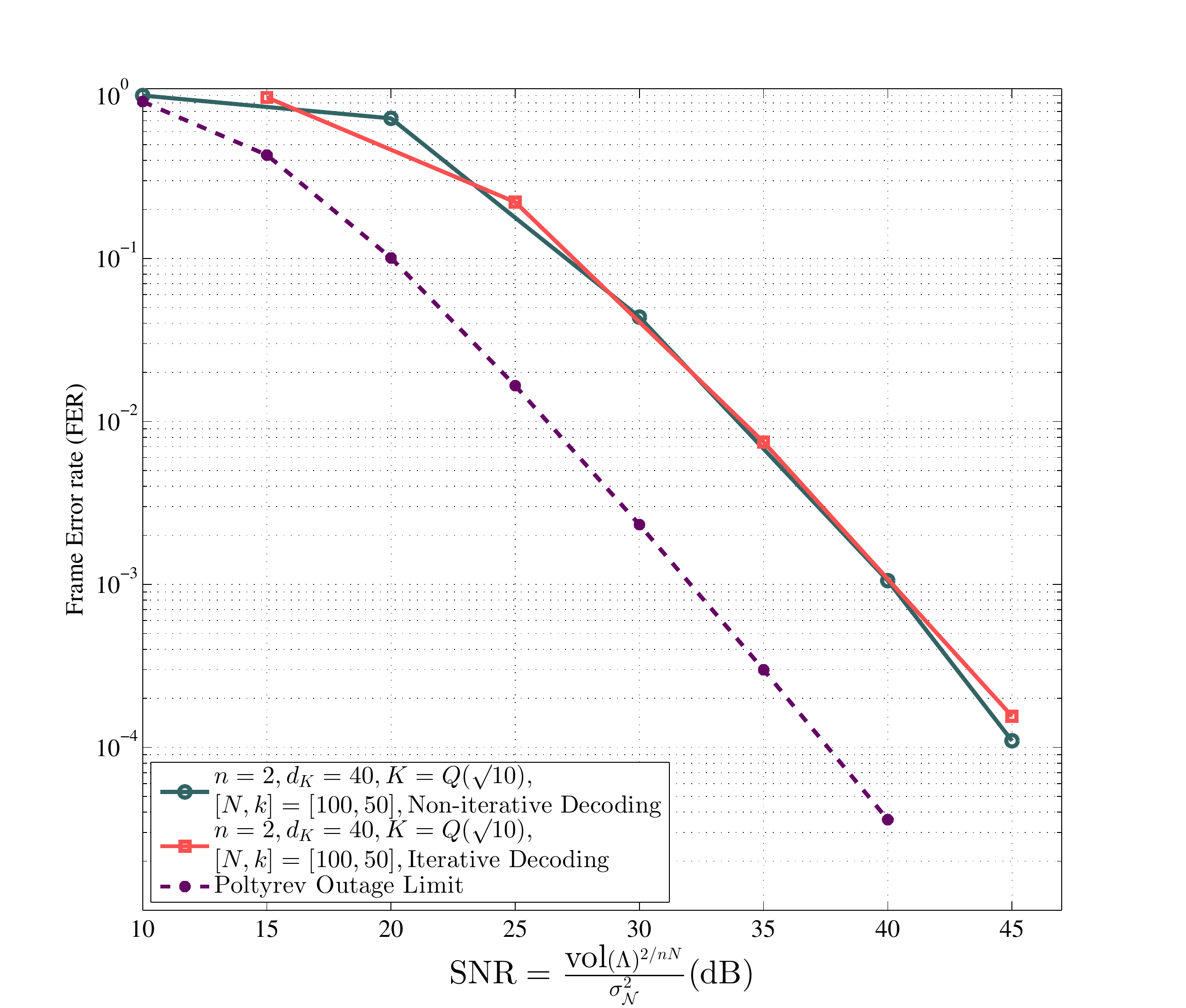}
%\caption{\small{Decoding performance of double-diversity   algebraic LDPC lattices in dimensions
%$n = 200$ and $n=1000$. The plot also shows Poltyrev outage limits.}}
\caption{\small{Comparison between error performance of a Construction A lattice with dimension $200$ and diversity order $2$ under iterative and non-iterative decoding. }}
\label{figsim4}
\vspace{-0.5cm}
\end{center}
\end{figure}

In~\figurename~\ref{figsim4}, the comparison between the FER of a double-diversity algebraic LDPC lattice under iterative decoding and non-iterative decoding is provided. The considered full-diversity algebraic LDPC lattice is based on $\mathbb{Q}(\sqrt{10})$ with $[N,k]=[100,50]$.  According to the results of \figurename~\ref{figsim4}, at FER of $10^{-4}$, iterative algorithm performs 8.45dB away from POL and non-iterative algorithm performs 7.7dB away from POL. Thus, non-iterative decoding outperforms the iterative decoding by 0.75dB.

\begin{figure}[ht]
\begin{center}
%\vspace{-1cm}
%\includegraphics[width=3.5in]{Performance3.pdf}
\includegraphics[trim={1.4cm 0 1.3cm 1cm},clip,width=5in]{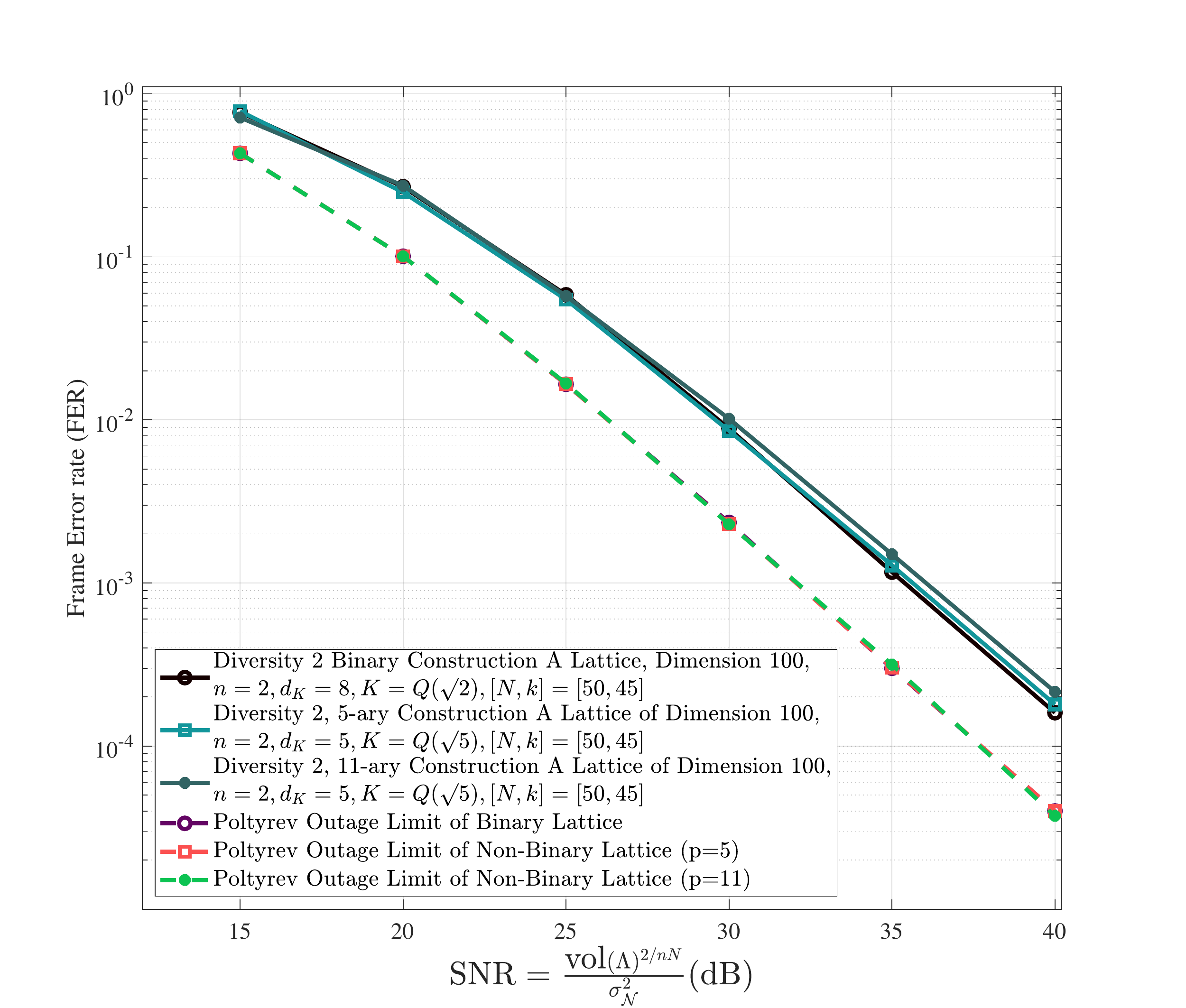}
%\caption{\small{Decoding performance of double-diversity   algebraic LDPC lattices in dimensions
%$n = 200$ and $n=1000$. The plot also shows Poltyrev outage limits.}}
\caption{\small{Comparison between error performance of non-binary and binary Construction A lattices with dimension $100$ and diversity order $2$ under non-iterative decoding algorithm.}}
\label{figsim5}
\vspace{-0.5cm}
\end{center}
\end{figure}

In~\figurename~\ref{figsim5}, the comparison between the FER of double-diversity binary and non-binary Construction A lattices under  non-iterative decoding is provided. \textcolor{mycolor3}{Both non-binary lattices are based on $\mathbb{Q}(\sqrt{5})$ with $[N,k]=[50,45]$ and they uses $5$-ary and $11$-ary linear codes as their underlying codes. Let $\theta=\frac{\sqrt{5}+1}{2}$. Then, the prime ideal considered to obtain the lattice based on the $5$-ary code is $\mathfrak{P}_1=5\mathcal{O}_K+(3-\theta)\mathcal{O}_K$ which has the $\mathbb{Z}$-basis $\left\{5,\theta+2\right\}$. The prime ideal considered to obtain the lattice based on the $11$-ary code is $\mathfrak{P}_2=11\mathcal{O}_K+(4-\theta)\mathcal{O}_K$ which has the $\mathbb{Z}$-basis $\left\{11,\theta+7\right\}$. The binary lattice is based on $\mathbb{Q}(\sqrt{2})$.  According to the results of \figurename~\ref{figsim5}, at FER of $10^{-4}$, the binary lattice performs 3.3dB away from POL, the non-binary $5$-ary lattice performs 3.68dB away from POL and the non-binary $11$-ary lattice performs 4.07dB away from POL. In order to make a fair comparison, the binary and non-binary lattices should be based on the same number field. The number field $\mathbb{Q}(\sqrt{2})$ which is employed to obtain binary lattice, has higher discriminant compared to $\mathbb{Q}(\sqrt{5})$ which makes its performance potentially weaker. Nevertheless, the binary Construction A lattice outperforms the non-binary $5$-ary one by 0.38dB. Moreover, the $5$-ary lattice outperforms the $11$-ary lattice about $0.4$dB.}


\begin{figure}[ht]
\begin{center}
%\vspace{-1cm}
%\includegraphics[width=3.5in]{Performance3.pdf}
\includegraphics[trim={1.4cm 0 1.3cm 1cm},clip,width=5in]{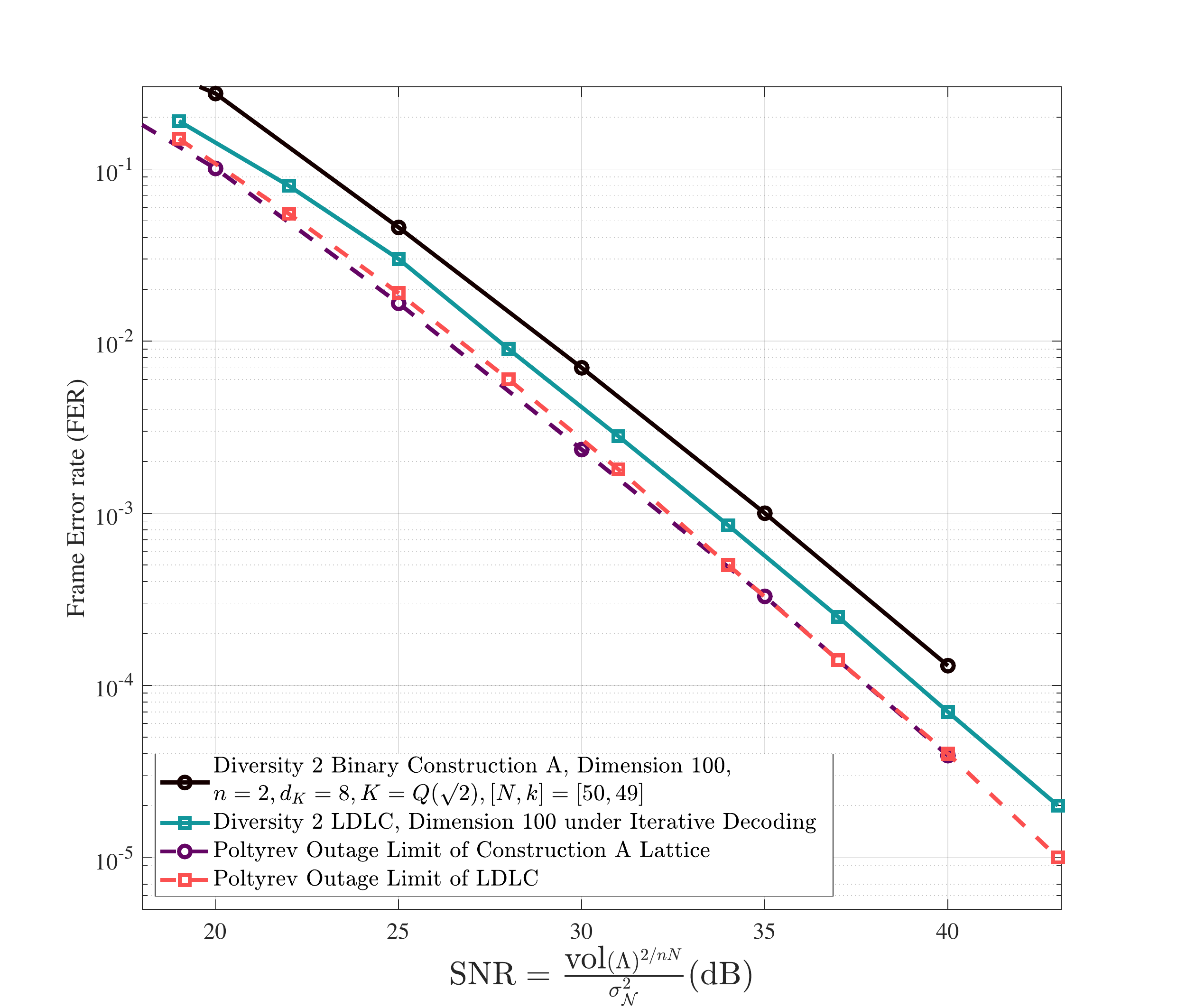}
%\caption{\small{Decoding performance of double-diversity   algebraic LDPC lattices in dimensions
%$n = 200$ and $n=1000$. The plot also shows Poltyrev outage limits.}}
\caption{\small{Comparison between error performance of a binary Construction A lattice under non-iterative decoding algorithm and an LDLC of dimension $100$ with diversity order $2$. }}
\label{figsim6}
\vspace{-0.5cm}
\end{center}
\end{figure}

In~\figurename~\ref{figsim6}, the comparison between the FER of a double-diversity binary Construction A lattice under non-iterative decoding and the FER of an LDLC of dimension $100$ with diversity order $2$ is provided \cite{punekar}. \textcolor{mycolor3}{The full-diversity Construction A lattice is based on $\mathbb{Q}(\sqrt{2})$ and its underlying code is a binary $[50,49]$ random code.  According to the results of \figurename~\ref{figsim6}, at FER of $10^{-4}$, LDLC performs 1.35dB away from its corresponding POL and Construction A lattice performs 2.82dB away from its corresponding POL.}

\begin{figure}[ht]
\begin{center}
\vspace{-1cm}
\includegraphics[trim={1.4cm 0 1.3cm 1cm},clip,width=5in]{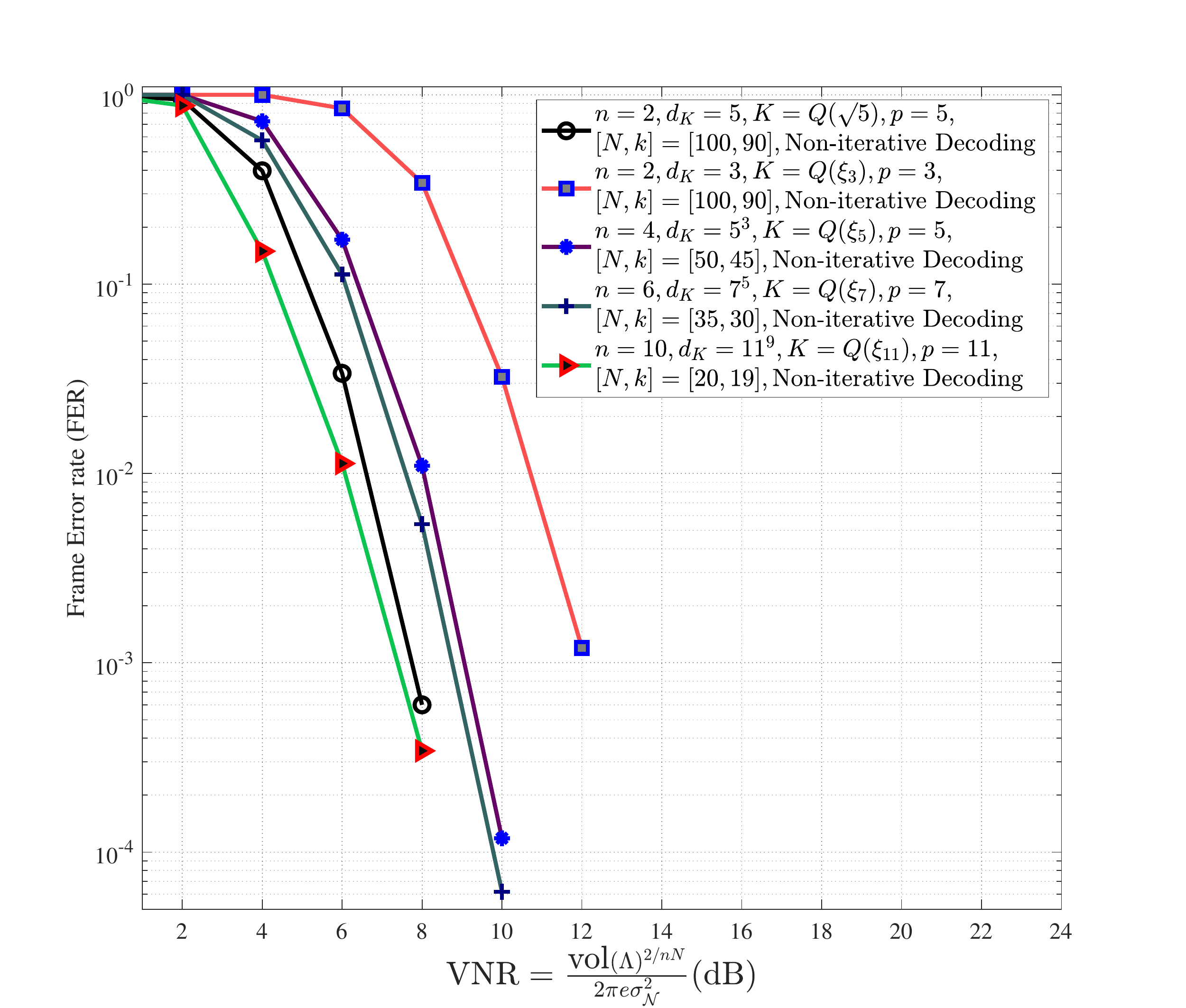}
%\caption{\small{Decoding performance of double-diversity   algebraic LDPC lattices in dimensions
%$n = 200$ and $n=1000$. The plot also shows Poltyrev outage limits.}}
\caption{\small{Comparison between frame error rate of Construction A lattices based on totally real and totally complex number fields under non-iterative decoding algorithm over AWGN channel. }}
\label{figsim7}
\vspace{-0.5cm}
\end{center}
\end{figure}

\textcolor{mycolor3}{In~\figurename~\ref{figsim7}, the comparison between the FER versus volume to noise ratio (VNR) performance of  Construction A lattices based on totally real and totally complex number fields under non-iterative decoding over AWGN channel is provided. The employed totally complex number fields are $\mathbb{Q}(\xi_3)$, $\mathbb{Q}(\xi_5)$, $\mathbb{Q}(\xi_7)$ and $\mathbb{Q}(\xi_{11})$. For each prime number $p$, the cyclotomic number field $K=\mathbb{Q}(\xi_p)$ is monogenic of degree $n=p-1$, with discriminant $p^{p-2}$ and its ring of integers is $\mathcal{O}_K=\mathbb{Z}[\xi_p]$. For $p=3,5,7,11$, we employ the prime ideals of the form $\mathfrak{P}=(1-\xi_p)$ and random $p$-ary linear codes to obtain the simulated examples in \figurename~\ref{figsim7}. All considered examples in this figure are roughly of dimension $200$. We observe that by increasing the discriminant of totally complex cyclotomic fields, one can obtain a better performance. This is natural since the dimension of the lattice based on the prime ideal $\mathfrak{P}$ is $p-1$ which increases by increasing the discriminant. Since the decoding of $\mathfrak{P}$ is somehow  an ML decoding, by keeping the dimension fixed, the overall behaviour of decoding tends to ML decoding when $p$ increases. Indeed, since the dimension $N(p-1)$ is assumed to be 200 in our simulations, by increasing $p$, $N$ approaches 1. The penalty of increasing $p$ will appear to be exponentially in the decoding complexity. It should be noted that the volume of a Construction A lattice $\Lambda$ based on $K=\mathbb{Q}(\xi_p)$ and an $[N,k]$, $p$-ary code $\mathcal{C}$ is $2^{-nN/2}d_K^{N/2}p^{N-k}$ \cite{Xialo}. Hence, the VNR in the case of using the totally complex cyclotomic number field  $\mathbb{Q}(\xi_p)$ over an AWGN channel with variance $\sigma_{\mathcal{N}}^2$ per dimension is
\begin{IEEEeqnarray*}{rCL}
\mathrm{VNR}&=&\frac{\mathrm{vol}(\Lambda)^{2/nN}}{2\pi e\sigma_{\mathcal{N}}^2}=\frac{0.5d_K^{1/n}p^{2(N-k)/nN}}{2\pi e\sigma_{\mathcal{N}}^2}\\
&=& \frac{p^{\frac{p-2+2(1-R)}{p-1}}}{4\pi e\sigma_{\mathcal{N}}^2}=\frac{p^{\frac{p-2R}{p-1}}}{4\pi e\sigma_{\mathcal{N}}^2},
\end{IEEEeqnarray*}
where $R=k/N$ is the rate of the underlying code $\mathcal{C}$. The FER performance of a $5$-ary Construction A lattice based on the totally real number field $\mathbb{Q}(\sqrt{5})$ is also provided in this figure. In terms of FER, the lattice based on totally real number field $\mathbb{Q}(\sqrt{5})$ outperforms all the ones based on totally complex number fields except the one based on $\mathbb{Q}(\xi_{11})$ which has 0.35dB better performance in the FER of $10^{-3}$.}

\begin{figure}[ht]
\begin{center}
%\vspace{-1cm}
%\includegraphics[width=3.5in]{Performance3.pdf}
\includegraphics[trim={1.4cm 0 1.3cm 1cm},clip,width=5in]{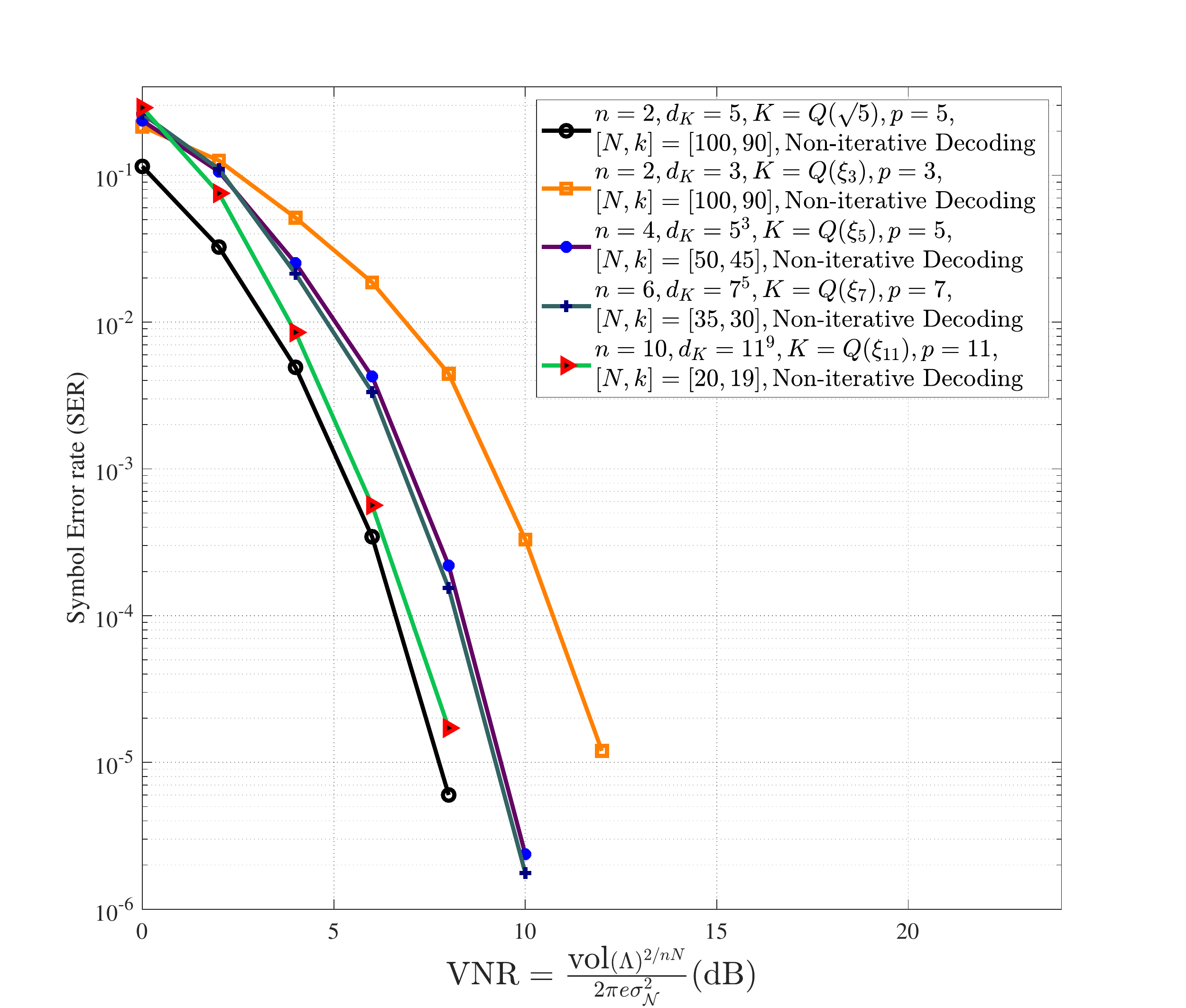}
%\caption{\small{Decoding performance of double-diversity   algebraic LDPC lattices in dimensions
%$n = 200$ and $n=1000$. The plot also shows Poltyrev outage limits.}}
\caption{\small{Comparison between symbol error rate of Construction A lattices based on totally real and totally complex number fields under non-iterative decoding algorithm over AWGN channel. }}
\label{figsim8}
\vspace{-0.5cm}
\end{center}
\end{figure}

\textcolor{mycolor3}{In~\figurename~\ref{figsim8}, the same comparisons are provided in terms of symbol error rate (SER). In this case, the lattice based on totally real number field outperforms all the ones based on totally complex number fields with much lower decoding complexity. More specifically, at the SER of $2\times 10^{-5}$, the lattice based on $\mathbb{Q}(\sqrt{5})$ has 0.5dB better performance compared to the one based on $\mathbb{Q}(\xi_{11})$.}

\textcolor{Mycolor2}{The authors of \cite{punekar} have employed the decoding algorithm  of LDLCs proposed in \cite{LDLC} which has complexity $O(n\cdot d\cdot t\cdot \frac{1}{\Delta}\cdot \log_2 (\frac{1}{\Delta}))$, where $\Delta$ is the resolution; its typical value using the considered parameters of \cite{punekar}  is $1/64$ (selected pdf length which is denoted by $L$ in \cite{LDLC}, is $2^{16}$ and FFT size which is denoted by $D$ in \cite{LDLC}, is $2^{10}$ and $\frac{1}{\Delta}=\frac{L}{D}$). Here, $n$ is the dimension of lattice, $t$ is the number of iterations and $d$ is the
average code degree.
Regarding the various parameters involved in estimating the complexity of this decoder  that complicates a  rigorous comparison, we give a rough comparison idea
by replacing the typical values of these parameters and looking at the  numerical values.
Using the parameters of \cite{punekar}, $n=100$, $d=4$, $\frac{1}{\Delta}=64$ and $t=50$. Computing
$n\cdot d\cdot t\cdot \frac{1}{\Delta}\cdot \log_2 (\frac{1}{\Delta})$ using these parameters estimates $7680000$ computational operations in the decoding of this lattice. The complexity of our non-iterative decoder is $O(pN \cdot f(n))$ in which $f(n)$ indicates the number of searches done by sphere decoder in dimension $n$. The expected total number of points visited by the sphere decoding is proportional to the total number of lattice points inside spheres of radius $d$ and of dimensions $i=1,\ldots ,n$ \cite{hassibi}:
\begin{IEEEeqnarray}{rCl}\label{SD}
f(n)=\sum_{i=1}^{n} \frac{\pi^{i/2}}{\Gamma(i/2+1)}d^i \geq \frac{1}{\sqrt{\pi}}\alpha^{\frac{n}{2\alpha}+\frac{1}{2}}n^{\frac{1}{2\alpha}-\frac{1}{2}},
%\approx \left(\frac{2\pi e d^2}{n}\right)^{n/2} \frac{1}{\sqrt{\pi n}},
\end{IEEEeqnarray}
where $\Gamma(x)\vcentcolon= \int_{0}^{+\infty}t^{x-1}e^{-t}dt $ denotes the Gamma function and $1<\alpha\leq n$ is a number defined in \cite{hassibi}; for example we can take $\alpha=2$.
The latter inequality in~(\ref{SD})  is obtained by using Stirling's formula for the Gamma function and considering  $2\pi e d^2 \approx n^{1+\frac{1}{n}}$  in such a way that the probability of the sphere decoder finding a lattice point does not vanish to zero. Considering $d=2$ and $n=2$ in (\ref{SD}) gives $f(n)=4\pi+4/3\approx14$. We also have $p=2$ and $N=50$ and our decoding involves  $pN \cdot f(n)\approx1400$ searches each one equivalent to multiplying a vector of length $n$ by an $n\times n$ matrix. Hence,  our decoding algorithm requires  $1400\times 2^2=5600$ computational operations which is $1371$ times lesser compared to  the number of operations in the decoder of LDLCs. We recall again that this is just a rough comparison and a deep comparison involving all parameters and situations is needed before we can claim that our decoding method is preferable.}
%\vspace{-1cm}
\section{Conclusions}\label{conclusion}
In this paper, we have proposed full-diversity   Construction A lattices on BF channels, based on totally real number fields.
The framework for obtaining any diversity order is provided and examples with diversity order $2,3$ and $4$ is discussed through the paper.  In order to apply these structures in practical implementations, we have proposed two new decoding methods which have complexity growing linearly  with the dimension of the lattice. It makes the decoding of high-dimension  Construction A lattices on the BF channels tractable.
The
first decoder is proposed for full-diversity algebraic LDPC lattices which
are generalized Construction A lattices with a binary LDPC code
as underlying code. This decoding method contains iterative and
non-iterative phases. In order to implement the iterative phase
of our decoding algorithm, we have proposed the definition of a parity-check matrix and Tanner graph for full-diversity Construction
A lattices. We have proved that the constructed algebraic LDPC lattices
together with the proposed decoding method admit full diversity
 over BF channels. In the second decoding method, the iterative phase has been removed which enables
full-diversity practical decoding of all generalized Construction
A lattices without any assumption about their underlying code.
\textcolor{mycolor3}{We have also provided some insights about the design criteria of lattices for BF channels. Our simulation results indicate that  Construction A lattices obtained from binary codes and the ones based on non-binary codes have comparable error performance in BF channels. In addition, the decoding complexity in the binary case is much lower compared to the non-binary case.}
%We have proved that Construction A lattices obtained from binary codes outperform the ones based on non-binary codes in BF channels.
Since available lattice construction methods
from totally real and complex multiplication (CM) fields does not
provide diversity in the binary case, we have generalized Construction
A lattices over a wider family of number fields namely monogenic
number fields.
\section*{Acknowledgements}
The authors are grateful to the referees for their very meticulous
reading of this manuscript. Their suggestions were very helpful in creating the
improved final version.
\bibliographystyle{IEEEtran}
\nocite{*}
%\hfill \today
%\bibliography{BIB} %Trans style bib
%=============================================
%\bibliography{BIB_alphabetic}
%=============================================
%----------------------------------------------------------------------------biblography

\end{document}